%% file: JASP_CHAARI_revision2.tex
\documentclass[11pt]{bmc_article}    
\usepackage{amsmath}
\usepackage{amssymb}
\usepackage{enumerate,url,multirow,colortbl}
\usepackage{theorem}
\usepackage{graphicx,color}
\usepackage{pifont}
\usepackage[noadjust]{cite}
\usepackage{citesort}

\usepackage{algorithm,algorithmic}
\pdfoutput=1

\include{Definitions}
\newtheorem{definition}{Definition}[section]
\usepackage{xcolor}
\usepackage[normalem]{ulem}
\newcommand{\ADDED}[1]{#1}

\newcommand{\ADDEDRevtwo}[1]{#1}
\def\mSENSE{\texttt{mSENSE}}
\def\hermit{\mbox{\tiny\textsf{H}}}

\usepackage{cite} 
\usepackage{url}  
\usepackage{ifthen}  
\usepackage{multicol}   
\usepackage[utf8]{inputenc} 
\newboolean{publ}

\newenvironment{bmcformat}{\baselineskip20pt\sloppy\setboolean{publ}{false}}{\baselineskip20pt\sloppy}

\usepackage{graphicx,color}
\graphicspath{{./Figures/}}

\begin{document}
\begin{bmcformat}

\title{Spatio-temporal wavelet regularization for parallel MRI reconstruction: application to functional MRI}

\author {Lotfi CHAARI\correspondingauthor$^1$
\email{lotfi.chaari@enseeiht.fr}
, S\'ebastien M\'ERIAUX$^2$
\email{sebastien.meriaux@cea.fr}, Jean-Christophe PESQUET$^3$\email{jean-christophe.pesquet@univ-paris-est.fr} and 
Philippe CIUCIU$^2$\email{philippe.ciuciu@cea.fr}
}

\address{%
    \iid(1)University of Toulouse, IRIT - INP-ENSEEIHT, France.\\
    \iid(2)LNAO, CEA-NeuroSpin, France.\\
    \iid(3)LIGM, University Paris-Est, France.    
}%

\maketitle

\begin{abstract}
\small
Parallel MRI is a fast imaging technique that \ADDEDRevtwo{helps} in acquiring highly resolved images in space or/and in time. 
The performance of parallel 
imaging strongly depends on the reconstruction algorithm, which can proceed either in the original $k$-space~(GRAPPA, SMASH) or in the image domain~(SENSE-like 
methods). To improve the performance of the widely used SENSE algorithm, 2D- or slice-specific regularization in the wavelet domain has 
been investigated. In this paper, we extend this approach using 3D-wavelet representations
in order to handle all slices together and address reconstruction artifacts which propagate across adjacent slices. 
The gain induced by such 
extension (3D-Unconstrained Wavelet Regularized -SENSE: 3D-UWR-SENSE) is validated on anatomical image reconstruction where no temporal acquisition is considered. Another important extension accounts 
for temporal correlations that exist between successive scans 
in functional MRI~(fMRI). In addition to the case of 2D+$t$ acquisition schemes addressed by some other methods like \emph{kt}-FOCUSS, 
our approach allows to
  deal with 3D+$t$ acquisition schemes which are widely used in neuroimaging. 
The resulting 3D-UWR-SENSE and 4D-UWR-SENSE 
reconstruction schemes 
are fully \emph{unsupervised} in the sense that all regularization parameters are 
estimated in the maximum likelihood sense on a reference scan. The gain induced by such extensions is illustrated on both 
anatomical and functional 
image reconstruction, and also measured in terms of statistical sensitivity for the 4D-UWR-SENSE approach during a fast event-related fMRI protocol. 
Our 4D-UWR-SENSE algorithm outperforms the SENSE reconstruction at the subject and group levels~(15 subjects) 
for different contrasts of interest~(e.g., motor or computation tasks) and using
different parallel acceleration factors~($R=2$ and $R=4$) on
$2\times2\times3$mm$^3$ EPI images.  
\end{abstract}
\normalsize

\ifthenelse{\boolean{publ}}{\begin{multicols}{2}}{}

\section{Introduction}\label{sec:intro}

Reducing scanning time in Magnetic Resonance Imaging~(MRI) exams remains a worldwide challenging issue since it has to be achieved
while maintaining high image quality~\cite{Kochunov05b,Rabrait07}.
The expected benefits
are {\em i.)} to limit patient's exposure to the MRI environment
either for safety or discomfort reasons, {\em ii.)} to  improve acquisition robustness against subject's motion artifacts and {\em iii.)} to limit geometric distortions.
One basic idea to make MRI acquisitions faster~(or to improve spatial resolution in a fixed scanning time) consists of reducing the amount of acquired samples in the $k$-space~(spatial Fourier domain) and developing dedicated reconstruction pipelines. To achieve this goal, three main research avenues have been developed so far:
\begin{itemize}
\item {\em parallel imaging} or parallel MRI that relies on a geometrical principle involving multiple receiver coils with complementary sensitivity profiles. 
This enables the $k$-space undersampling along the phase encoding direction without degrading spatial resolution or truncating the Field-Of-View~(FOV). pMRI requires the unfolding of reduced FOV coil-specific images to reconstruct the full FOV 
image~\cite{Sodickson_D_97,pruessmann_99,griswold_02}.

\item {\em Compressed Sensing~(CS) MRI} that exploits three ingredients: \emph{sparsity} of MR images in wavelet bases, the \emph{incoherence} between Fourier and inverse wavelet bases which allows to randomly undersample $k$-space and the \emph{nonlinear recovery} of MR images by solving a convex but nonsmooth $\ell_1$ minimization 
problem~\cite{Candes_06,Lustig07}.
This approach remains usable with classical receiver coil but
can also be combined with parallel MRI~\cite{Liang09,Boyer12}.

\item In the dynamic MRI context, fast parallel acquisition schemes have been proposed to increase the acquisition rate by reducing the amount of acquired $k$-space samples in each frame using interleaved partial $k$-space sampling between successive frames~(UNFOLD 
approach~\cite{Madore99}). To further reduce the scanning time, a strategy named $kt$-BLAST taking advantage of both the spatial~(actually in the $k$-space) and temporal correlations between successive scans in the dataset has been pushed 
forward~\cite{Tsao03}. 
\end{itemize}

In parallel MRI~(pMRI), many reconstruction methods like SMASH~(Simultaneous Acquisition of Spatial Harmonics)~\cite{Sodickson_D_97}, GRAPPA (Generalized Autocalibrating Partially Parallel 
Acquisitions)~\cite{griswold_02} and SENSE~(Sensitivity Encoding)~\cite{pruessmann_99} have been proposed in the literature to reconstruct a full FOV image from multiple $k$-space undersampled images acquired on separate channels. The main difference between them lies in the space on which they operate. GRAPPA performs multichannel full FOV reconstruction in the $k$-space domain whereas SENSE carries out the unfolding process in the image domain: all undersampled images are first reconstructed by inverse Fourier transform before combining them to unwrap the full FOV image. Also, GRAPPA is autocalibrated, whereas SENSE needs a separate coil sensitivity estimation step based on a reference scan. Note however that autocalibrated versions of SENSE are now available such that the \mSENSE algorithm on Siemens scanners.

In the dynamic MRI context, combined strategies mixing parallel imaging and accelerated sampling schemes along the temporal axis have also been investigated. The corresponding reconstruction algorithms have been referenced to as 
\emph{kt}-SENSE~\cite{Tsao03,Tsao05}, \emph{kt}-GRAPPA~\cite{Huang05}. Compared to \mSENSE where the centre of the $k$-space is acquired only once at the beginning, these methods have to acquire the central $k$-space area at each repetition time, which decreases the acceleration factor. More recently, optimized versions of \emph{kt}-BLAST and \emph{kt}-SENSE reconstruction algorithms referenced to as 
\emph{kt}-FOCUSS~\cite{Jung07,Jung09} have been designed to combine the CS theory in space with Fourier or alternative transforms along the time axis. They enable to further reduce data acquisition time without significantly compromising image quality, provided that the image sequence exhibits a high degree of spatio-temporal correlation, either by nature or by design. Typical examples that enter in this context are \emph{i.)} dynamic MRI capturing an organ~(liver, kidney, heart) during a quasi-periodic motion due to the respiratory cycle and cardiac beat and \emph{ii.)} functional MRI based on periodic blocked design.


However, this interleaved partial $k$-space sampling cannot be exploited in aperiodic dynamic acquisition schemes like in resting state fMRI (rs-fMRI) or during fast-event related fMRI
paradigms~\cite{Damoiseaux06,Dale99}. In rs-fMRI, spontaneous brain activity is recorded without any experimental design in order to probe intrinsic functional 
connectivity~\cite{Damoiseaux06,Varoquaux10,Ciuciu12}. In fast event-related designs, the presence of jittering combined with random delivery of stimuli introduces a trial-varying delay between the stimulus and acquisition time 
points~\cite{Birn02}. This prevents the use of an interleaved $k$-space sampling strategy between successive scans since there is no guarantee that the BOLD response is quasi-periodic.
Because the vast majority of fMRI studies in neurosciences make use either of rs-fMRI or fast event-related designs~\cite{Birn02,Logothetis08}, the most reliable acquisition strategy in such contexts remains the ``scan and
repeat'' approach, although it is suboptimal. To our knowledge, only one \emph{kt-contribution}~($kt$-GRAPPA~\cite{Huang05}) has claimed its ability to accurately reconstruct fMRI images in aperiodic paradigms. 

\subsection*{Overview of our contribution}
 
The present paper therefore aims at proposing a new 3D/(3D+t)-dimensional pMRI reconstruction algorithm that can be adopted irrespective of the nature of the encoding scheme or the fMRI paradigm. In particular, we show that our approach outperforms its SENSE-like alternatives not only in terms of artifact removal for
anatomical image reconstruction, but also in terms of statistical sensitivity 
at the subject and group-levels in fast event-related fMRI. 

In the fMRI literature, few studies have been conducted to measure the impact of the parallel imaging reconstruction algorithm on EPI volumes and subsequent statistical sensitivity for detecting evoked brain 
activity~\cite{deZwart02,Preibisch03,deZwart06,Rabrait07}.
In these works, reliable activations were detected for an acceleration factor up to~3. More recently, a special attention has been paid in~\cite{Utting10} to assess the performance of
dynamic MRI reconstruction algorithms on BOLD fMRI sensitivity. In~\cite{Utting10}, the authors have reported that $kt$-based
approaches perform better than conventional SENSE for BOLD fMRI in the sense that reliable sensitivity may be achieved at higher undersampling factors~(up to 5). However, most of the time, these comparisons are made on a small group of individuals and statistical analysis is only performed at the subject level. Here, we perform the comparison of several parallel MRI reconstruction algorithms both at the subject and group levels for different acceleration factors.

To remove reconstruction artifacts that occur at high
acceleration factors, regularized SENSE methods have been proposed in the literature~\cite{Liang_02,Ying_L_04,Liu_08_1,chaari_08,Liu_08_2}. Some of them apply quadratic or Total Variation~(TV) regularizations while others resort to 2D regularization 
in the wavelet transform domain~(e.g. UWR-SENSE: Unconstrained Wavelet Regularized SENSE)~\cite{Chaari_MEDIA_2011}). 
The latter strategy has proved its efficiency on the reconstruction of anatomical or functional~(resting-state only) data, compared to standard SENSE and TV-based 
regularization~\cite{chaari_08,Chaari_MEDIA_2011}. 
More recently, unconstrained Wavelet Regularized SENSE~(or UWR-SENSE) has been assessed on EPI images and compared with \mSENSE on a brain activation 
fMRI dataset~\cite{Chaari10e}. This comparison was performed at the subject level only.
Besides, except some non-regularized contributions like 3D GRAPPA~\cite{Jakob_06}, most of the available reconstruction methods in the literature operate slice by slice and thus reconstruct each slice irrespective of its neighbours. Iterating over slices is thus necessary to recover the whole 3D volume. This observation led us to consider 3D or full FOV image reconstruction as a single step in which all slices are treated together. For doing so, we introduce 3D wavelet transform and 3D sparsity-promoting regularization term 
in the wavelet domain. This approach can still apply even if the acquisition is performed in 2D instead of 3D. 
Following the same principle, an fMRI run usually consists of 
several tens of successive scans that are reconstructed independently one to another. Iterating over all acquired 3D volumes remains the classical approach to reconstruct the 4D or 3D~+~$t$ dataset. However, it has been shown for a long while that fMRI data are serially correlated in time even under the null hypothesis~(i.e., ongoing activity 
only)~\cite{Aguirre97,Zarahn97,Purdon98}. To capture this dependence between successive time points, an autoregressive model has demonstrated its 
relevance~\cite{Woolrich01,Worsley02,Penny03,Chaari_TMI_2012}. 
Hence, we propose to account for this temporal structure at the reconstruction step.


These two key ideas have played a central role to extend the UWR-SENSE
approach~\cite{Chaari_MEDIA_2011} through a more general regularization scheme that relies on a convex but nonsmooth 
criterion to be minimized. 
This criterion is made up of three terms. The first one (data fidelity) accounts for 3D spatial and temporal dependencies between 
successive slices and repetitions~(i.e., scans) by combining all repetitions and involving a 3D wavelet transform. 
The second and third terms promote sparsity in the 3D wavelet domain as well as the temporal smoothness of 
the sought (3D~+~$t$) image sequence, respectively.
The minimization of this criterion relies on the Parallel ProXimal Algorithm~(PPXA)~\cite{Combettes_PL_08} which can address a 
broader scope of optimization problems than the forward-backward and Douglas-Rachford methods employed 
in \cite{Chaari_MEDIA_2011}, or even FISTA as used in~\cite{Unser_TMI_2011}. All these algorithms are only able to optimize 
the sum of two convex functions, whereas PPXA deals with the optimization of any sum of convex functions. 
Our work can also be viewed as a \emph{dynamic} extension of the \emph{static} wavelet-based approach proposed 
in \cite{Unser_TMI_2011}.



The rest of this paper is organized as follows. Section~\ref{sec:parallel} recalls the principle of parallel MRI and describes 
the proposed reconstruction algorithms and optimization aspects. In Section~\ref{sec:validation}, experimental validation of the 
3D/4D-UWR-SENSE approaches is performed 
on anatomical $T_1$ MRI and BOLD fMRI data, respectively. In Section~\ref{sec:discussion}, we discuss the pros and cons of our 
method. Finally, conclusions and perspectives are drawn in Section~\ref{sec:conclusion}.

\section{Materials and Methods}\label{sec:parallel}
\subsection{Parallel imaging in MRI}
In parallel MRI, an array of $L$ coils  is employed to \ADDED{indirectly} measure the spin density 
$\overline{\rho}$~\cite{Sodickson_D_00} into the object under investigation\footnote{The overbar is used to distinguish the ``true'' data from a generic variable.}. The signal $\widetilde{d}_\ell$ received by each coil $\ell$ ($1\leq \ell \leq L$) is the Fourier 
transform of the desired 2D 
field\footnote{\ADDED{For simplicity, we address here the multislice acquisition context.}} $\overline{\rho}\in\RR^{X\times Y}$ on the specified FOV weighted by the coil sensitivity profile $s_\ell$, 
evaluated at some location $\vect{k}=(k_x,k_y)^{\trans}$ in the $k$-space:

\begin{equation}
\widetilde{d}_\ell(\vect{k})=\int\overline{\rho}(\vect{r})s_\ell(\vect{r})e^{-\imath 2\pi
\vect{k}^{\trans}\vect{r}}\,d\vect{r} +\widetilde{n}_\ell(\vect{k}), \label{eq:signal}
\end{equation} 

\noindent where $\widetilde{n}_\ell(\vect{k})$ is a coil-dependent additive zero-mean Gaussian noise, 
which is independent and identically distributed~(iid) in the $k$-space, and $\vect{r}=(x,y)^{\trans}\in X\times Y$ 
is the spatial position in the image domain ($\cdot^{\trans}$ being the transpose operator). 
The size of the reduced FOV acquired data $\widetilde{d}_\ell$ in the $k$-space clearly depends on the sampling 
scheme.  

In parallel MRI, the sampling period along the phase encoding direction is $R$ times larger than the one used for conventional acquisition, $R \leq L$ being the reduction factor. To recover full FOV images, many algorithms have been proposed but only 
SENSE-like~\cite{pruessmann_99} and GRAPPA-like~\cite{griswold_02} methods are provided by scanner manufacturers. In what follows, we focus on SENSE-like methods operating in the image domain.

Let $\Delta y=\frac{Y}{R}$ be the aliasing period and $y$ the position in the image domain along the phase encoding direction. Let $x$ be the position in the image domain along the frequency encoding direction. A 2D inverse Fourier transform allows us to recover the measured signal in the image domain. By accounting for the $k$-space undersampling at $R$-rate, the inverse Fourier transform gives us the spatial counterpart of Eq.~\eqref{eq:signal} in matrix form:

\begin{align}\label{eq:matriciel}
\vect{d}(\vect{r}) &= \vect{S}(\vect{r}) \overline{\vect{\rho}}(\vect{r}) + \vect{n}(\vect{r}),
\end{align}
where

\begin{align}
\vect{S}(\vect{r})\,\eqdef\,\left[
\begin{array}{ccc}
s_1(x,y)&\ldots&s_1(x,y+(R-1)\Delta y)\\
\vdots&\vdots&\vdots\\
s_L(x,y)&\ldots&s_L(x,y+(R-1)\Delta y)\\
\end{array}
\right]
, \quad&
\vect{n}(\vect{r})\,\eqdef\,\left[
\begin{array}{c}
n_1(x,y)\\
n_2(x,y)\\
\vdots\\
n_L(x,y)\\
\end{array}
\right]
\nonumber
\end{align}

\begin{align}
\label{eq:defvrho}
& \overline{\vect{\rho}}(\vect{r})\,\eqdef\,
\left[
\begin{array}{c}
\overline{\rho}(x,y)\\
\overline{\rho}(x,y+\Delta y)\\
\vdots\\
\overline{\rho}(x,y+(R-1)\Delta y)\\
\end{array}
\right] \quad \mathrm{and} 
 \quad 
\vect{d}(\vect{r})\,\eqdef\,
\left[
\begin{array}{c}
d_1(x,y)\\
d_2(x,y)\\
\vdots\\
d_L(x,y)\\
\end{array}
\right]. 
\end{align}

Based upon this model, the reconstruction step consists of solving Eq.~\eqref{eq:matriciel} so as to recover $\overline{\vect{\rho}}(\vect{r})$ from $\vect{d}(\vect{r})$ and an estimate of $\vect{S}(\vect{r})$ at each spatial position $\vect{r}=(x,y)^{\trans}$. The spatial mixture or \emph{sensitivity} matrix $\vect{S}(\vect{r})$ is estimated using a reference scan and varies according to the coil geometry. Note that the coil images $(d_\ell)_{1\leq l \leq L}$ as well as the sought image $\overline{\rho}$ are complex-valued, although $|\overline{\rho}|$ is only considered for visualization purposes. The next section describes the widely used SENSE algorithm as well as its regularized extensions.

\vspace*{-0.5cm}
\subsection{Reconstruction algorithms} \label{sec:algos}

\subsubsection{1D-SENSE}
In its simplest form, SENSE imaging amounts to solving a one-dimensional inversion problem due to the separability of the Fourier transform. Note however that this inverse problem admits a two-dimensional extension in 3D imaging sequences like Echo Volume 
Imaging~(EVI)~\cite{Rabrait07} where undersampling occurs in two $k$-space directions. The 1D-SENSE reconstruction 
method~\cite{pruessmann_99} actually minimizes a Weighted Least Squares~(WLS) criterion $\mathcal{J}_{\rm WLS}$ given by:

\begin{equation}
\label{eq:crit_WLS}
\mathcal{J}_{\rm WLS}(\rho) = \sum_{\mathbf{r}\in \{1,\ldots,X\} \times \{1,\ldots,Y/R\}} \parallel \vect{d}(\vect{r})-\vect{S}(\vect{r})\vect{\rho}(\vect{r}) \parallel^2_{\vect{\Psi}^{-1}},  
\end{equation}

\noindent where $\|\cdot\|_{\vect{\Psi}^{-1}}= \sqrt{(\cdot)^{\hermit}\vect{\Psi}^{-1}(\cdot)}$, and the noise covariance matrix $\vect{\Psi}$ is usually estimated based on $L$ acquired images $(\underbar{d}_{\ell})_{1\leq \ell \leq L}$ from all coils without radio frequency pulse.  Hence, the SENSE full FOV image is nothing but the maximum likelihood estimate under Gaussian noise assumption, which admits the following closed-form expression at each spatial position $\vect{r}$:

\begin{equation}
 \widehat{\vect{\rho}}_{\rm WLS}(\vect{r}) = \pth{\vect{S}^{\hermit}(\vect{r})\vect{\Psi}^{-1}\vect{S}(\vect{r})}^{\sharp}\vect{S}^{\hermit}(\vect{r})
\vect{\Psi}^{-1}\vect{d}(\vect{r}),
\end{equation}

\noindent where $(\cdot)^{\hermit}$~(respectively $(\cdot)^{\sharp}$) stands for the transposed complex 
conjugate~(respectively pseudo-inverse). It should be noticed here that the described 1D-SENSE reconstruction method has been designed to reconstruct one slice~(2D image). To reconstruct a full volume, the 1D-SENSE reconstruction algorithm has to be iterated over all slices.
\ADDED{
In practice, the performance of the SENSE method is limited because of {\em i)} different sources of noise such as 
distortions in the measurements $\vect{d}(\vect{r})$, and 
{\em ii)} distortions in estimation and ill-conditioning of $\vect{S}(\vect{r})$ mainly at brain/air interfaces. }
To enhance the robustness of the solution to this ill-posed problem, a regularization is usually introduced in the reconstruction process. To improve results obtained with quadratic regularization techniques \cite{Liang_02,Ying_L_04}, edge-preserving regularization has been widely investigated in the pMRI reconstruction literature. For instance, reconstruction methods based on Total Variation~(TV) regularization have been proposed in a number of recent works 
like \cite{keeling_03,Liu_08}. However, TV is mostly adapted to piecewise constant images, which are not always accurate models in MRI, especially in fMRI. 
As investigated by \textit{Chaari et al.}~\cite{Chaari_MEDIA_2011},
\textit{Liu et al.}~\cite{Liu_08_2} and \textit{Guerquin-Kern et al.}~\cite{Unser_TMI_2011}, regularization in the Wavelet Transform (WT) domain is a powerful tool to improve SENSE reconstruction.  In what follows, we summarize the principles of the wavelet-based regularization approach.

\subsubsection{Proposed wavelet-based regularized SENSE}\label{subsec:WRSENSE}

Akin to \cite{Chaari_MEDIA_2011} where a regularized reconstruction algorithm relying on 2D separable WTs was investigated, to the best of our knowledge, all the existing approaches in the pMRI regularization literature proceed slice by slice. 
The drawback of this strategy is that no spatial continuity between adjacent slices is taken into account since the slices are processed 
independently. Moreover, since the whole brain volume has to be acquired several times in an fMRI study, 
separately iterating over all the acquired 
3D volumes is then necessary in order to reconstruct a 4D data volume corresponding to a fMRI session.

Consequently, the 3D volumes are supposed independent whereas fMRI time-series are serially correlated in time because of two distinct 
effects: the BOLD signal itself is a low-pass filtered version of the neural activity, and physiological artifacts make the fMRI time 
series strongly dependent. For these reasons, modeling temporal dependence across scans at the reconstruction step may impact subsequent 
statistical analysis. 
This has motivated the extension of the wavelet regularized reconstruction approach in \cite{Chaari_MEDIA_2011} in order to:
\begin{itemize}
 \item account for 3D spatial dependencies between adjacent slices by using 3D WTs,
 \item exploit the temporal dependency between acquired 3D volumes by applying an additional regularization term along the temporal dimension of the 4D dataset.
\end{itemize}

This additional regularization will help us in increasing the Signal to Noise Ratio~(SNR) through the acquired volumes, and therefore 
enhance the reliability of the statistical analysis in fMRI. These temporal dependencies have also been used in the dynamic MRI 
literature in order to improve the reconstruction quality in conventional MRI \cite{Sumbul_2009}. However, since the imaged object 
geometry in the latter context generally changes during the acquisition, taking into account the temporal regularization in the 
reconstruction process is very difficult. 
An optimal design of 3D reconstruction should integrate slice-timing and motion correction in the reconstruction pipeline. 
For the sake of computational efficiency, our approach only performs 3D reconstruction before considering slice-timing
and motion correction.


To deal with a 4D reconstruction of the $N_r$ acquired volumes, we will first rewrite the observation model in Eq.~\eqref{eq:matriciel}
as follows:

\begin{equation}
\vect{d}^t(\vect{r}) = \vect{S}(\vect{r})\vect{\rho}^t(\vect{r}) + \vect{n}^t(\vect{r}),
\end{equation}

\noindent where $t \in \{1,\ldots,N_r\}$ is the acquisition time and $\vect{r} = (x,y,z)$ is the 3D spatial position, $z \in \{1, \ldots, Z\}$ being the position along the third direction (slice selection one).

At a given time $t$, the full FOV 3D complex-valued image $\overline{\rho}^t$ of size $X \times Y \times Z$ can be seen as an element of the Euclidean space $\mathbb{C}^K$ with
 $K = X \times Y \times Z$ endowed with the standard inner product $\scal{\,\cdot\,}{\,\cdot\,}$ and norm 
$\|\cdot \|$. We employ a dyadic 3D orthonormal wavelet decomposition operator $T$ over $j_\mathrm{max}$ 
resolution levels. The coefficient field resulting from the wavelet decomposition of a target image $\rho^t$ 
is defined as $\zeta^t =\big(\zetab^t_{a}, (\zetab^t_{o,j})_{o\in \mathbb{O},1 \le j \le j_\mathrm{max}}\big)$
 with $o \in \mathbb{O} =\{0,1\}^3\setminus \{(0,0,0)\}$, $\zetab^t_{a} = (\zeta^t_{a,k})_{1 \le k\le K_{j_\mathrm{max}}}$ 
and $\zetab^t_{o,j}=(\zeta^t_{o,j,k})_{1\le k \le K_j}$ where $K_{j}= K2^{-3j}$ is the number of wavelet 
coefficients in a given subband at resolution $j$ (by assuming that $X$, $Y$ and $Z$ are multiple of 
$2^{j_{\mathrm{max}}}$). Adopting such a notation, the wavelet coefficients have been reindexed so that
 $\zetab^t_{a}$ denotes the approximation coefficient vector at the resolution level  $j_\mathrm{max}$, 
while $\zetab^t_{o,j}$ denotes the detail coefficient vector at the orientation $o$ and resolution level $j$.
 Using 3D dyadic WTs allows us to smooth reconstruction artifacts along the slice selection direction 
 that may appear at the same spatial position, which 
is not possible using a slice by slice processing. Also, even if reconstruction artifacts 
do not exactly appear in the 
same positions, the proposed method allows us to incorporate reliable information from adjacent 
slices in the reconstruction model.

The proposed regularization procedure relies on the introduction of two penalty terms. The first penalty term describes the prior 3D spatial knowledge about the wavelet coefficients of the target solution and it is expressed as:
\begin{equation}
g(\zeta) = \sum_{t = 1}^{N_r} \Big[\sum_{k=1}^{K_{j_\mathrm{max}}} \Phi_{a}(\zeta^t_{a,k}) + \sum_{o\in \mathbb{O}} \sum_{j=1}^{j_{\mathrm{max}}} \sum_{k=1}^{K_j}  
\Phi_{o,j}(\zeta^t_{o,j,k}) \Big], 
\end{equation}
where $\zeta = (\zeta^1,\zeta^2,\ldots,\zeta^{N_r})$ and we have, for every $o \in \mathbb{O}$ and $j \in \{1,\ldots,j_{\rm max}\}$ 
(and similarly for $\Phi_{a}$ relative to the approximation coefficients), 
\begin{equation}
\forall \xi \in \CC,\quad \; \Phi_{o,j}(\xi) = \Phi^{\rm Re}_{o,j}(\xi) + \Phi^{\rm Im}_{o,j}(\xi) 
\end{equation} 

\noindent where $\Phi^{\rm Re}_{o,j}(\xi) = \alpha_{o,j}^{\mathrm{Re}}|\mathrm{Re}(\xi - \mu_{o,j})| + \frac{\beta_{o,j}^{\mathrm{Re}}}{2}|\mathrm{Re}(\xi - \mu_{o,j})|^2$ 
and $\Phi^{\rm Im}_{o,j}(\xi) = \alpha_{o,j}^{\mathrm{Im}}|\mathrm{Im}(\xi - \mu_{o,j})| + \frac{\beta_{o,j}^{\mathrm{Im}}}{2}|\mathrm{Im}(\xi - \mu_{o,j})|^2$ 
with $\mu_{o,j} = \mu^{\rm Re}_{o,j} + \imath \mu^{\rm Im}_{o,j}\in \CC$, and 
$\alpha_{o,j}^{\mathrm{Re}}$, $\beta_{o,j}^{\mathrm{Re}}$, $\alpha_{o,j}^{\mathrm{Im}}$, 
$\beta_{o,j}^{\mathrm{Im}}$ are some positive real constants. Hereabove, $\mathrm{Re}(\cdot)$ and 
$\mathrm{Im}(\cdot)$ (or $\cdot^{\mathrm{Re}}$ and $\cdot^{\mathrm{Im}}$) stand for the real and imaginary 
parts, respectively. For both real and imaginary parts, this regularization term allows us to keep a compromise between 
sparsity and smoothness of the wavelet coefficients due to the $\ell_1$ and $\ell_2$ terms, respectively. \\
The second regularization term penalizes the temporal variation between successive 3D volumes: 

\begin{equation}
 h(\zeta) = \kappa \sum_{t = 2}^{N_r} \Vert T^*\zeta^{t} - T^*\zeta^{t-1} \Vert_p^p
\end{equation}

\noindent where $T^*$ is the 3D wavelet reconstruction operator. The prior parameters 
$\vect{\alpha}_{o,j} = (\alpha_{o,j}^{\mathrm{Re}},
\alpha_{o,j}^{\mathrm{Im}})$,
$\vect{\beta}_{o,j}=(\beta_{o,j}^{\mathrm{Re}},\beta_{o,j}^{\mathrm{Im}})$,
$\mub_{o,j}=(\mu^{\mathrm{Re}}_{o,j},\mu^{\mathrm{Im}}_{o,j})$,
$\kappa \in [0,+\infty[$ and $p \in [1,+\infty[$ are unknown and they need to be estimated\linebreak 
(see~Appendix~\ref{append:a2}). The used $\ell_p$ norm gives more flexibility to the temporal penalization term by allowing it 
to promote different levels of sparsity depending on the value of $p$. Such a penalization has been chosen based on  
empirical studies that have been conducted on the time-course of the BOLD signal at the voxel level. 

The operator $T^*$ is then applied to each component $\zeta^t$ of $\zeta$ to obtain the reconstructed 3D volume $\rho^t$ related to 
acquisition time $t$. It should be noticed here that other choices for the penalty functions are also possible provided that the 
convexity of the resulting optimality criterion is ensured. This condition enables the use of fast and efficient convex optimization algorithms. 
Adopting this formulation, the minimization procedure plays a prominent role in the reconstruction process. 
The proposed optimization procedure is detailed in Appendix~\ref{append:a1}. 
\section{Results}\label{sec:validation}
This section is dedicated to the experimental validation of the reconstruction algorithm we proposed in Section~\ref{subsec:WRSENSE}.
Experiments have been conducted on both anatomical and functional data which was acquired on a 3T Siemens Trio magnet.
For fMRI acquisition, ethics approval was given by the local research ethics committee~(Kremlin-Bic\^etre, \texttt{CPP}: 08 032) and fifteen subjects gave written informed consent for participation.

For anatomical data, the proposed 3D-UWR-SENSE algorithm (4D-UWR-SENSE without temporal regularization) is compared to the Siemens reconstruction pipeline. As regards fMRI validation, results of subject and group-level fMRI statistical analyses are compared for two reconstruction pipelines:
the one available on the Siemens workstation and our own pipeline which, for the sake of completeness, involves  either the early UWR-SENSE~\cite{Chaari_MEDIA_2011} or the 4D-UWR-SENSE version of the proposed pMRI reconstruction algorithm. 

\subsection{Anatomical data}\label{subsec:anat}

Anatomical data has been acquired using a 3D $T_1$-weighted MP-RAGE pulse sequence at a $1\times 1 \times 1.1~\rm{mm}^3$ spatial resolution ($\text{TE}=2.98~\rm ms$, $\text{TR}=2300~\rm ms$, $TI=900~\rm ms$, flip angle $= 9 ^\circ$, slice thickness = 1.1~mm, transversal orientation, FOV = $256\times 240\times 176~\mathrm{mm}^3$, TR between two RF pulses: $7.1$~ms, antero-posterior phase encoding). Data has been collected using a 32-channel receiver coil~(no parallel transmission has been used) at two different acceleration factors, $R=2$ and $R=4$.

To compare the proposed approach to 
the \mSENSE\footnote{SENSE reconstruction implemented by the Siemens scanner, software ICE, VB 17.} one,
  Fig.~\ref{fig:anats} illustrates coronal anatomical slices
  reconstructed with both algorithms while turning off the temporal
  regularization in 4D-UWR-SENSE, so resulting in
  the so-called 3D-UWR-SENSE approach. Red circles clearly
  show reconstruction artifacts and noise in the \mSENSE~reconstruction, which
  have been removed using our 3D-UWR-SENSE approach. Comparison may also be made through reconstructed slices 
  for $R=2$ and $R=4$, as well as with the conventional acquisition ($R=1$). This figure shows that increasing $R$
  generates more noise and artifacts in \mSENSE~results whereas the impact on our results is attenuated. 
Artifacts are smoothed by using the continuity of spatial information across contiguous slices in the wavelet space.
Depending on the used wavelet basis and the number of vanishing moments, more or less~(4 or 8 for instance) adjacent
slices  are involved in the reconstruction of a given slice. For instance, using Symmlet filters of length 8 
(4 vanishing moments) as in the conducted experiments here, 8 adjacent slices are involved in reconstructing a 
given slice. However, it is worth noticing that the introduced smoothing is anisotropic, in contrast to standard 
Gaussian smoothing that could be applied to anatomical data.
Fig.~\ref{fig:anats} also compares 3D-UWR-SENSE and \mSENSE~reconstructed slices when applying additional spatial smoothing 
to the later with a $2\times 2 \times 2$~mm$^3$ Gaussian kernel. Comparisons 
clearly show that, even at such low spatial smoothing level, \mSENSE~images suffer from a significant blur. 
Moreover, the artifact present at $R=4$ for \mSENSE~(left red circle) is spread out but not fully removed by 
applying isotropic spatial smoothing. 

Even for slice-selective acquisition schemes where the signal 
is supposed to be independent 
between adjacent slices, the proposed algorithm still allows us to exploit information continuity across slices which results 
from the imaged anatomy. 
Moreover, the smoothing level strongly depends on the regularization parameters that 
are used to set the thresholding level of wavelet coefficients. 
Images reconstructed using our algorithm present higher 
smoothing level than \mSENSE~without altering key information in the images. When carefully analysing the image background, 
one can notice the presence of motion-like artifacts that only affect the background and do not alter the brain mask. 
Such artifacts are nothing but boundary effects that are due to the use of wavelet transforms. 

\begin{figure}[!ht]
\centering
\begin{tabular}{cc||ccc}
&$R=1$&$R=2$&$R=4$\\
\raisebox{1.cm}{SOS}&\includegraphics[width=2.2cm, height=2.2cm]{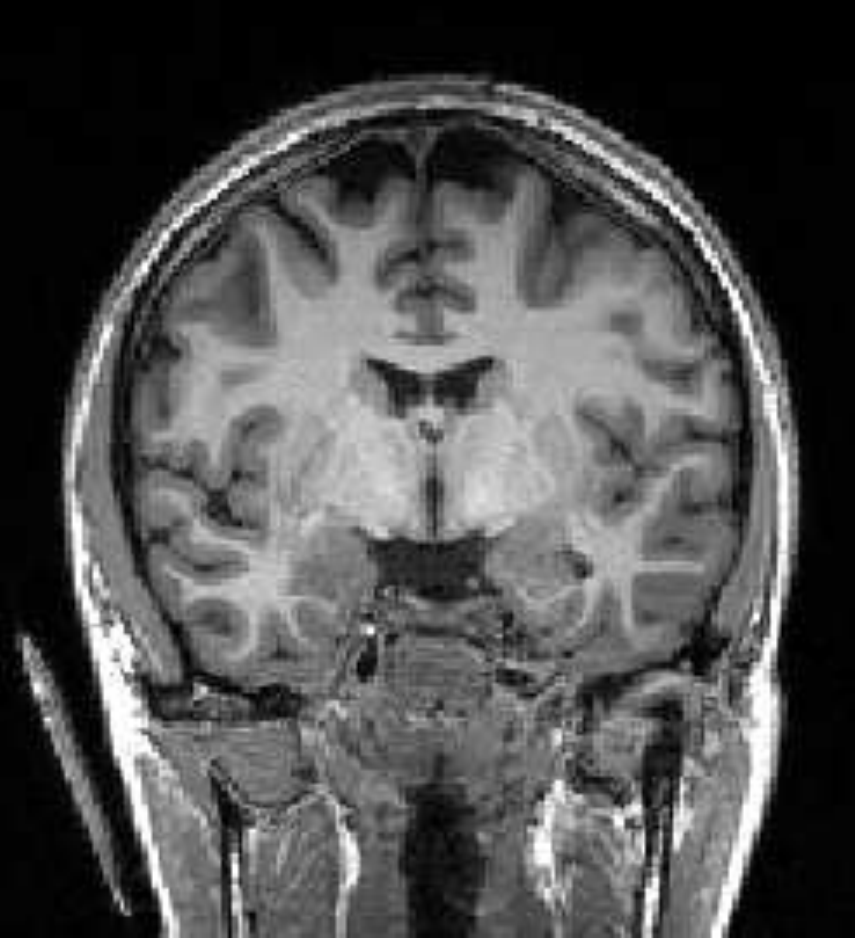}&
\includegraphics[width=2.2cm, height=2.2cm]{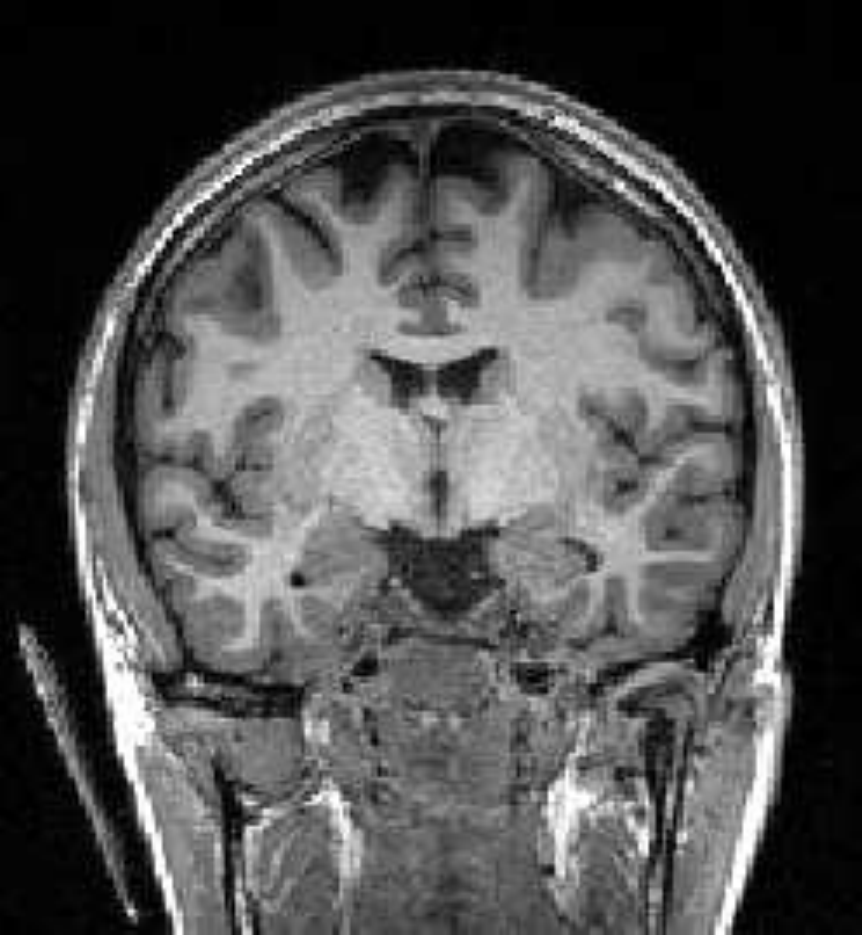}&
\includegraphics[width=2.2cm, height=2.2cm]{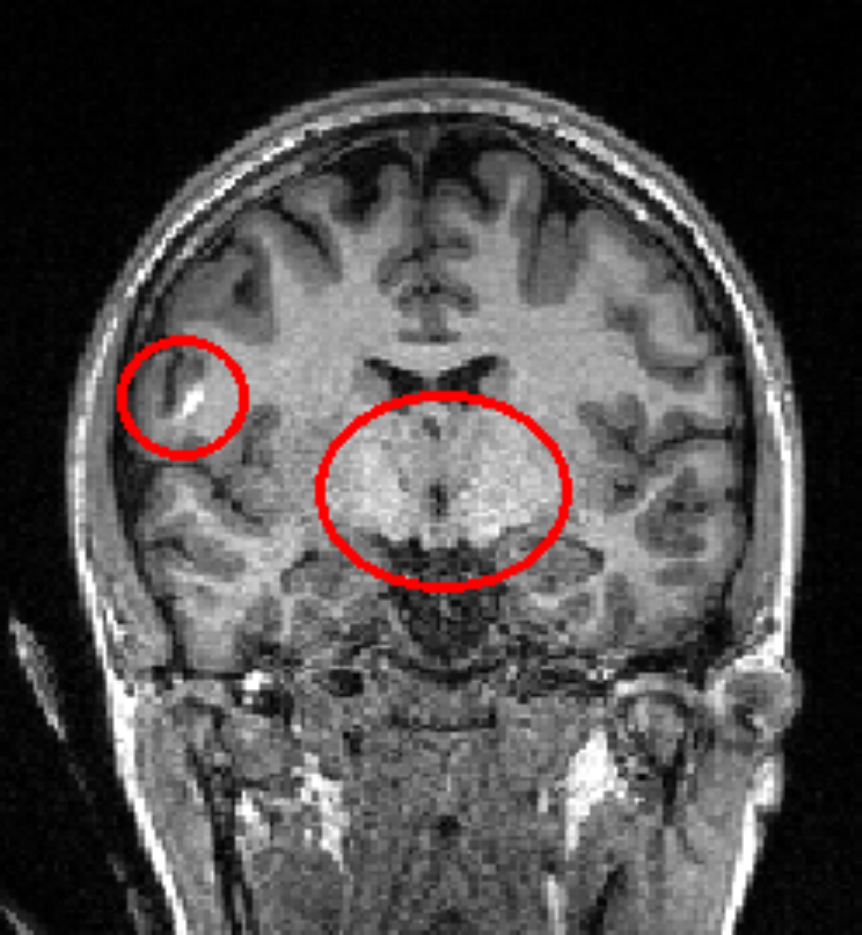}&\raisebox{1.cm}{\mSENSE}\\
\raisebox{1.cm}{SOS}&\includegraphics[width=2.2cm, height=2.2cm]{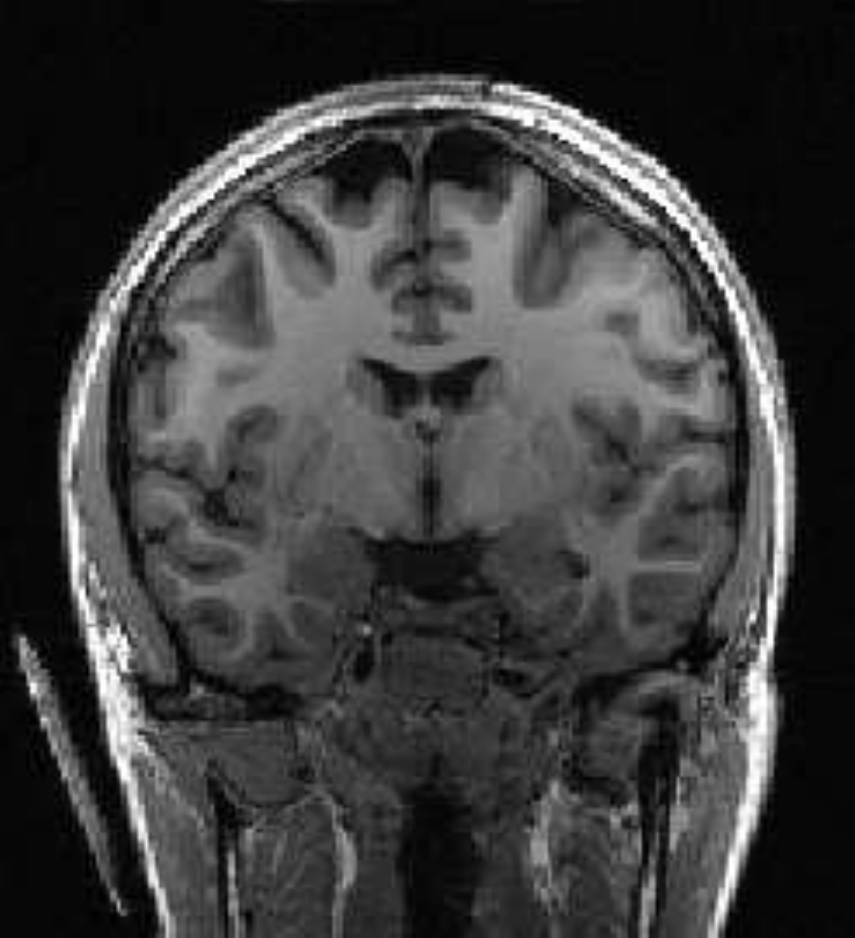}&
\includegraphics[width=2.2cm, height=2.2cm]{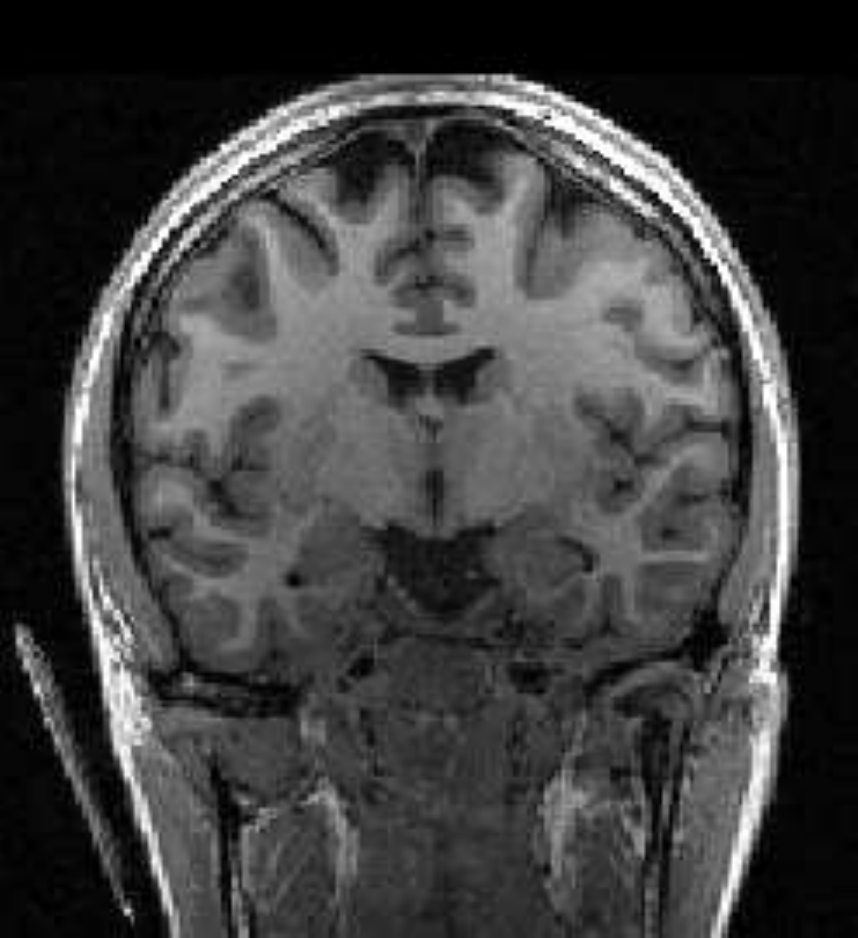}&
\includegraphics[width=2.2cm, height=2.2cm]{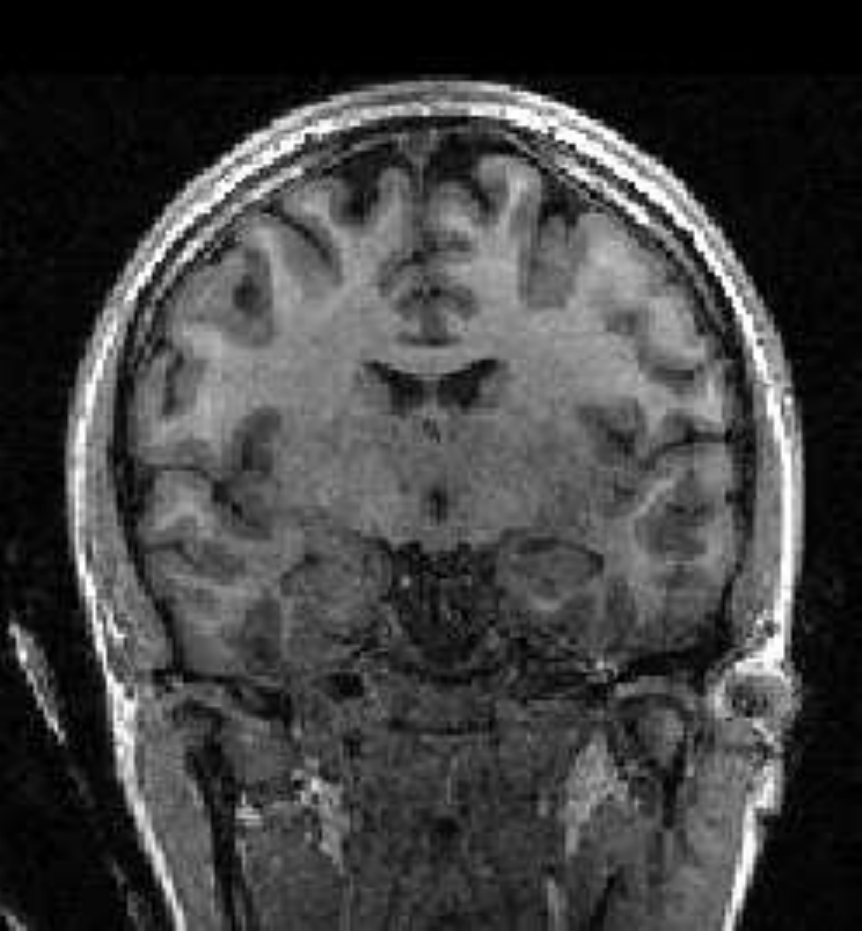}&\raisebox{1.cm}{3D-UWR-SENSE}\\
\raisebox{1.cm}{SOS}&\includegraphics[width=2.2cm, height=2.2cm]{newT1MRI_HC080251_siemens_R1_cor132.pdf}&
\includegraphics[width=2.2cm, height=2.2cm]{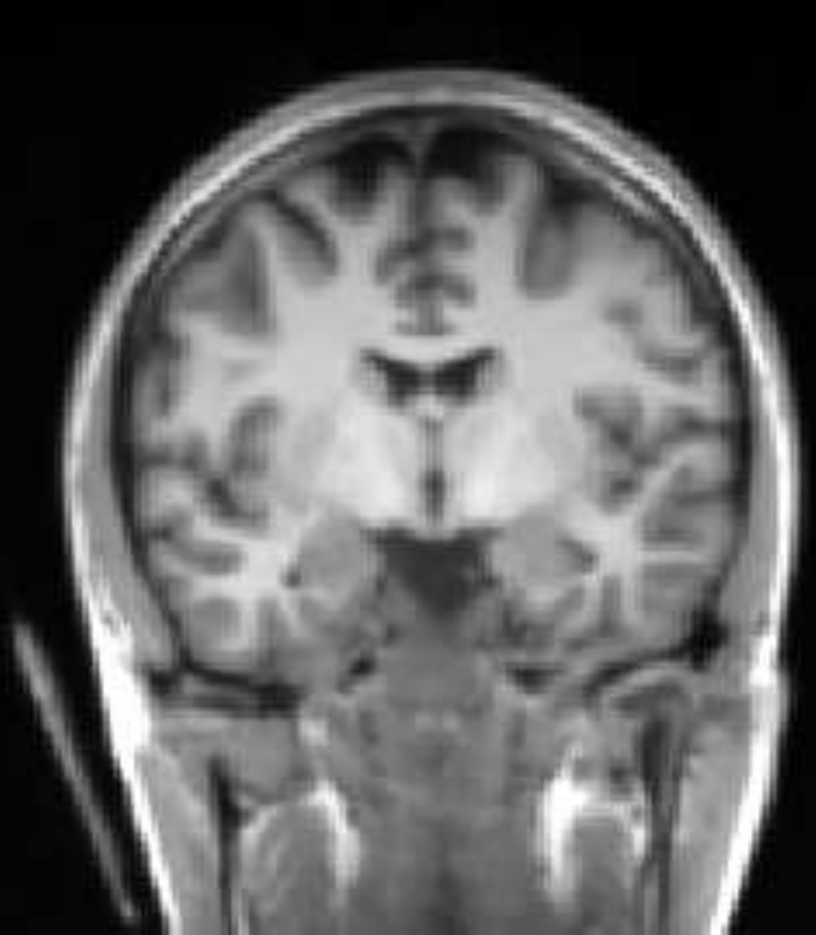}&
\includegraphics[width=2.2cm, height=2.2cm]{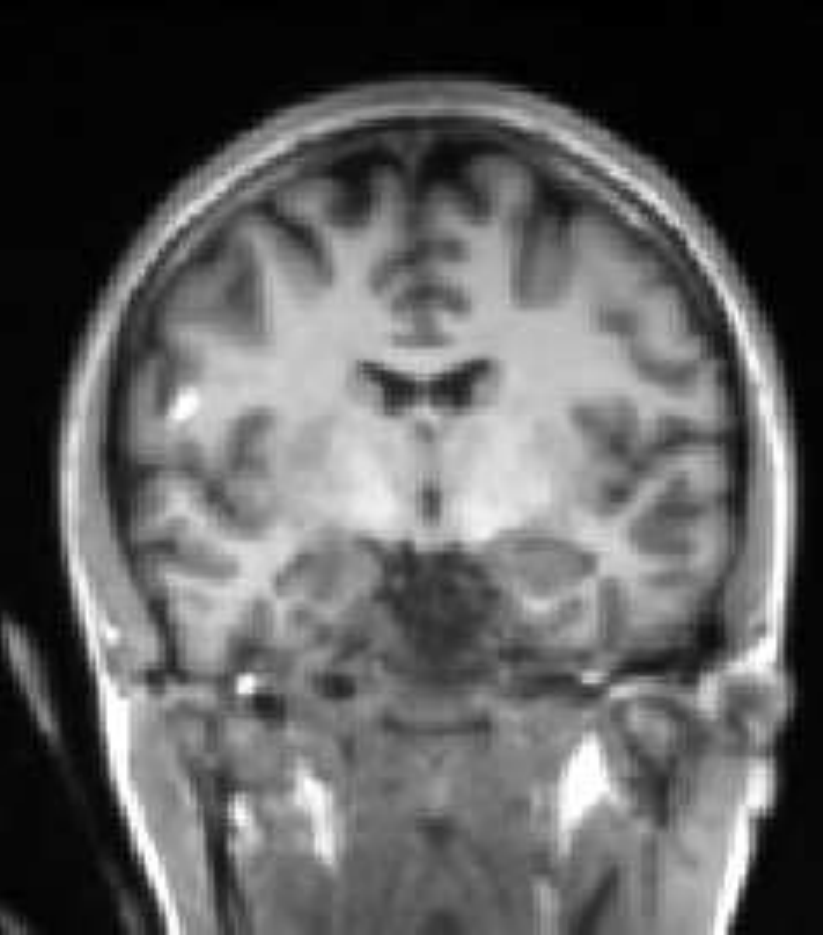}&
\raisebox{1.cm}{smoothed \mSENSE}\\
\end{tabular}
\caption{\textbf{Coronal} reconstructed slices using \mSENSE~(without and with $2\times 2 \times 2$~mm$^3$ 
spatial Gaussian smoothing) and 3D-UWR-SENSE (4D-UWR-SENSE without temporal regularization) 
for $R=2$ and $R=4$ with $1\times 1 \times 1.1~\rm{mm}^3$ spatial resolution. Reconstructed slices are also provided for 
a conventional acquisition (non accelerated with $R=1$) as the Sum Of Squares (SOS). Red ellipsoids indicate the position of reconstruction artifacts using \mSENSE. 
\label{fig:anats}}
\end{figure}

In order to evaluate the impact of such smoothing, classification 
tests have been conducted based on images reconstructed with both methods. 
Gray and white matter classification results using the Morphologist 2012 pipeline of $T_1$-MRI toolbox of Brainvisa 
software\footnote{http://brainvisa.info} at $R=2$ and $R=4$ 
are compared to those obtained without acceleration (i.e. at $R=1$), considered as the ground truth. 
Displayed results in Fig.~\ref{fig:classif} show that classification errors occur due to reconstruction artifacts for \mSENSE, 
especially at $R=4$. Results show that the gray matter is better classified using our 3D-UWR-SENSE algorithm especially next to the 
artifact into the red circle (Fig.~\ref{fig:classif}~[$R=4$]), which lies at the frontier between the white and gray matters. 
Moreover, reconstruction noise with \mSENSE~in the centre of the white matter (left red circle in
Fig~\ref{fig:classif}~[$R=4$]) also causes miss-classification errors far from the gray/while matter frontier. 
However, at $R=1$ and $R=2$ classification performance is rather similar for both methods, which confirms the
ability of the proposed method to attenuate reconstruction artifacts while keeping classification results unbiased.


\begin{figure}[!ht]
\centering
\begin{tabular}{cc||ccc}
&$R=1$&$R=2$&$R=4$\\
\raisebox{1.cm}{SOS}&
\includegraphics[width=2.2cm, height=2.2cm]{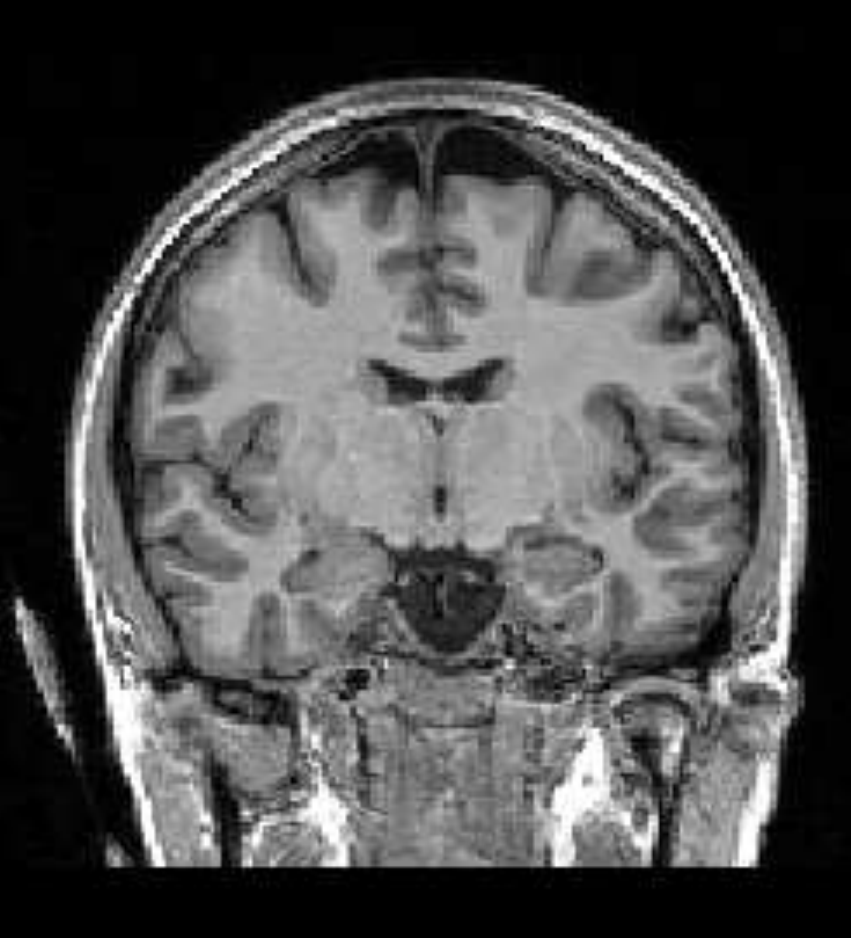}&
\includegraphics[width=2.2cm, height=2.2cm]{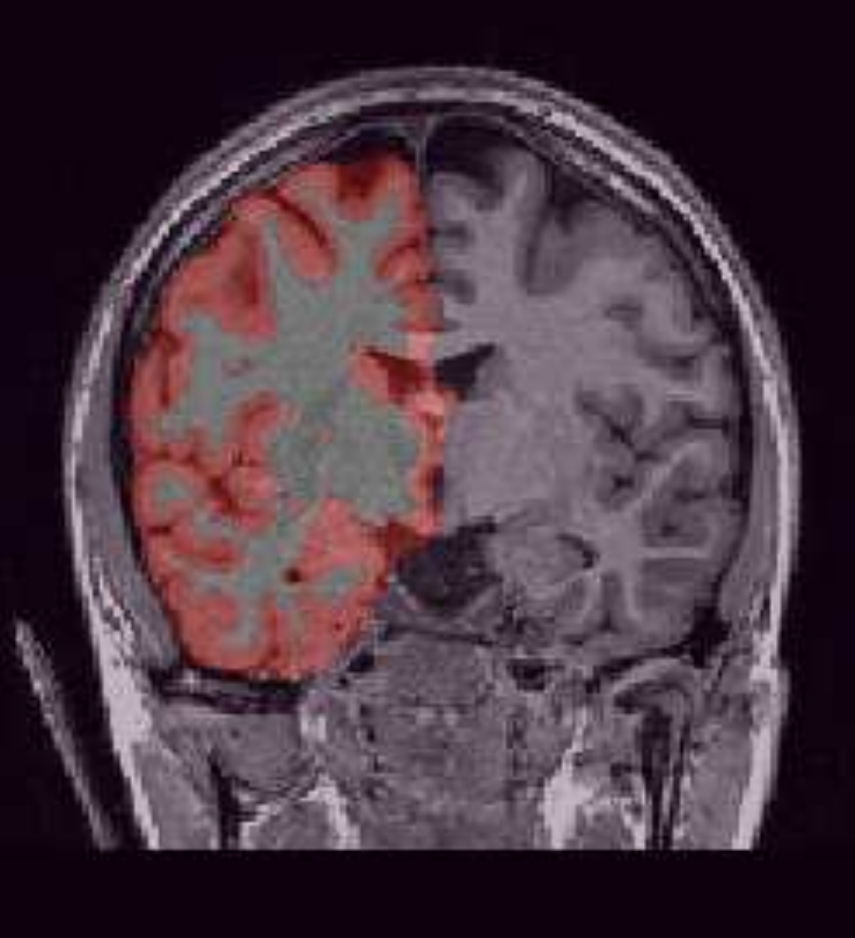}&
\includegraphics[width=2.2cm, height=2.2cm]{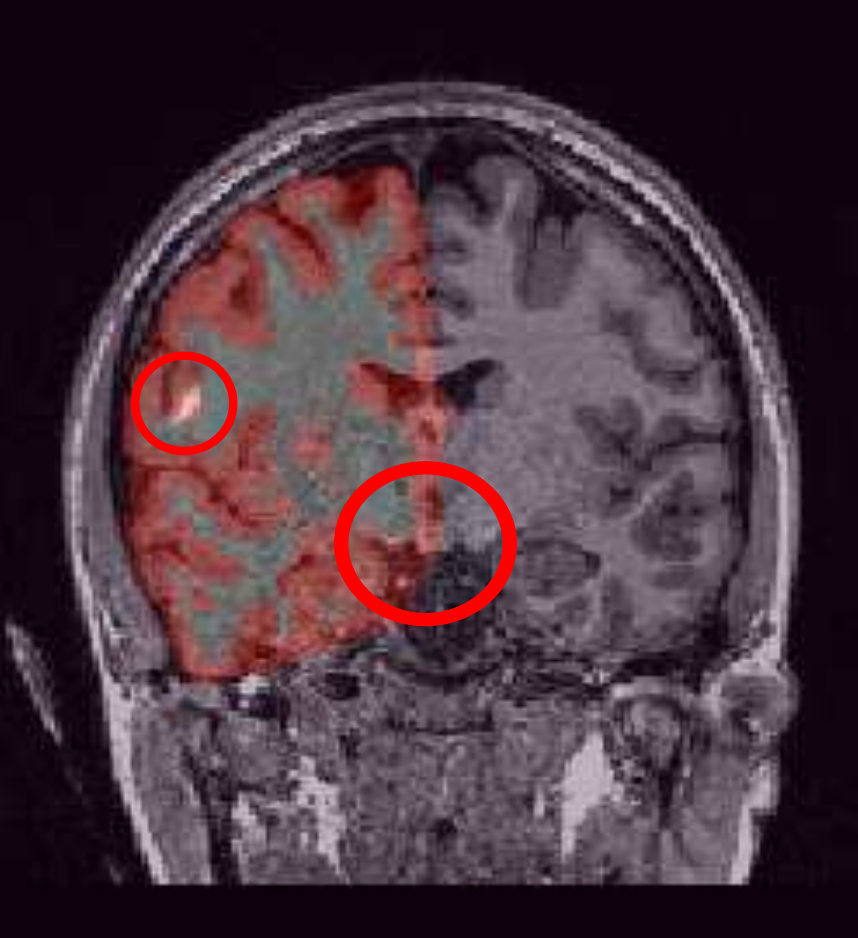}&\raisebox{1.cm}{\mSENSE}\\
\raisebox{1.cm}{SOS}&
\includegraphics[width=2.2cm, height=2.2cm]{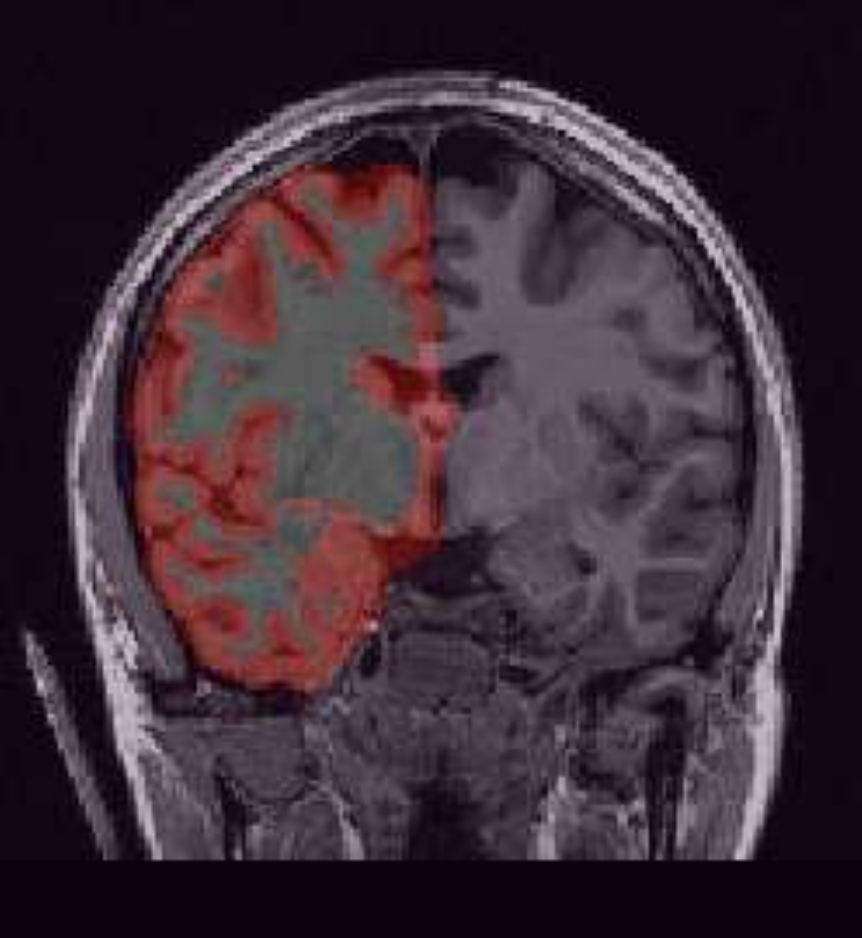}&
\includegraphics[width=2.2cm, height=2.2cm]{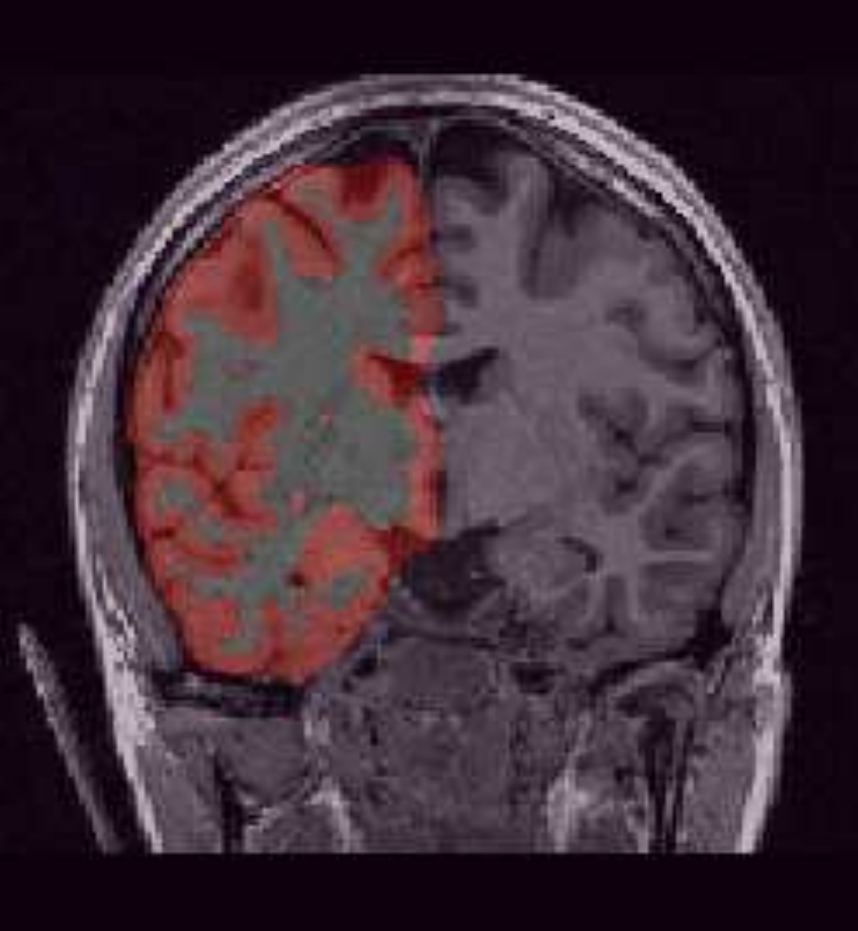}&
\includegraphics[width=2.2cm, height=2.2cm]{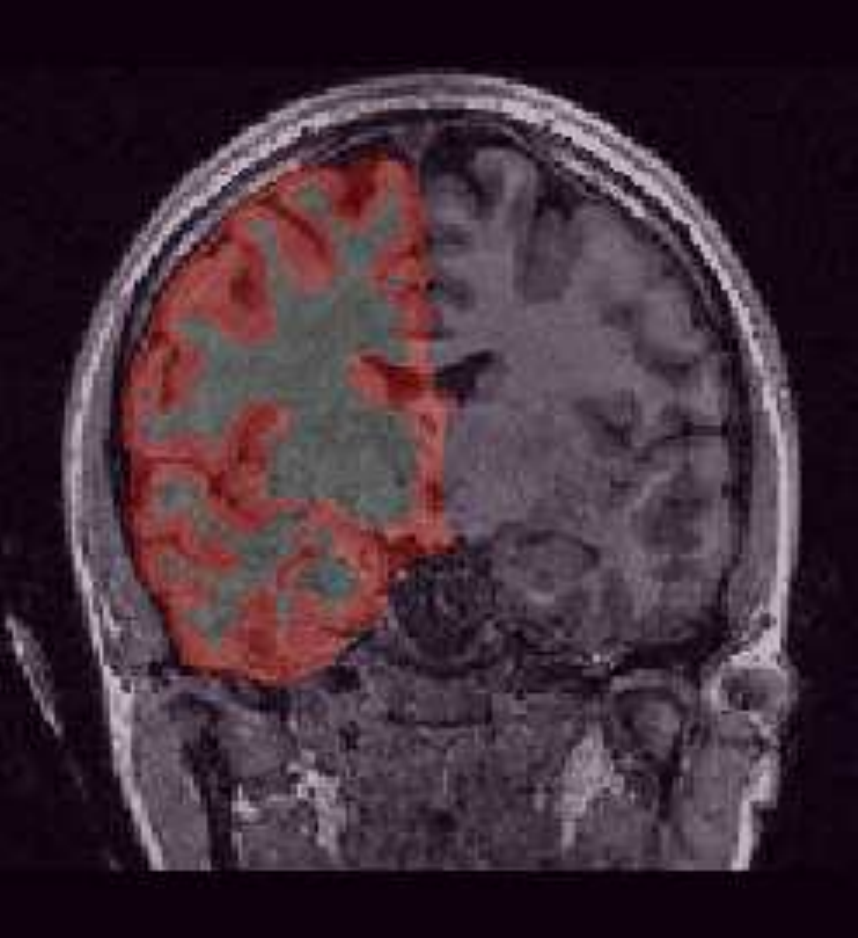}&\raisebox{1.cm}{3D-UWR-SENSE}\\
\end{tabular}
\caption{Classification results based on reconstructed slices using \mSENSE~and 3D-UWR-SENSE 
for $R=2$ and $R=4$ with $1\times 1 \times 1.1~\rm{mm}^3$ spatial resolution (\textbf{Coronal} view). 
Classification results based on the SOS of a non-accelerated acquisition ($R=1$) 
are also provided as a ground truth. Red circles indicate the position of 
reconstruction artifacts using \mSENSE~for $R=4$. 
\label{fig:classif}}
\end{figure}

To further investigate the smoothing effect of our reconstruction algorithm, gray matter interface of the 
cortical surface has bee extracted using the above mentioned BrainVISA pipeline. Extracted surfaces~(medial and
lateral views) from \mSENSE~ and 3D-UWR-SENSE images are show in Fig.~\ref{fig:classifsurf} for $R=4$. 
For comparison purpose, we provide results with \mSENSE~at $R=1$ as ground truth.

\begin{figure}[!ht]
\centering
\begin{tabular}{cccc}
&Ground truth: $R=1$&\mSENSE&3D-UWR-SENSE\\
  \raisebox{1.1cm}{medial view}&
\includegraphics[width=2.8cm, height=2cm]{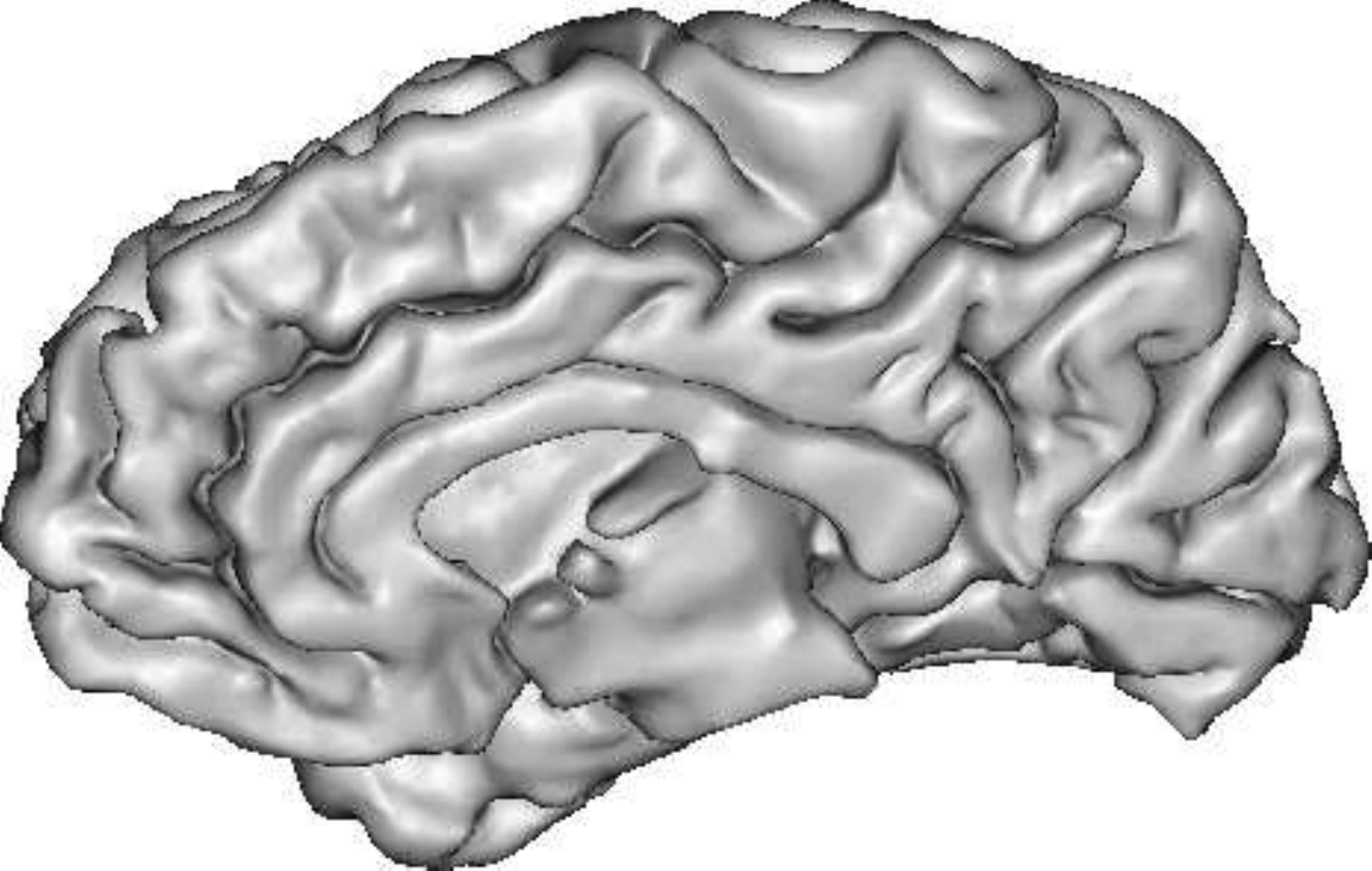}&
\includegraphics[width=2.8cm, height=2cm]{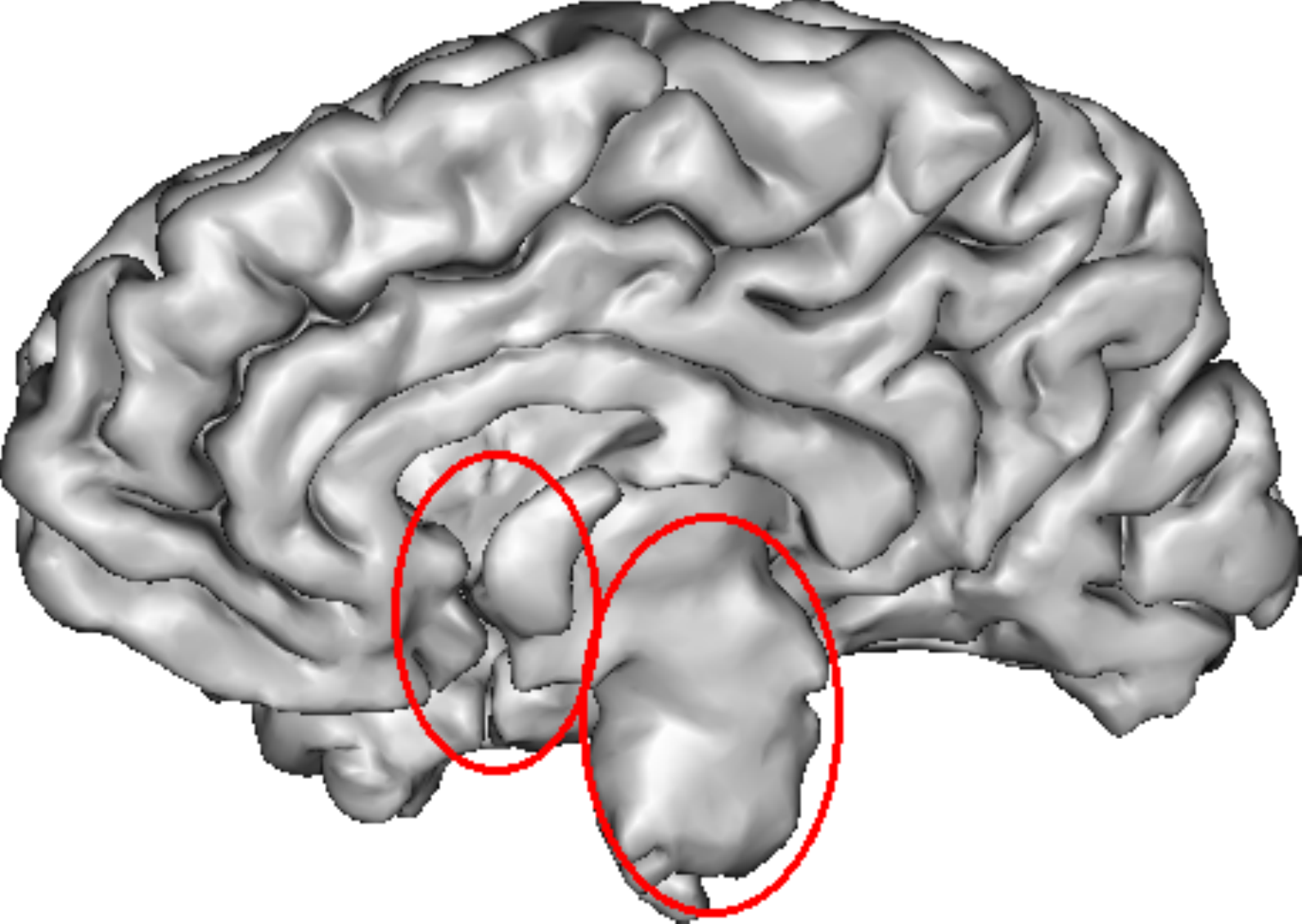}&
\includegraphics[width=2.8cm, height=2cm]{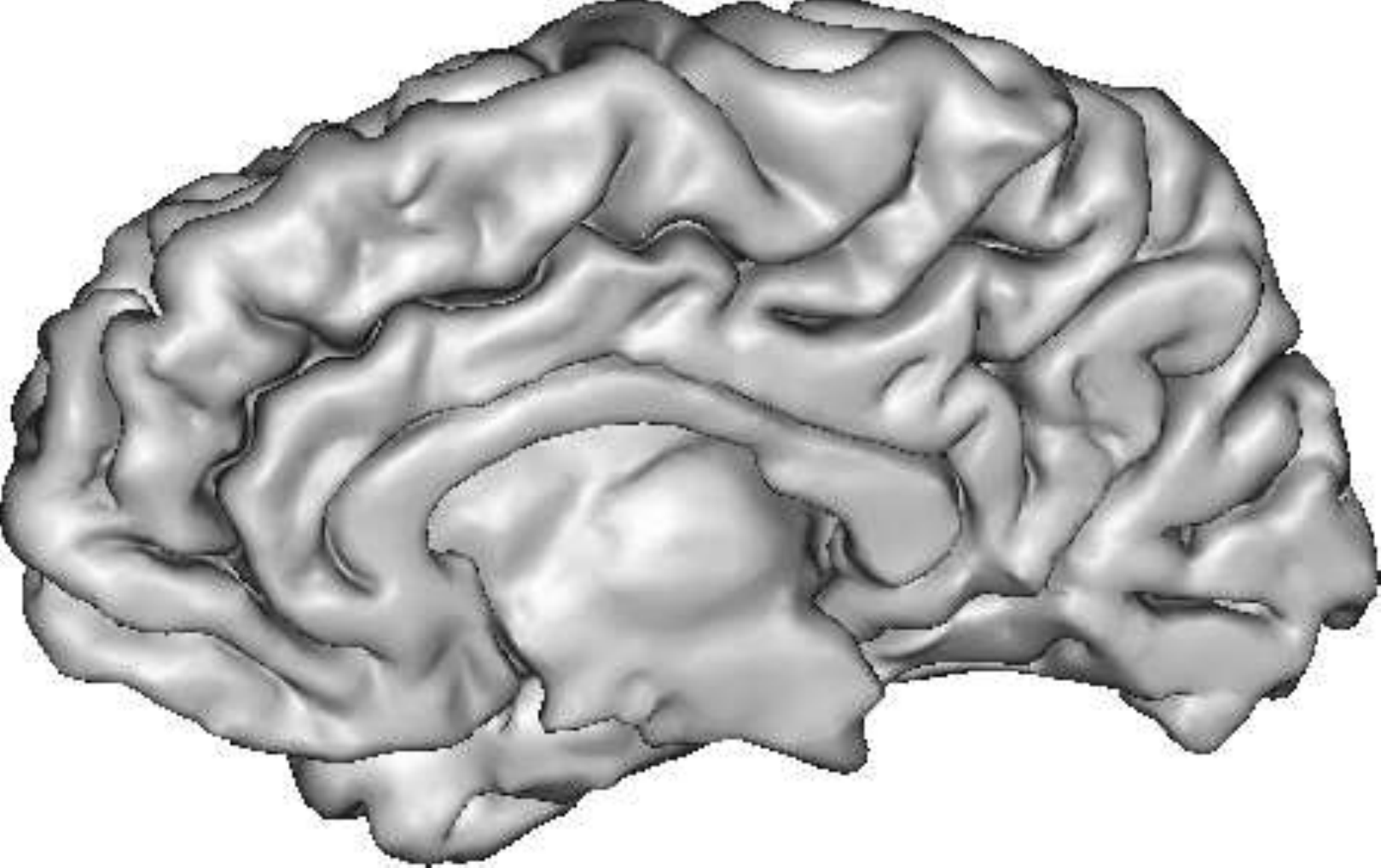}\\
  \raisebox{1.1cm}{lateral view}&
\includegraphics[width=2.8cm, height=2cm]{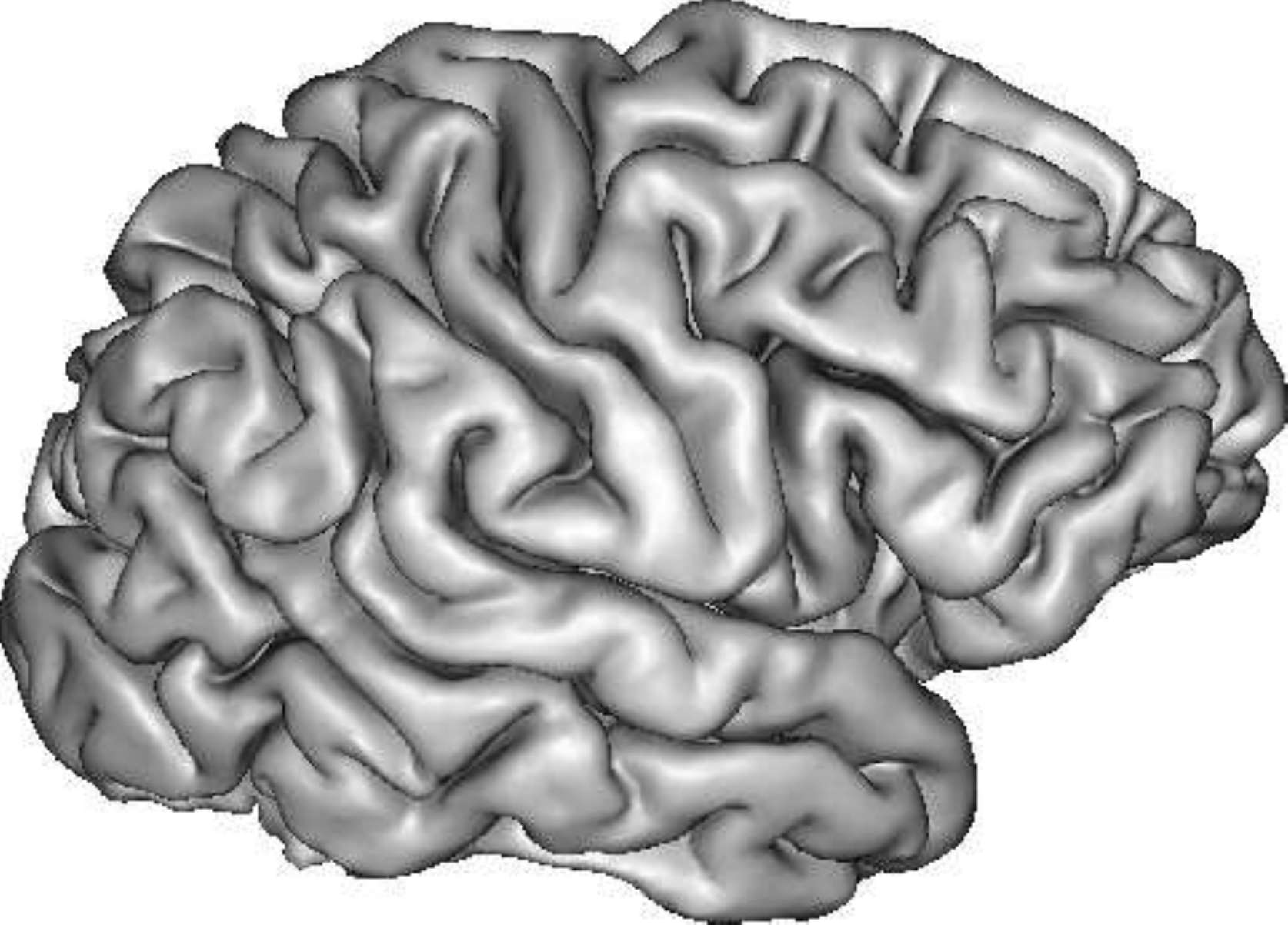}&
\includegraphics[width=2.8cm, height=2cm]{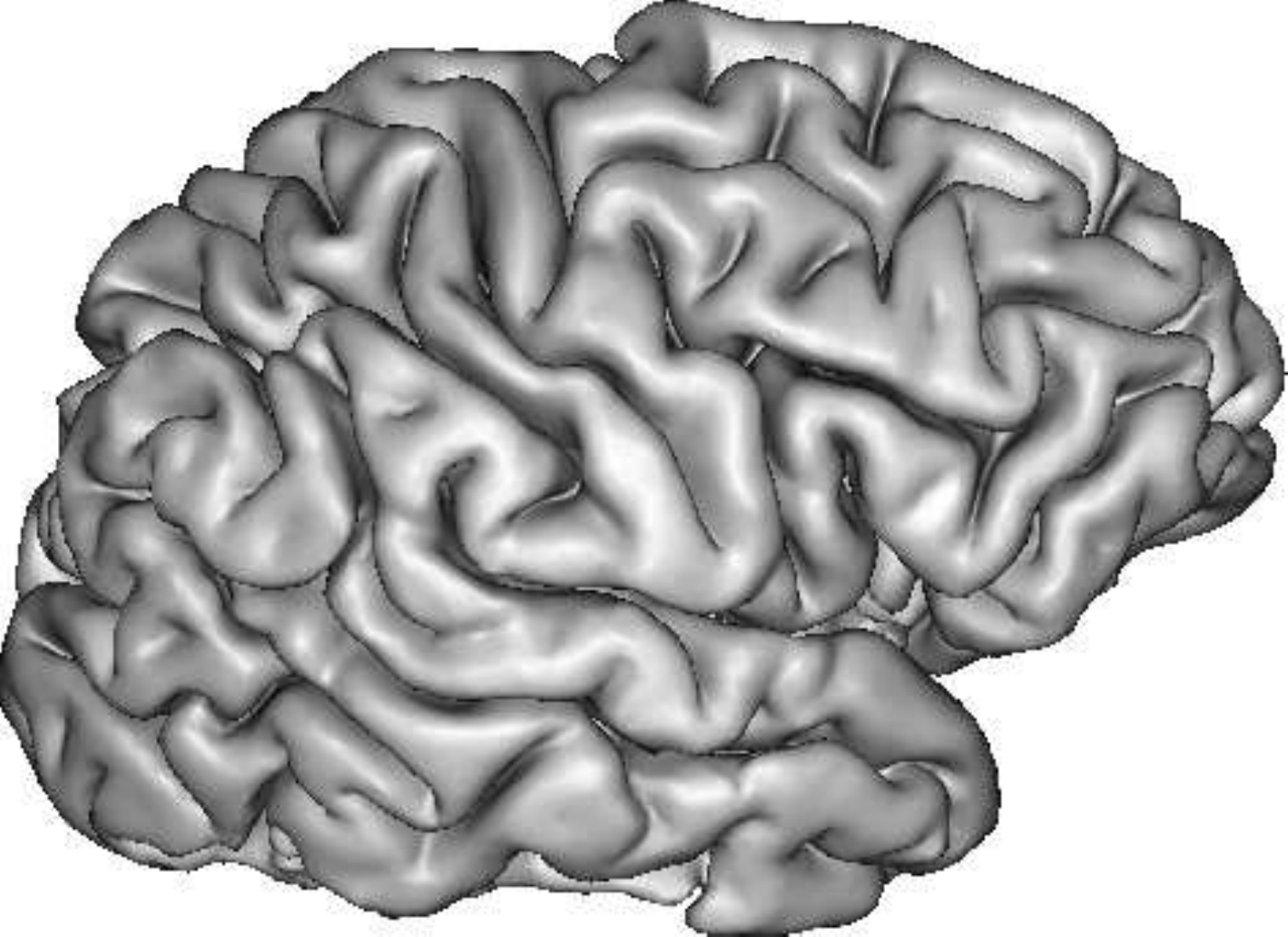}&
\includegraphics[width=2.8cm, height=2cm]{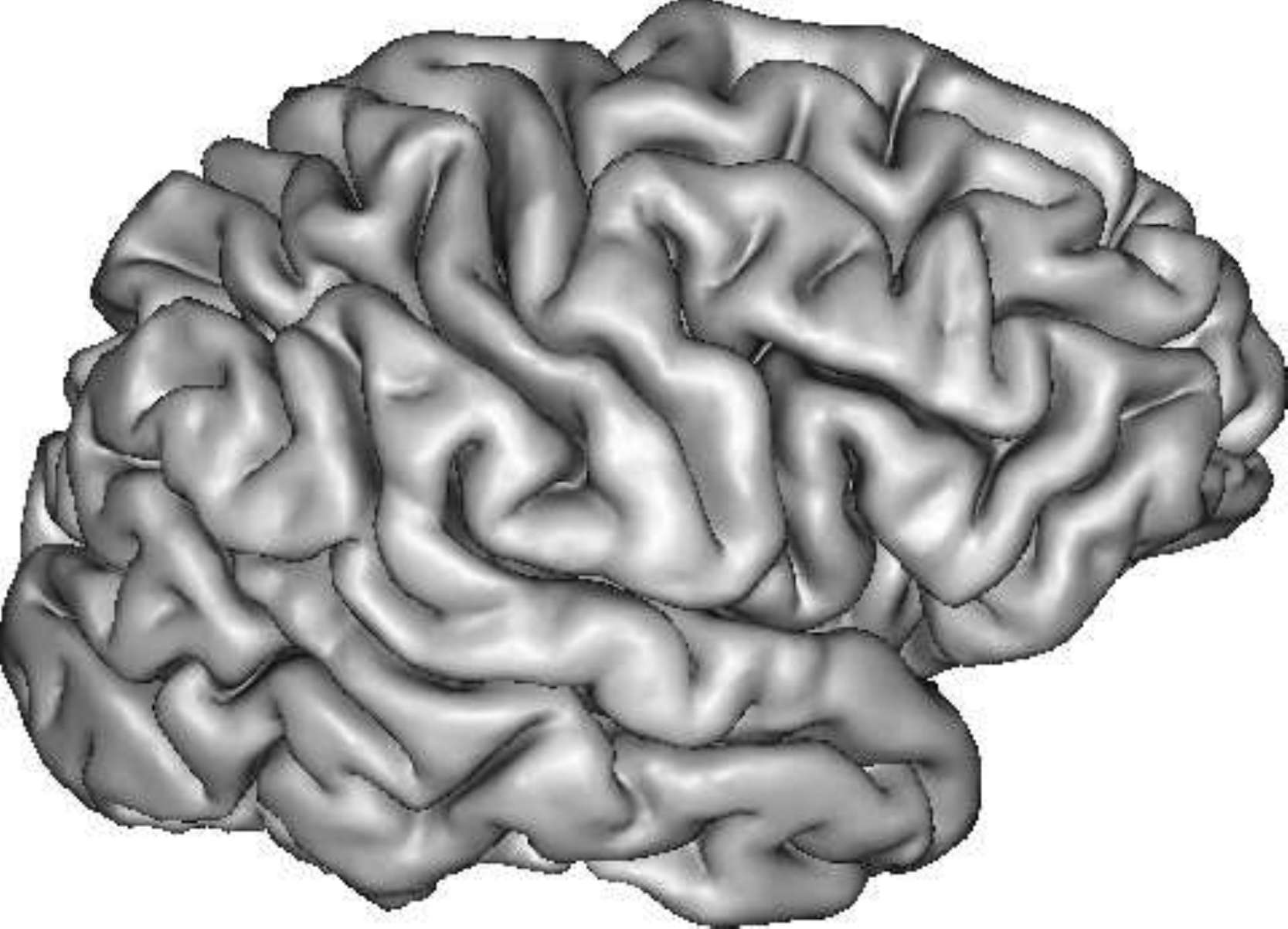}\\
\end{tabular}
\caption{Gray matter surface extraction based on reconstructed slices using \mSENSE~and 3D-UWR-SENSE 
for $R=4$. Results obtained with $R=1$ are also provided as a ground truth. 
\label{fig:classifsurf}}
\end{figure}

For the lateral view, one can easily conclude that extracted surfaces are very similar. 
However, the medial view shows that \mSENSE~is not able to correctly segment the brainstem (see right red ellipsoid in
the \mSENSE~medial view). Moreover, results with \mSENSE~are more noisy compared to 3D-UWR-SENSE (see left red ellipsoid 
in the \mSENSE~medial view). In contrast, the calcarine sulcus is slightly less accurately extracted with
our approach.\\
It is also worth noticing that similar results have been obtained \ADDED{on 14 other subjects}.

\subsection{Functional datasets}\label{subsec:func}

For fMRI data, a Gradient-Echo EPI~(GE-EPI) sequence has been used 
($\text{TE}=30~\rm ms$, $\text{TR}=2400~\rm ms$, slice thickness = 3~mm, transversal orientation, 
FOV = $192 \times 192~\mathrm{mm}^2$, flip angle $= 81 ^\circ$) during a cognitive \textit{localizer}~\cite{Pinel_07}
protocol. 
Slices have been collected in a sequential order (slice n$^\circ$1 in feet, last slice to head) using the
same 32-channel receiver coil to cover the whole brain in 39 slices for the two acceleration factors $R=2$ and $R=4$.
This leads to a spatial resolution of $2 \times 2 \times 3~\rm{mm}^3$ and a data matrix size of
$96 \times 96 \times 39$ for accelerated acquisitions.

This experiment has been designed to map auditory, visual and motor brain functions as well as higher cognitive
tasks such as number processing and language comprehension~(listening and reading). It consisted of a single session of $N_r = 128$ scans. The paradigm was a fast event-related design comprising 
sixty auditory, visual and motor stimuli, defined in ten experimental conditions~(auditory and visual sentences,
auditory and visual calculations, left/right auditory and visual clicks, horizontal and vertical checkerboards). 
Since data at $R=1$, $R=2$ and $R=4$ were acquired for each subject, acquisition orders have been equally balanced
between these three reduction factors over the fifteen subjects.

\subsubsection{FMRI reconstruction pipeline}

For each subject, fMRI data were collected at the $2\times2~\mathrm{mm}^2$ spatial in-plane resolution using different reduction factors~($R = 2$ or $R = 4$). 
Based on the raw data files delivered by the scanner, reduced FOV EPI images were reconstructed as detailed in Fig.~\ref{fig:reading_data}. This reconstruction 
is performed in two stages:
\begin{itemize}
\item[\textit{i)}] \ADDED{the \textit{1D $k$-space regridding} (blip gradients along phase encoding direction applied 
in-between readout gradients) }
to account for the
non-uniform $k$-space sampling during readout gradient ramp, which occurs in fast MRI sequences like GE-EPI;
 \item[\textit{ii)}] the \textit{Nyquist ghosting correction} to remove the odd-even echo inconsistencies during $k$-space acquisition of EPI images.
\end{itemize}

\begin{figure}[!ht]
\centering
 \includegraphics[width=15cm, height=1.3cm]{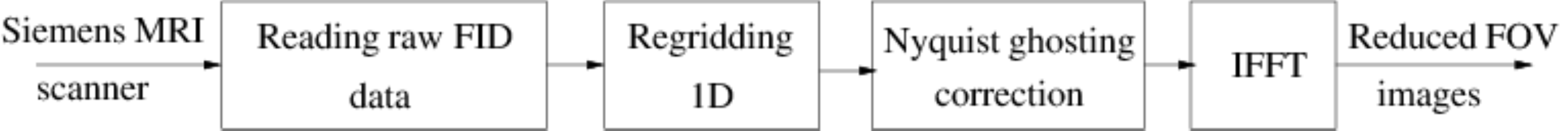}
\caption{Reconstruction pipeline of reduced FOV EPI images from the raw FID data.\label{fig:reading_data}}
\end{figure}

It must be emphasized here that since no interleaved $k$-space sampling is performed during the acquisition, 
and since the central lines of the $k$-space are not acquired for each TR due to the available imaging sequences on the Siemens scanner, \emph{kt}-FOCUSS-like methods are not applicable on the available dataset.

\noindent Once the reduced FOV images are available, the proposed pMRI 4D-UWR-SENSE algorithm and its early 
UWR-SENSE version have been
utilized in a final step to reconstruct the full FOV EPI images and compared to the \mSENSE~Siemens solution. 
For the wavelet-based regularization, dyadic \textit{Symmlet} orthonormal wavelet 
bases~\cite{daubechies_92} associated with filters of length 8 have been used over $j_{\mathrm{max}}=3$ 
resolution levels. The reconstructed EPI images then enter in our fMRI study in order to measure the impact 
of the reconstruction method choice on brain activity detection. Note also that the proposed reconstruction algorithm 
requires the estimation of the coil sensitivity maps~(matrix $\Sb(\cdot)$ in Eq.~\eqref{eq:matriciel}). 
As proposed in~\cite{pruessmann_99}, the latter were estimated by dividing the coil-specific images by the 
module of the Sum Of Squares~(SOS) images, which are computed from the specific acquisition of the $k$-space 
centre~(24 lines) before the $N_r$ scans. The same sensitivity map estimation is then used for all the compared 
methods. 
Fig.~\ref{fig:slice_Axial} compares the two pMRI reconstruction algorithms to illustrate on axial, coronal 
and sagittal EPI slices how the \mSENSE~reconstruction artifacts have been removed using the 4D-UWR-SENSE 
approach. Reconstructed \mSENSE~images actually present large artifacts located both at the centre and 
boundaries of the brain in sensory and cognitive regions~(temporal lobes, frontal and motor cortices, ...). 
This results in SNR loss and thus may have a dramatic impact for activation detection in these brain regions. 
Note that these conclusions are reproducible across subjects although the artifacts may appear on different 
slices~(see red circles in Fig.~\ref{fig:slice_Axial}). 
 One can also notice that some residual artifacts still exist in the reconstructed images with our 
 pipeline especially for $R=4$. Such strong artifacts are only attenuated and not fully removed 
 because of the high level of information loss at $R=4$.
\begin{figure}[!ht]
\centering
\begin{tabular}{c| c c |c }
\cline{3-4}
&&\mSENSE&4D-UWR-SENSE\\
&\raisebox{1.1cm}{Axial}&
\includegraphics[width=2.2cm, height=2cm]{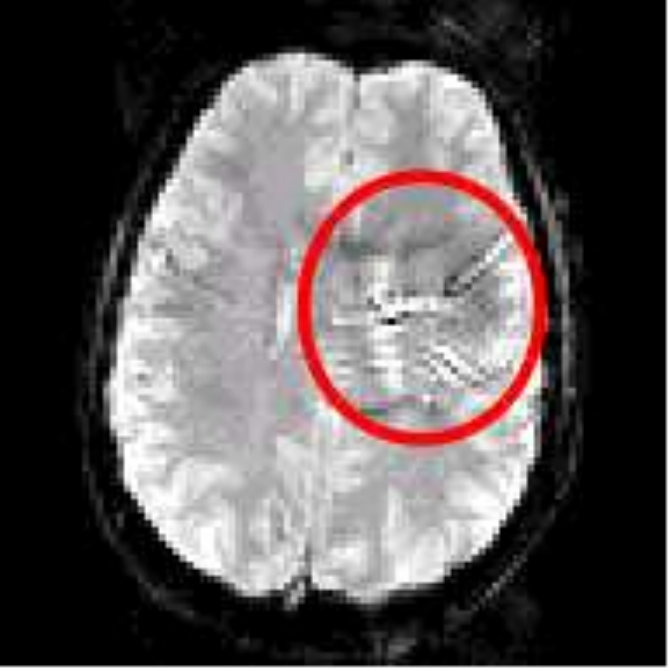}&
\includegraphics[width=2.2cm, height=2cm]{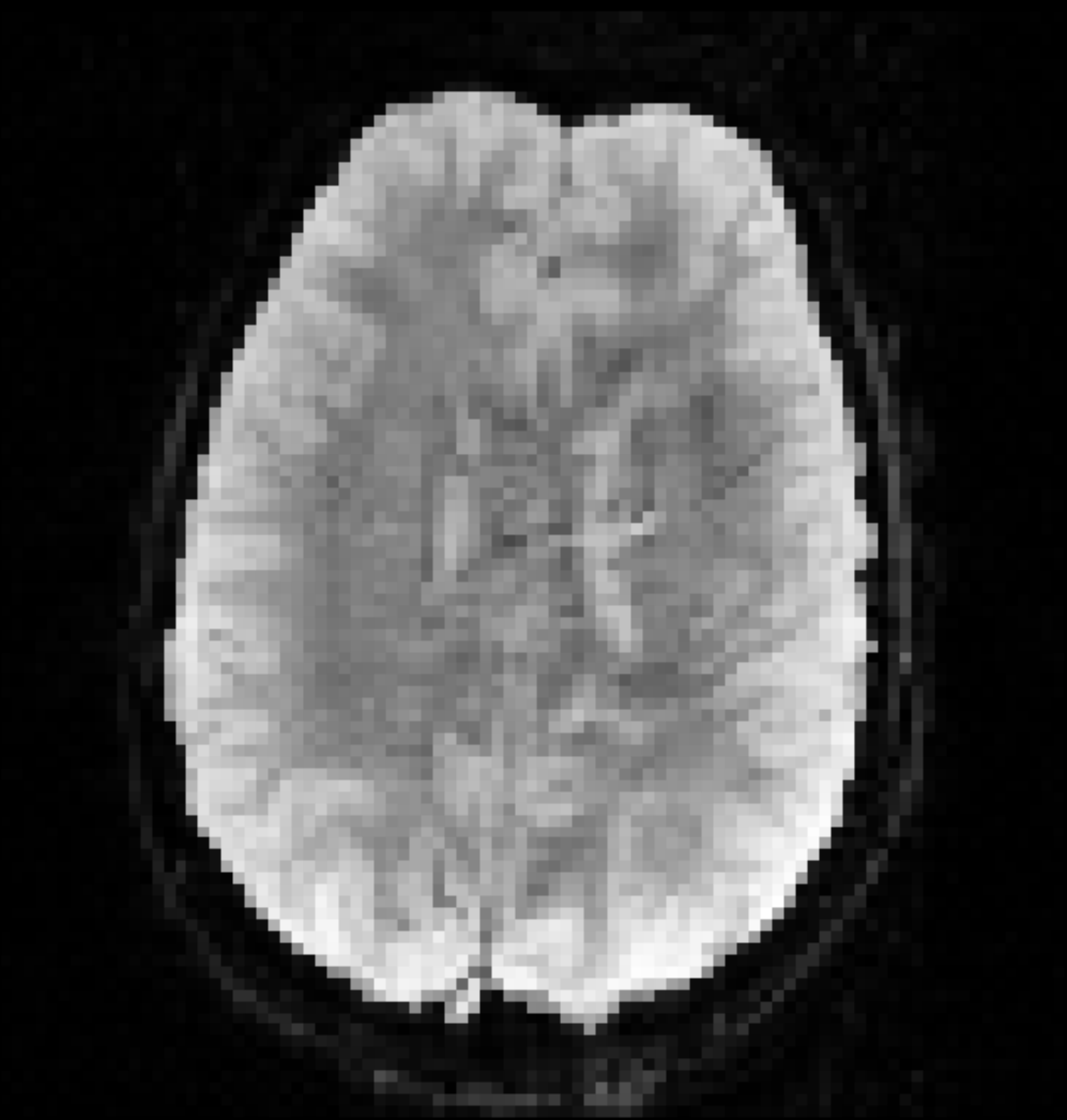}\\
\raisebox{1.1cm}{$R=2$}&
\raisebox{1.1cm}{Coronal}&\includegraphics[width=2.2cm, height=1.5cm]{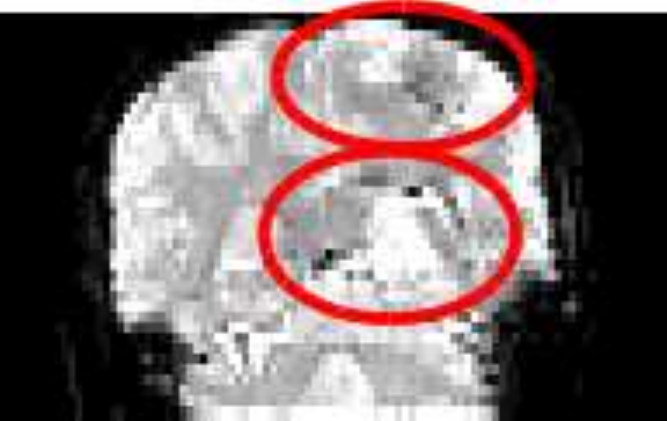}&
\includegraphics[width=2.2cm, height=1.5cm]{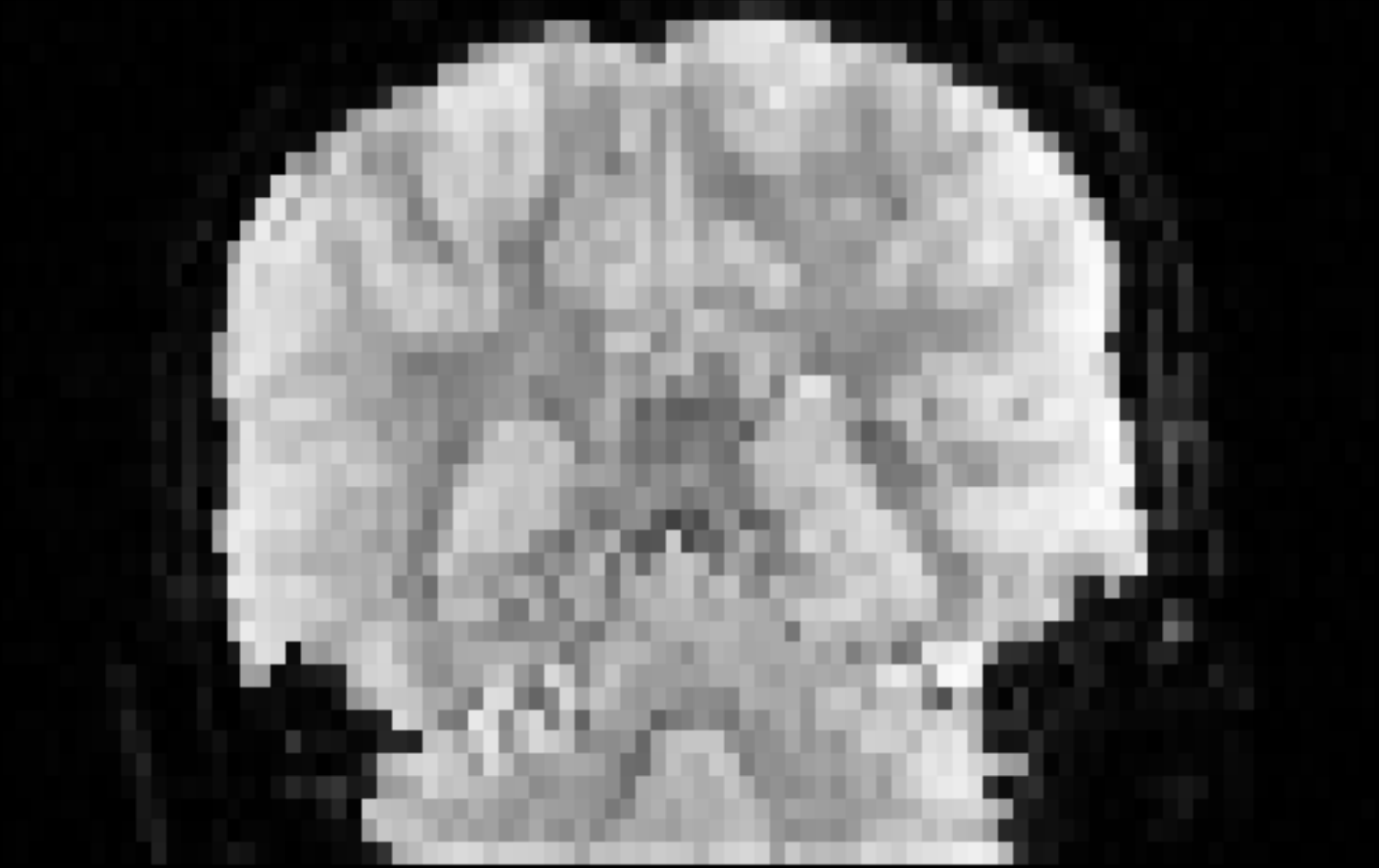}\\
&\raisebox{1.1cm}{Sagittal}&\includegraphics[width=2.2cm, height=1.5cm]{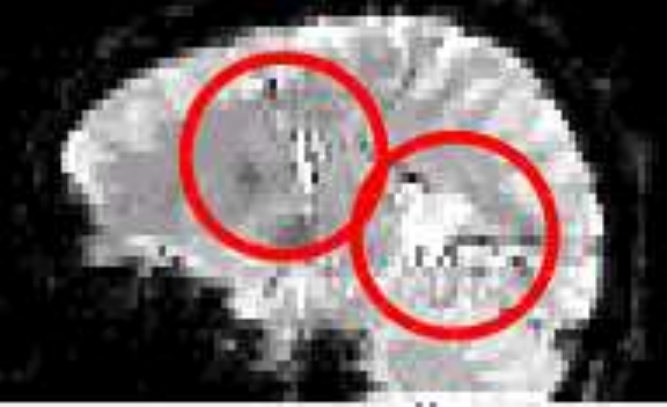}&
\includegraphics[width=2.2cm, height=1.5cm]{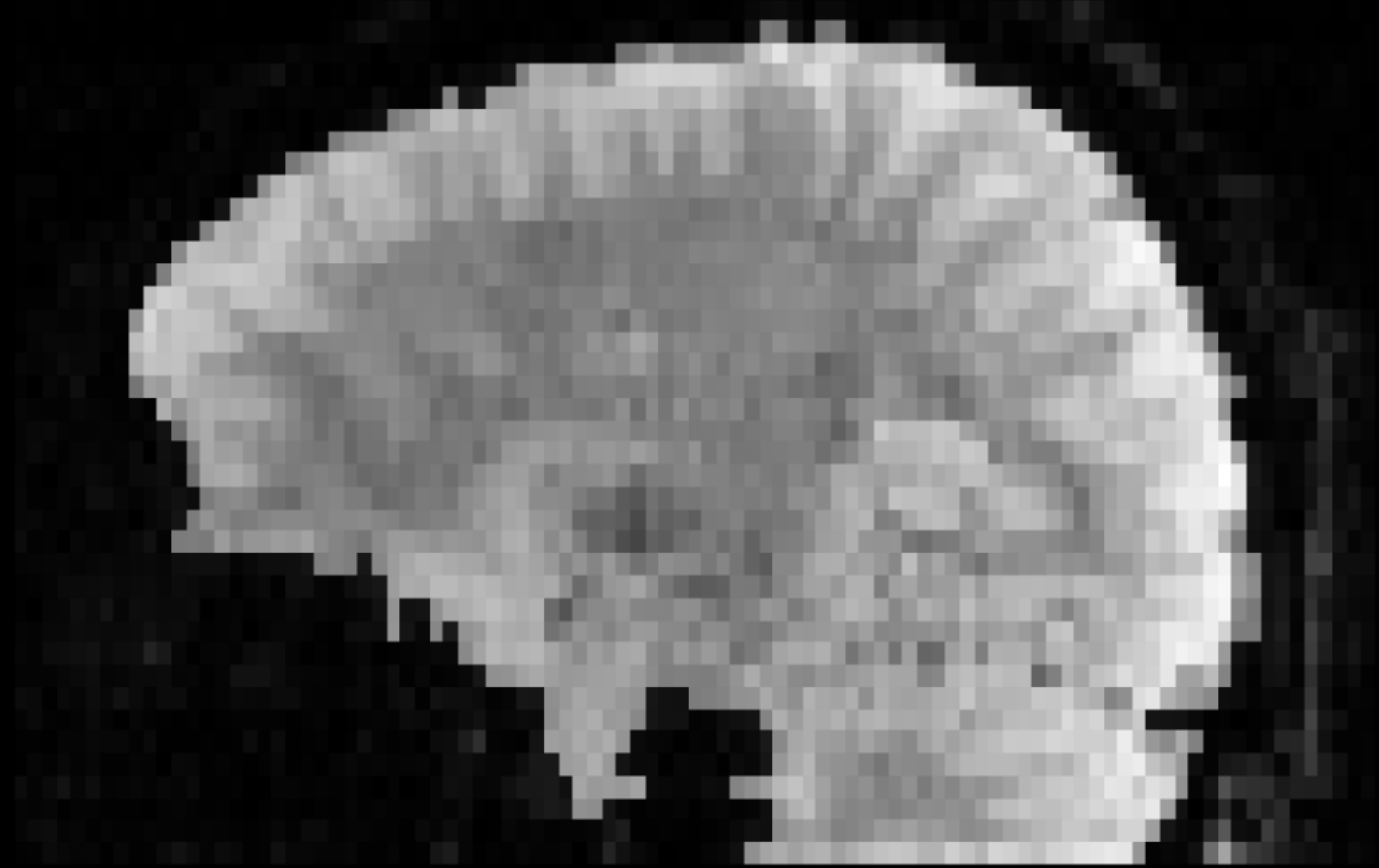}\\
\hline
&\raisebox{1.1cm}{Axial}&\includegraphics[width=2.2cm, height=2cm]{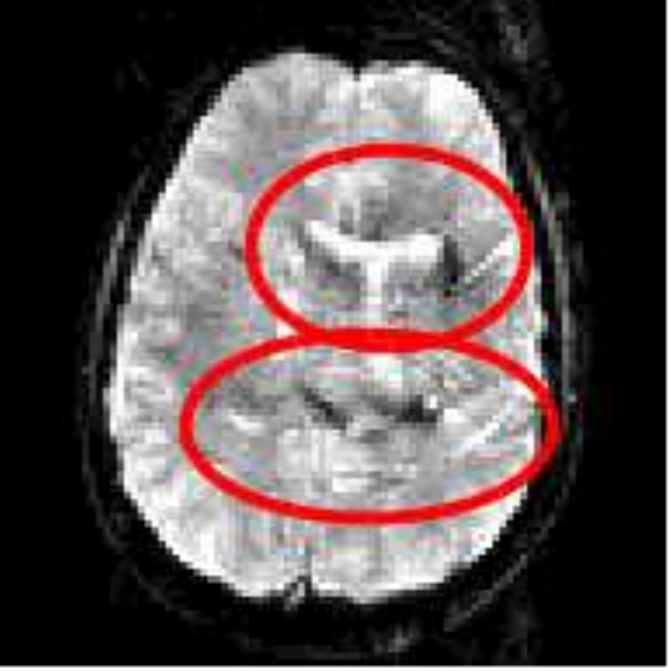}&
\includegraphics[width=2.2cm, height=2cm]{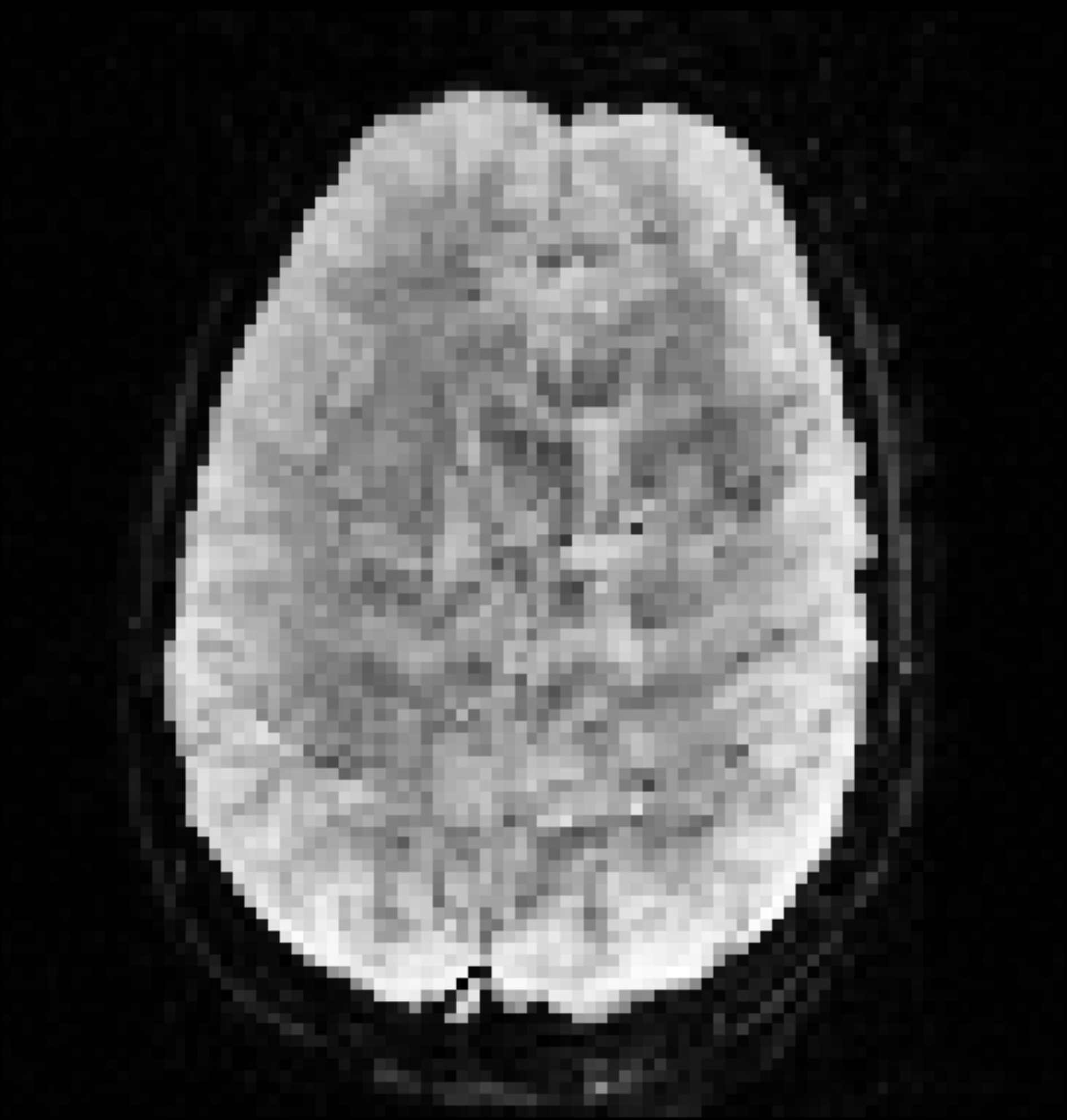}\\
\raisebox{1.1cm}{$R=4$}&
\raisebox{1.1cm}{Coronal}&
\includegraphics[width=2.2cm, height=1.5cm]{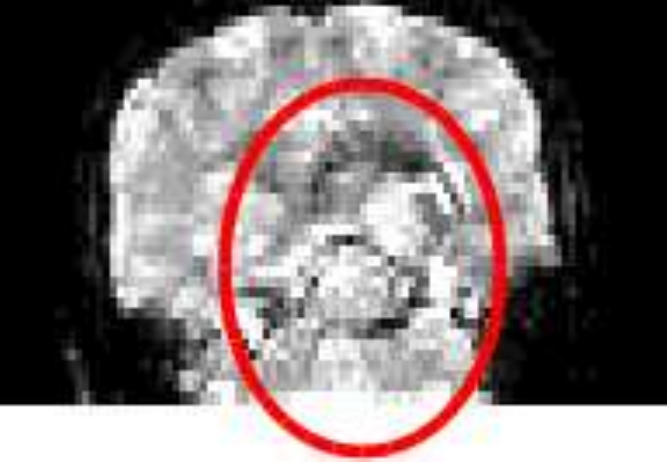}&
\includegraphics[width=2.2cm, height=1.5cm]{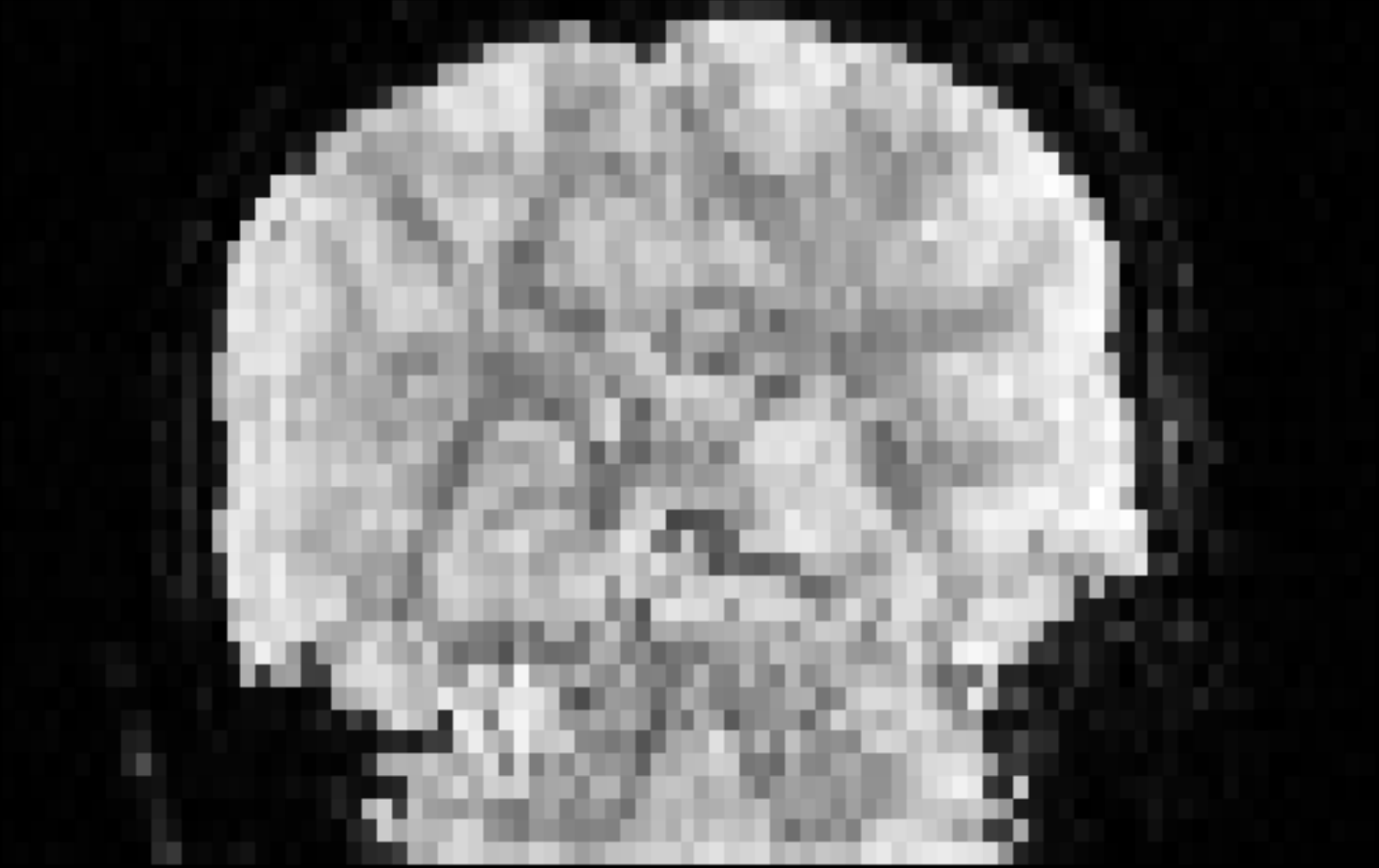}\\
&\raisebox{1.1cm}{Sagittal}&
\includegraphics[width=2.2cm, height=1.5cm]{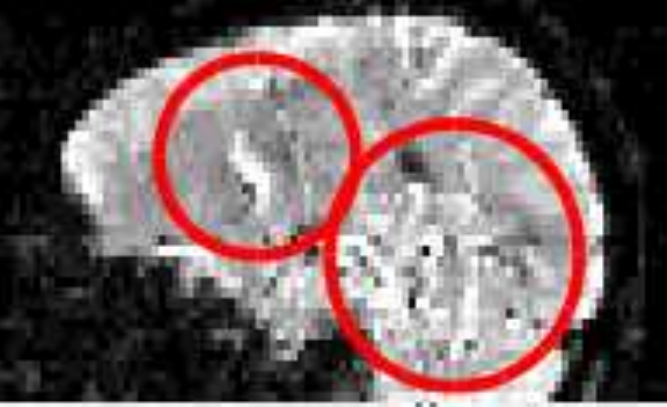}&
\includegraphics[width=2.2cm, height=1.5cm]{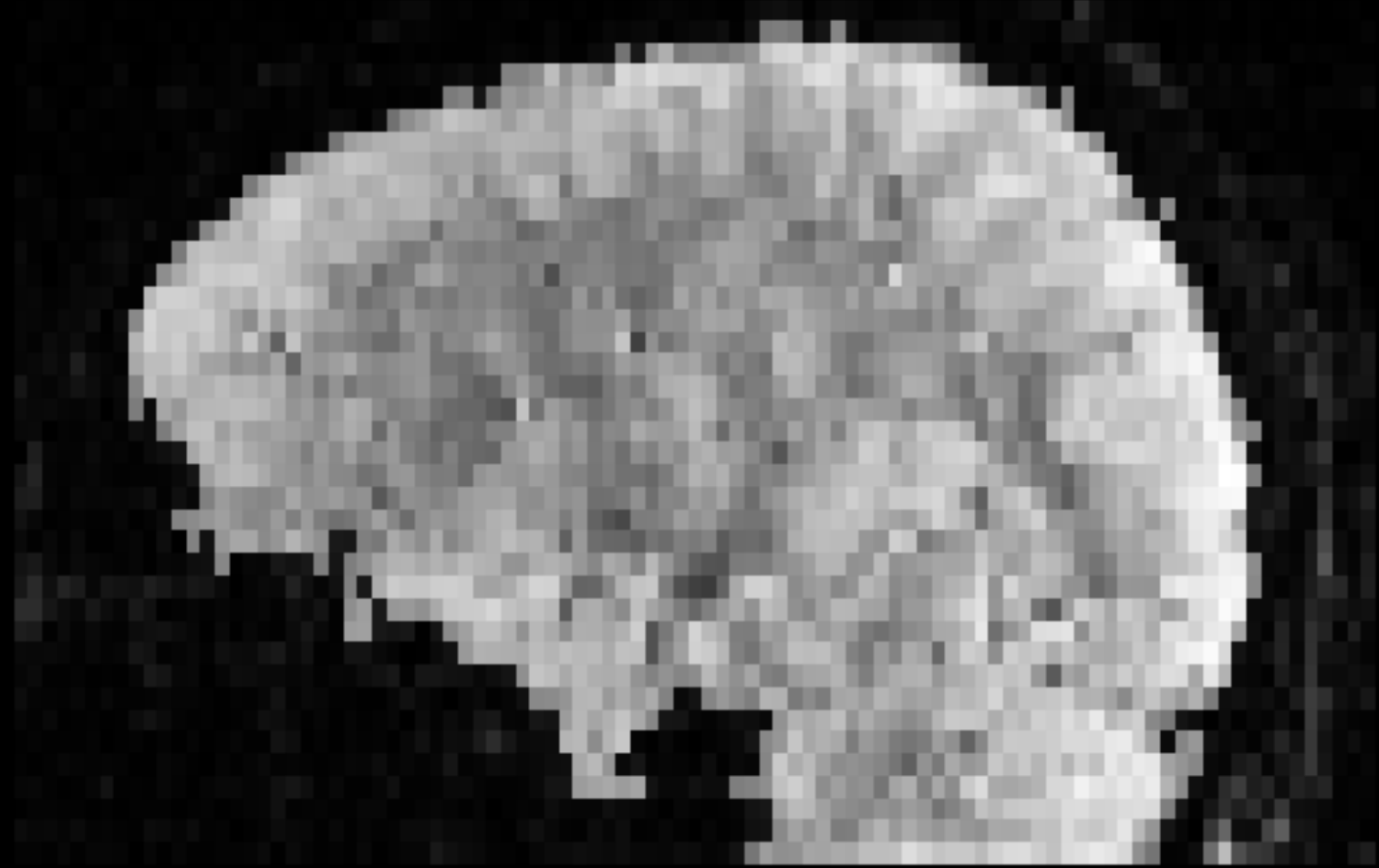}
\end{tabular}
\vspace{-.1cm}
\caption{\ADDED{ \textbf{Axial}, \textbf{Coronal} and \textbf{Sagittal} reconstructed slices using \mSENSE~and 4D-UWR-SENSE for $R=2$ and $R=4$ with 
$2\times 2~\rm{mm}^2$ in-plane spatial resolution. Red circles and ellipsoids indicate the position of reconstruction artifacts using \mSENSE. 
} \label{fig:slice_Axial}}
\end{figure}

Regarding computational load, the \mSENSE~algorithm is carried out on-line and remains compatible with real time processing. On the other hand, our pipeline is carried out off-line and requires more computations. For illustration purpose, on a biprocessor quadcore {\tt Intel  Xeon CPU}@~2.67GHz, one EPI slice is reconstructed in 4~s using the UWR-SENSE algorithm. Using parallel computing strategy and multithreading~(through the \texttt{OMP} library), each EPI volume consisting of 40 slices is reconstruced in 22~s. This makes the whole series of 128 EPI images available in about 47~min. By contrast, the proposed 4D-UWR-SENSE achieves the reconstruction of the series in about 40 min, but requires larger memory space 
due to large data volume processed simultaneously. 
 \subsubsection{fMRI data pre-processing}
\label{subsec:Preprocs}

Irrespective to the reconstruction pipeline, the full FOV fMRI images were then preprocessed using the SPM5 
software\footnote{\url{http://www.fil.ion.ucl.ac.uk/spm/software/spm5}}: preprocessing involves realignment, 
correction for motion and differences in slice acquisition time, spatial normalization, and smoothing with an isotropic Gaussian kernel of $4$mm full-width at half-maximum. 
Anatomical normalization to MNI space was performed by coregistration of the functional images with the anatomical
$T_1$ scan acquired with the thirty-two channels head coil. 
Parameters for the normalization to MNI space were estimated by normalizing this scan to the $T_1$ MNI 
template provided by SPM5, and were subsequently 
applied to all functional images.  
Tab.~\ref{tab:mvt} illustrates the mean over scans of the absolute maximum motion 
parameters (translation and rotation) for each subject, as well as their group-level average value. 
One can notice through this table that, across the 15 subjects, motion parameters estimated on 
images reconstructed using \mSENSE~and 4D-UWR-SENSE are quite similar even if the mean values are slightly 
higher with 4D-UWR-SENSE. Two-tailed statistical tests conducted on the absolute displacement maxima for the 15 subjects
(Student-t), after a Bonferroni correction for multiple comparisons, confirm that the difference is not significant between 
\mSENSE~and our algorithm at a p-value threshold $\alpha=0.05$.  
Tab.~\ref{tab:mvt} illustrates obtained p-values for each motion parameter.

\begin{table}[!ht]
\centering 
\caption{Estimated maximum absolute motion parameters over time for each subject in terms of translation (in~mm) and rotation (in~$^\circ$) along the three spatial axes for 
$R=4$.}
\begin{tabular}{|c|c|c|c|c||c|c|c|}
\cline{3-8}
\multicolumn{2}{c|}{}&\multicolumn{3}{|c||}{Translation}&
\multicolumn{3}{c|}{Rotation}\\
\cline{3-8}
\multicolumn{2}{c|}{}&$x$&$y$&$z$&roll&pitch&yow\\
\hline
\multirow{16}{*}{\mSENSE}
&Subj. 1&0.24 & 0.02&0.05&  0.14& 0.10 &0.16\\
&Subj. 2&0.26 & 0.09&0.18&  0.48 & 0.12&0.25\\
&Subj. 3&0.21 & 0.2 &0.02&  0.50 & 0.07&0.18\\
&Subj. 4&0.12 & 0.21&0.33&  0.51 & 0.21&0.23\\
&Subj. 5&0.21 & 0.07&0.18&  0.18 & 0.22&0.10\\
&Subj. 6&0.24 & 0.11&0.07&  0.17 & 0.05&0.15\\
&Subj. 7&0.18 & 0.08&0.16&  0.32 & 0.31&0.34\\
&Subj. 8&0.10  & 0.06&0.21&  0.28 & 0.44&0.22\\
&Subj. 9&0.38 & 0.16&0.92&  0.29 & 0.43&0.17\\
&Subj. 10&0.19& 0.09&0.11&  0.18 & 0.18&0.22\\
&Subj. 11&0.03& 0.05&0.16&  0.18 & 0.17&0.05\\
&Subj. 12&0.02& 0.27&0.09&  0.54 & 0.18&0.13\\
&Subj. 13&0.10& 0.12&0.22&  0.30 & 0.04&0.04\\
&Subj. 14&0.06& 0.18&0.38&  0.06 & 0.06&0.07\\
&Subj. 15&0.09& 0.07&0.11&  0.11 & 0.15&0.09\\
\cline{2-8}
&\textbf{Mean}&\textbf{0.16}& \textbf{0.12}&\textbf{0.22}& \textbf{ 0.28} & \textbf{0.18}&\textbf{0.16}\\
\cline{1-8}
\multirow{16}{*}{4D-UWR-SENSE}
&Subj. 1 &0.18 & 0.02&0.05&  0.16 & 0.14&0.21\\
&Subj. 2 &0.20 & 0.07&0.21&  0.51 & 0.13&0.24\\
&Subj. 3 &0.20 & 0.27&0.02&  0.50 & 0.09&0.16\\
&Subj. 4 &0.20 & 0.2 &0.4 &  0.70 & 0.37&0.28\\
&Subj. 5 &0.05 & 0.27&0.3 &  0.20 & 0.22&0.17\\
&Subj. 6 &0.04 & 0.06&0.06&  0.17 & 0.02&0.17\\
&Subj. 7 &0.12 & 0.13&0.20&  0.46 & 0.31&0.34\\
&Subj. 8 &0.08 & 0.10&0.20&  0.27 & 0.40&0.20\\
&Subj. 9 &0.33 & 0.27&1.00&  0.25 & 0.34&0.20\\
&Subj. 10&0.13 & 0.18&0.09&  0.22 & 0.20&0.26\\
&Subj. 11&0.04 & 0.11&0.18&  0.18 & 0.17&0.03\\
&Subj. 12&0.02 & 0.26&0.09&  0.56 & 0.20&0.27\\
&Subj. 13&0.07 & 0.12&0.24&  0.30 & 0.05&0.05\\
&Subj. 14&0.07 & 0.32&0.6 &  0.16 & 0.07&0.1\\
&Subj. 15&0.11 & 0.09 &0.18&  0.20 & 0.13&0.12\\
\cline{2-8}
&\textbf{Mean}&\textbf{0.16}& \textbf{0.14}&\textbf{0.26}& \textbf{ 0.32} & \textbf{0.19}&\textbf{0.21}\\
\hline
\multicolumn{2}{|c|}{p-values}& 0.3852   & 0.3101   & 0.3539  &  0.3772   & 0.8712    &0.5558\\
\hline
\end{tabular}
\label{tab:mvt}
\end{table}

It is worth noting here that for the SPM pipeline, the reconstruction step is supposed to be a kind of black box that delivers images for both methods, and we aim at comparing statistical results based on these images. Despite the applied 3D spatial and temporal regularizations that may introduce a smoothing effect in our algorithm, the same slice-timing correction is for instance applied 
to both image sets. As discussed in Section~\ref{sec:discussion}, accounting for the sequence acquisition parameters~(interleaved or not, 2D or 
3D,...) during the reconstruction step is beyond the scope of the present paper.  

\subsubsection{Subject-level analysis}

A General Linear Model~(GLM) was constructed to capture stimulus-related BOLD response. As shown in Fig.~\ref{fig:design}, 
the design matrix relies on ten experimental conditions and is thus made up of twenty one regressors corresponding to stick 
functions convolved with the canonical Haemodynamic Response Function~(HRF) and its first temporal derivative, the last regressor 
modelling the baseline. This GLM was then fitted to the same acquired images but reconstructed using either the Siemens reconstructor 
or our own pipeline, which in the following is derived from the early UWR-SENSE method~\cite{Chaari_MEDIA_2011} and from its 
4D-UWR-SENSE extension we propose here.
\begin{figure}[!ht]
\centering
\includegraphics[width=6cm, height=4cm]{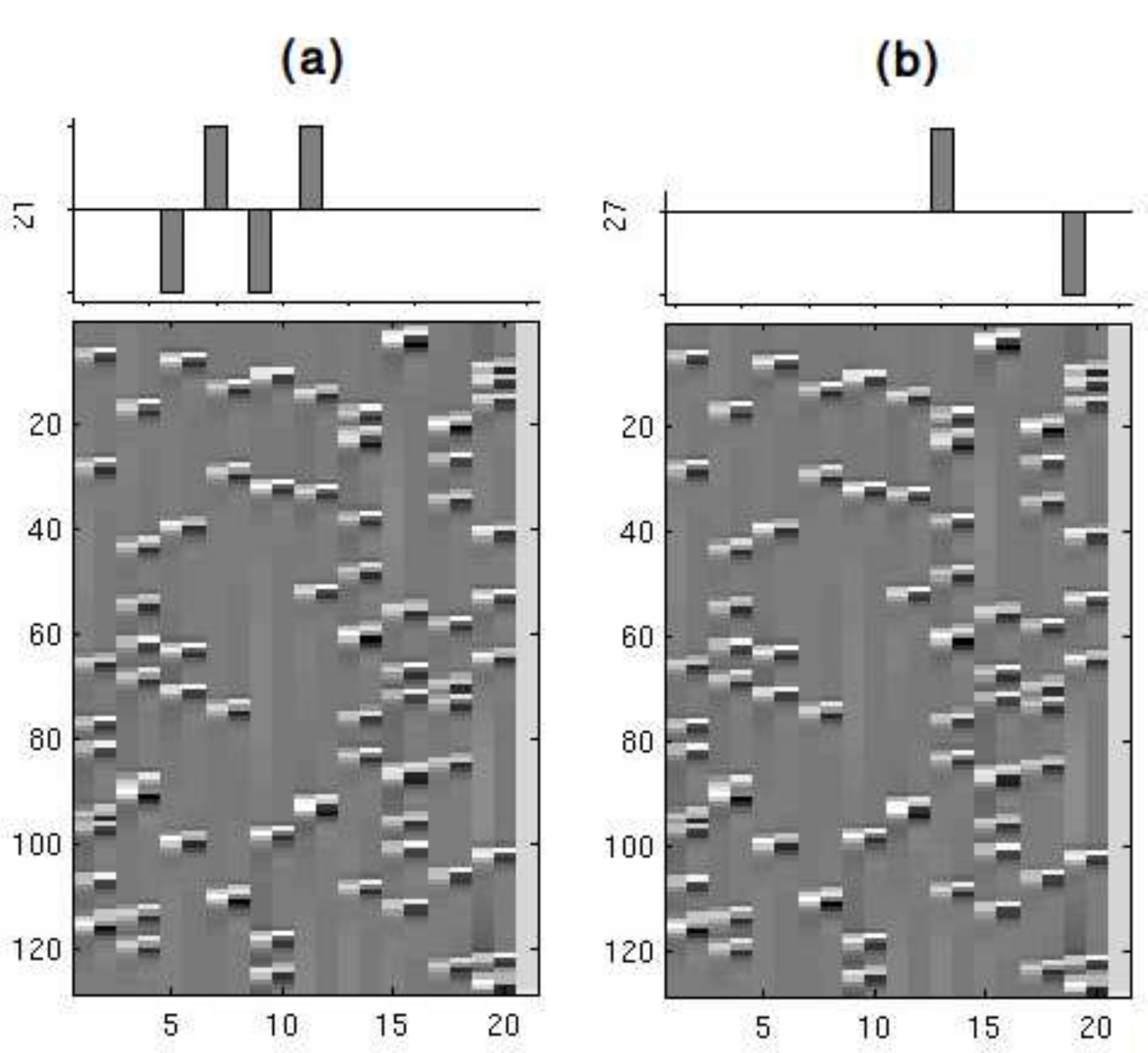}
\vspace{-.3cm}
\caption{(a): Design matrix and the \LcRc contrast involving two conditions~(grouping auditory and visual modalities); (b): design matrix and the \ACAS contrast involving four conditions~(sentence, computation, left click, right click).\label{fig:design}}
\end{figure}
Here, estimated contrast  images for motor responses and higher cognitive functions~(computation, language) were subjected to further analyses at the subject and group levels. These two analyses are complementary since the expected activations lie in different brain regions and thus can be differentially corrupted by reconstruction artifacts as outlined in Fig.~\ref{fig:slice_Axial}. More precisely, we studied:
\begin{itemize}

\item the {\bf Auditory computation vs. Auditory sentence}~(\ACAS) contrast which is supposed to elicit evoked activity in the frontal and parietal lobes, since solving mental arithmetic task involves working memory and more specifically the intra-parietal 
sulcus~\cite{Dehaene99}: see Fig.~\ref{fig:design}(b);

\item the {\bf Left click vs. Right click}~(\LcRc) contrast for which we expect evoked activity in the right motor cortex~(precentral gyrus, middle frontal gyrus). Indeed, the \LcRc contrast defines a compound comparison involving two motor stimuli which are presented either in the visual or auditory modality. This comparison aims therefore at detecting lateralization effect in the motor cortex: see Fig.~\ref{fig:design}(a). 

\end{itemize}

Interestingly, these two contrasts were chosen because they summarized well different situations~(large vs small activation 
clusters, distributed vs focal activation pattern, bilateral vs unilateral activity) that occurred for this paradigm when 
looking at sensory areas~(visual, auditory, motor) or regions involved in higher cognitive functions~(reading, calculation). 
In the following, our results are reported in terms of Student's $t$-maps thresholded at a cluster-level $p=0.05$ corrected for 
multiple comparisons according to the FamilyWise Error Rate~(FWER)~\cite{Nichols03,Brett_04}. Complementary statistical tables 
provide corrected cluster and voxel-level $p$-values, maximal $t$-scores and corresponding peak positions both for $R=2$ and $R=4$. 
Note that clusters are listed in a decreasing order of significance.

Concerning the \LcRc contrast on the data acquired with $R=2$,
Fig.~\ref{fig:res_T_Lc-Rc}~[top] shows that all reconstruction methods
enable to retrieve the expected activation in the right precentral gyrus. However, when looking more carefully 
at the statistical results~(see Tab.~\ref{tab:StatRes2allRcLc}), our pipeline and especially the 
4D-UWR-SENSE algorithm retrieves an additional cluster in the right middle frontal gyrus. On data acquired 
with $R=4$, the same \LcRc contrast elicits similar activations, i.e. in the same region. As demonstrated in 
Fig.~\ref{fig:res_T_Lc-Rc}~[bottom], this activity is enhanced when pMRI reconstruction is performed with our 
pipeline. Quantitative results in Tab.~\ref{tab:StatRes2allRcLc} confirm numerically what can be observed 
in Fig.~\ref{fig:res_T_Lc-Rc}: larger clusters with higher local $t$-scores are detected using the 4D-UWR-SENSE 
algorithm, both for $R=2$ and $R=4$. Also, a larger number of clusters is retrieved for $R=2$ using wavelet-based 
regularization.

In order to investigate the smoothing effect introduced by our algorithm, spatial smoothing in the 
pre-processing pipeline 
has been turned off and statistical results are illustrated in Fig.~\ref{fig:res_T_Lc-Rc}~[right] and 
Tab.~\ref{tab:StatRes2allRcLc} (Unsmoothed 4D-UWR-SENSE). As expected, qualitative and quantitative results 
show that deactivating the spatial smoothing gives slightly higher $t$-score values for activation maxima. However, 
smaller activated clusters are detected compared to results obtained based on smoothed data. 
As regards the temporal regularization effect, statistical results (not shown here) obtained with 3D-UWR-SENSE 
reconstructed images show intermediate performance which lies between those of the 2D (UWR-SENSE) and 4D (4D-UWR-SENSE) 
versions. Indeed, such a regularization helps improving the BOLD signal contrast which allows us to retrieve higher 
activation peaks.

\begin{figure}[!ht]
\centering
\begin{tabular}{c c c c || c}
&\mSENSE&UWR-SENSE&4D-UWR-SENSE&
\begin{minipage}{3cm}
\hspace{.4cm}Unsmoothed \\ 4D-UWR-SENSE\\
\end{minipage}
\\
\hspace*{-0.4cm}\raisebox{1.2cm}{$R=2$}&
 \includegraphics[width=2.5cm, height=2.5cm]{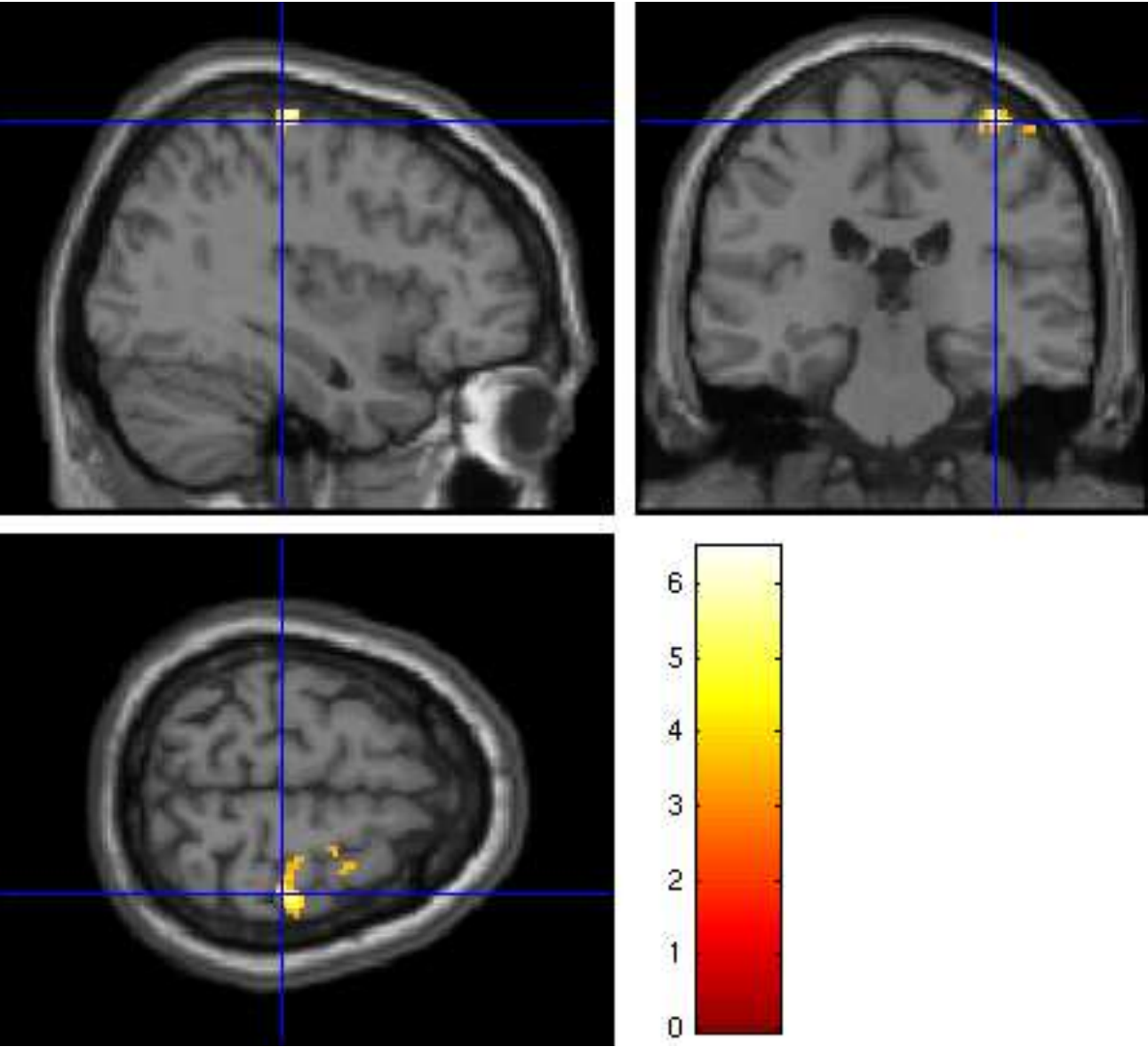}&
\hspace*{-0.3cm}\includegraphics[width=2.5cm, height=2.5cm]{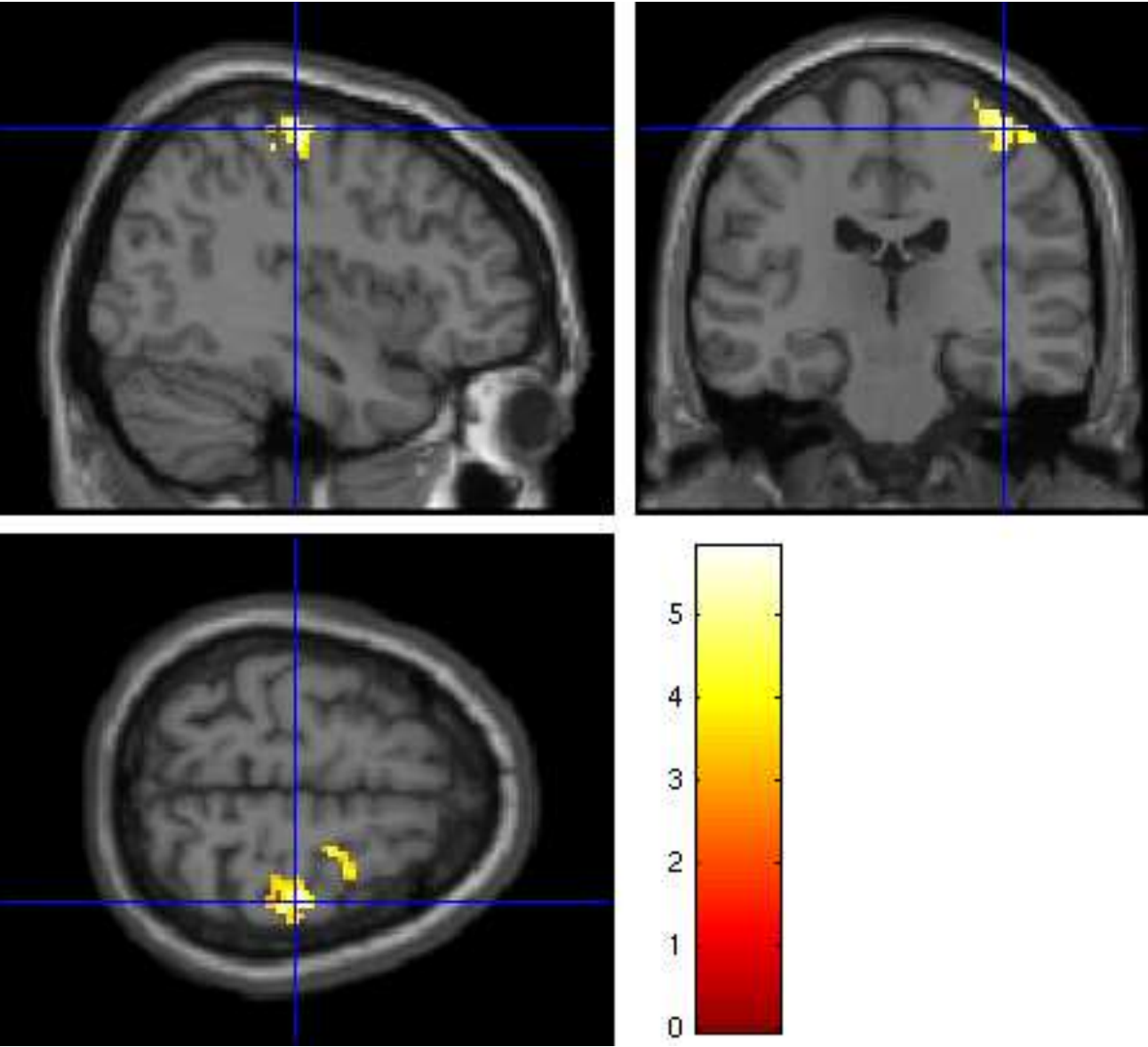}&
\hspace*{-0.3cm}\includegraphics[width=2.5cm, height=2.5cm]{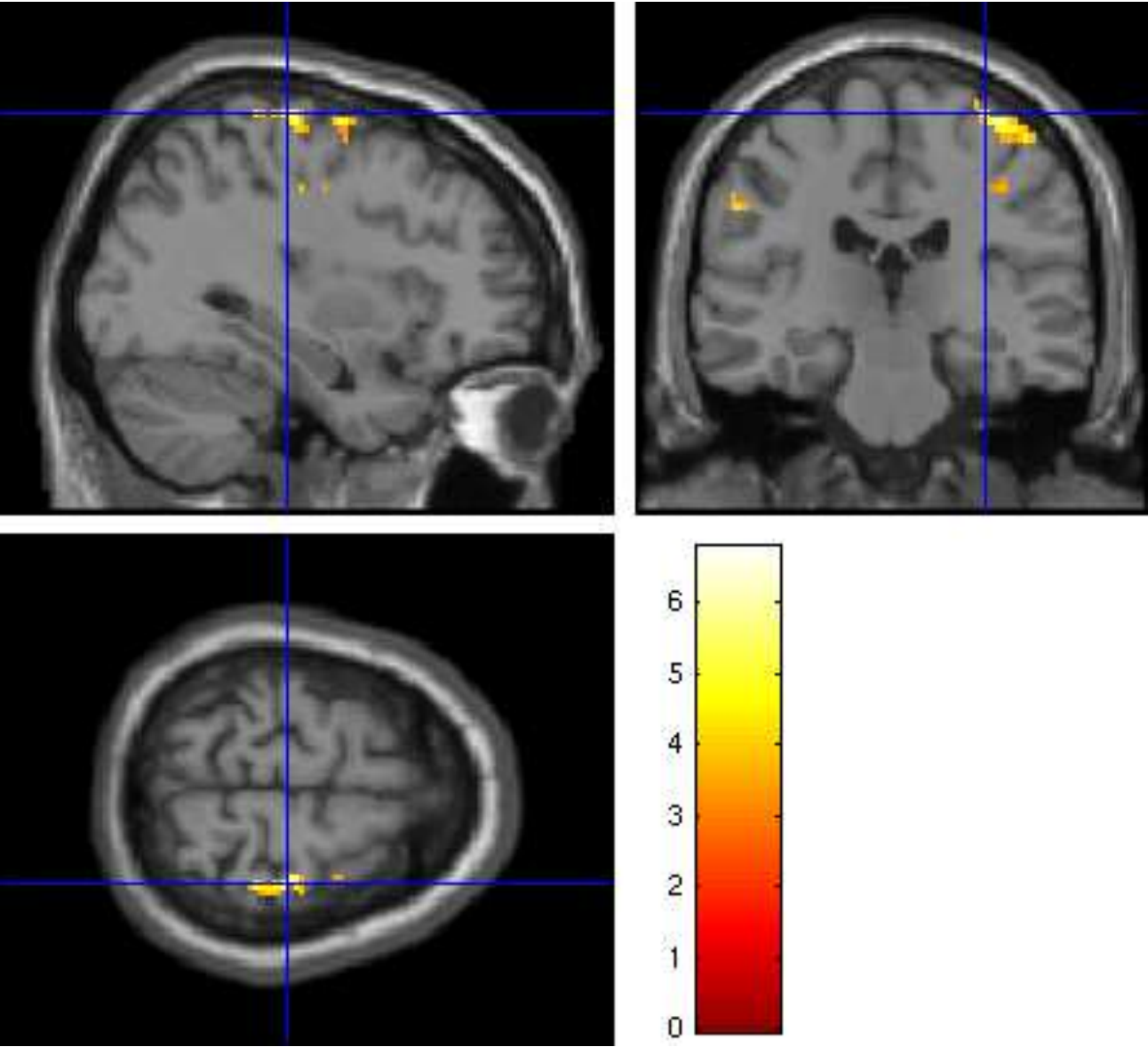}&
\hspace*{-0.1cm}\includegraphics[width=2.5cm, height=2.5cm]{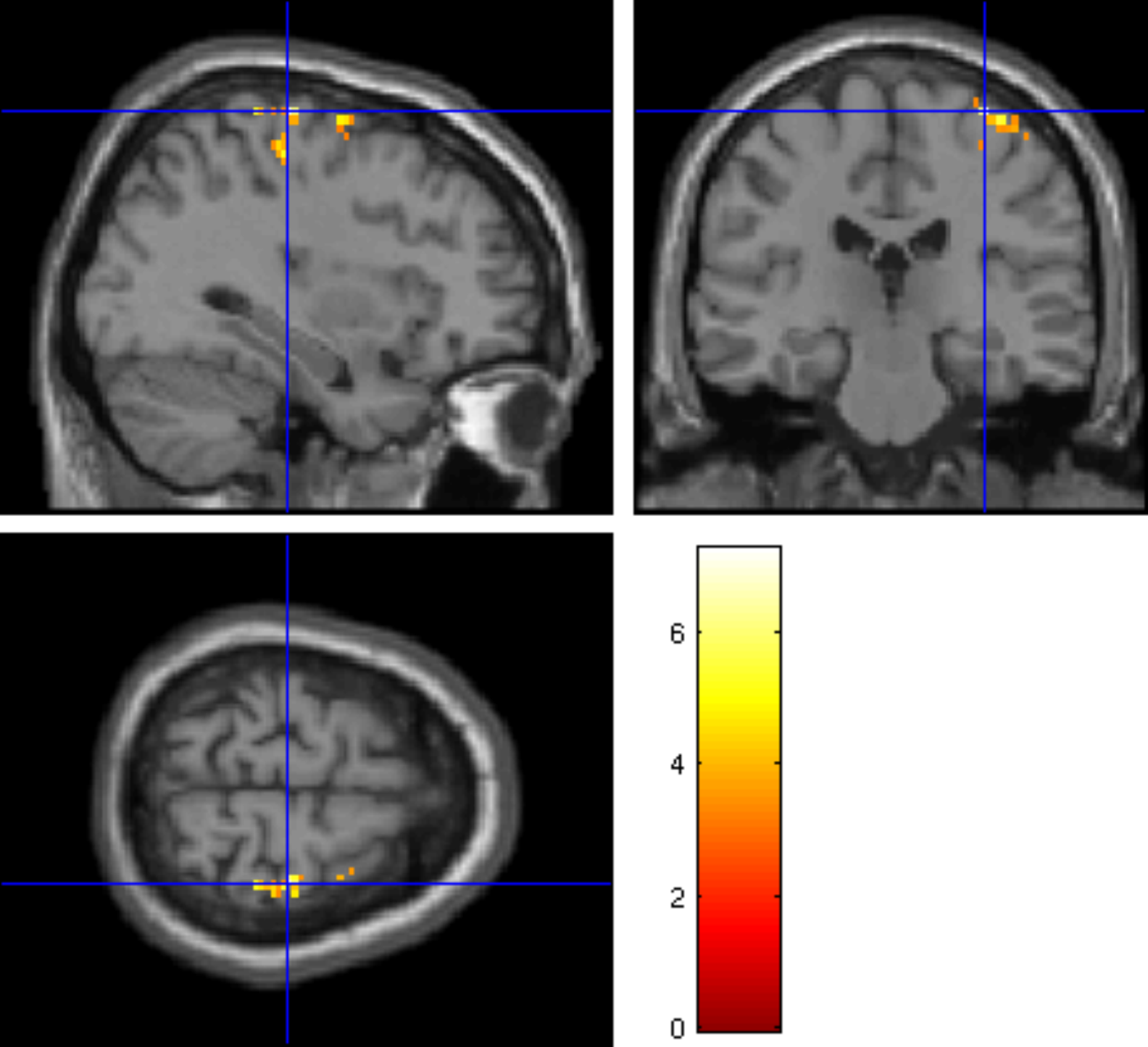}\\
\hspace*{-0.4cm}\raisebox{1.2cm}{$R=4$}&
\includegraphics[width=2.5cm, height=2.5cm]{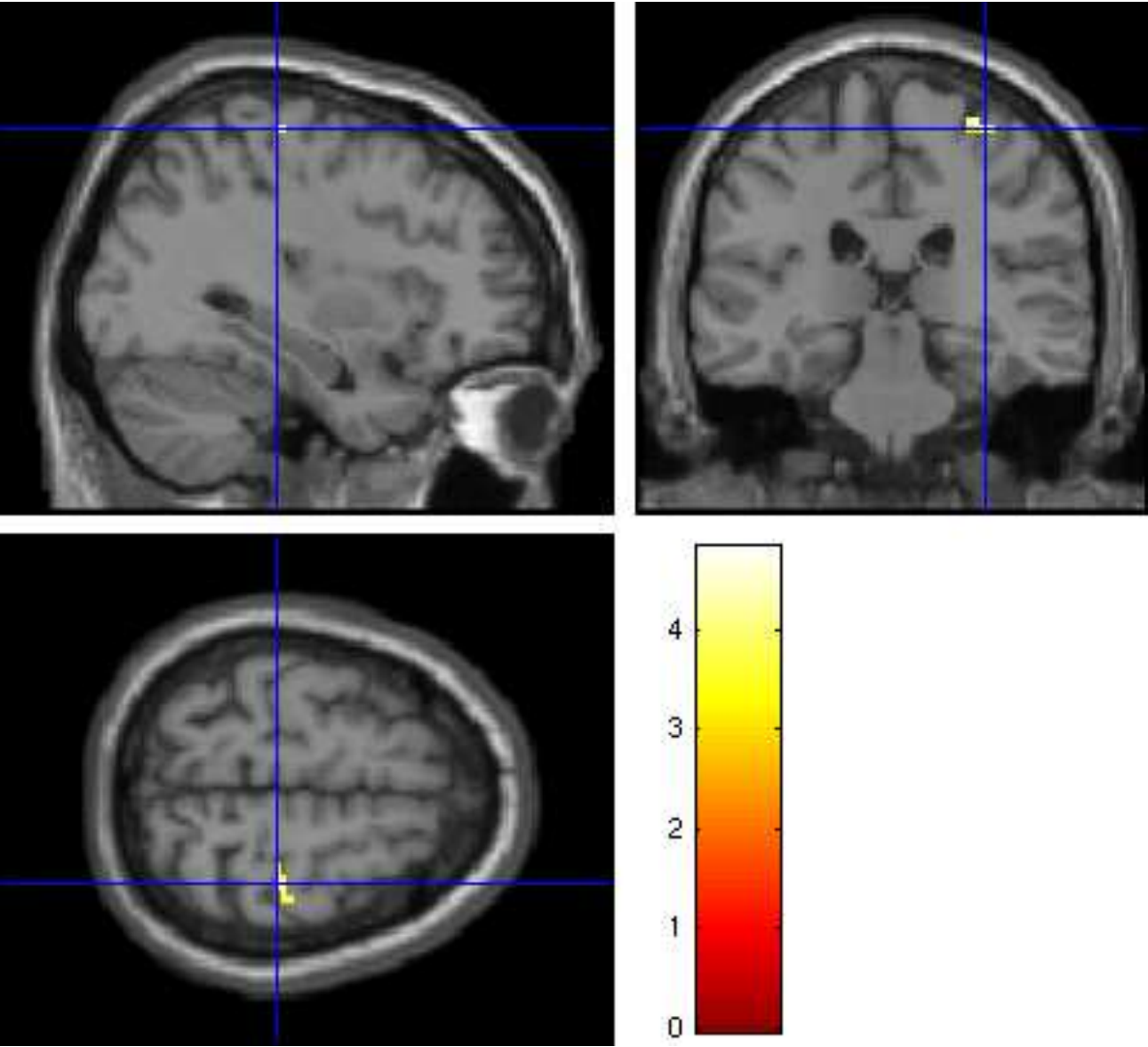}&
\hspace*{-0.3cm}\includegraphics[width=2.5cm, height=2.5cm]{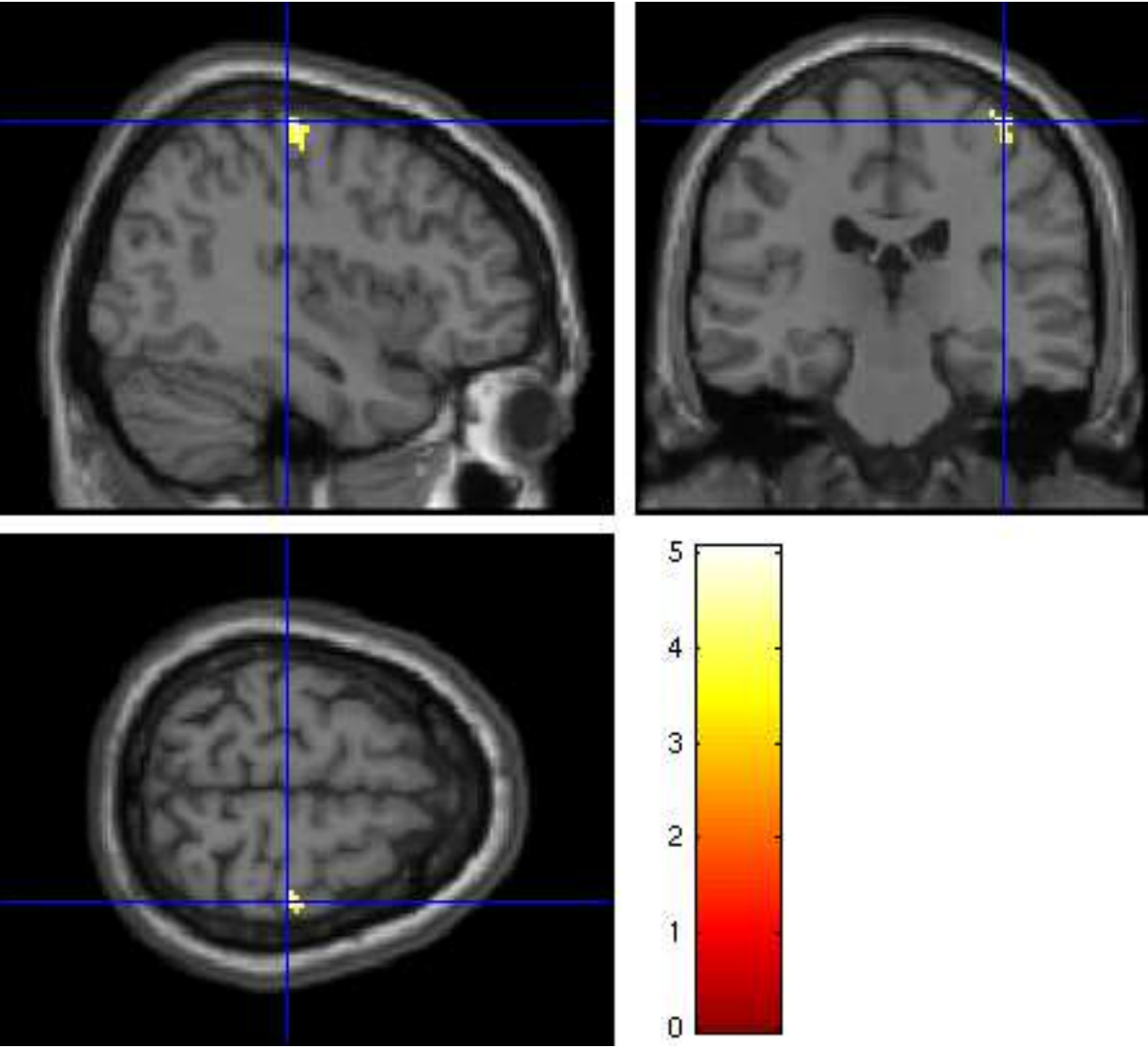}&
\hspace*{-0.3cm}\includegraphics[width=2.5cm, height=2.5cm]{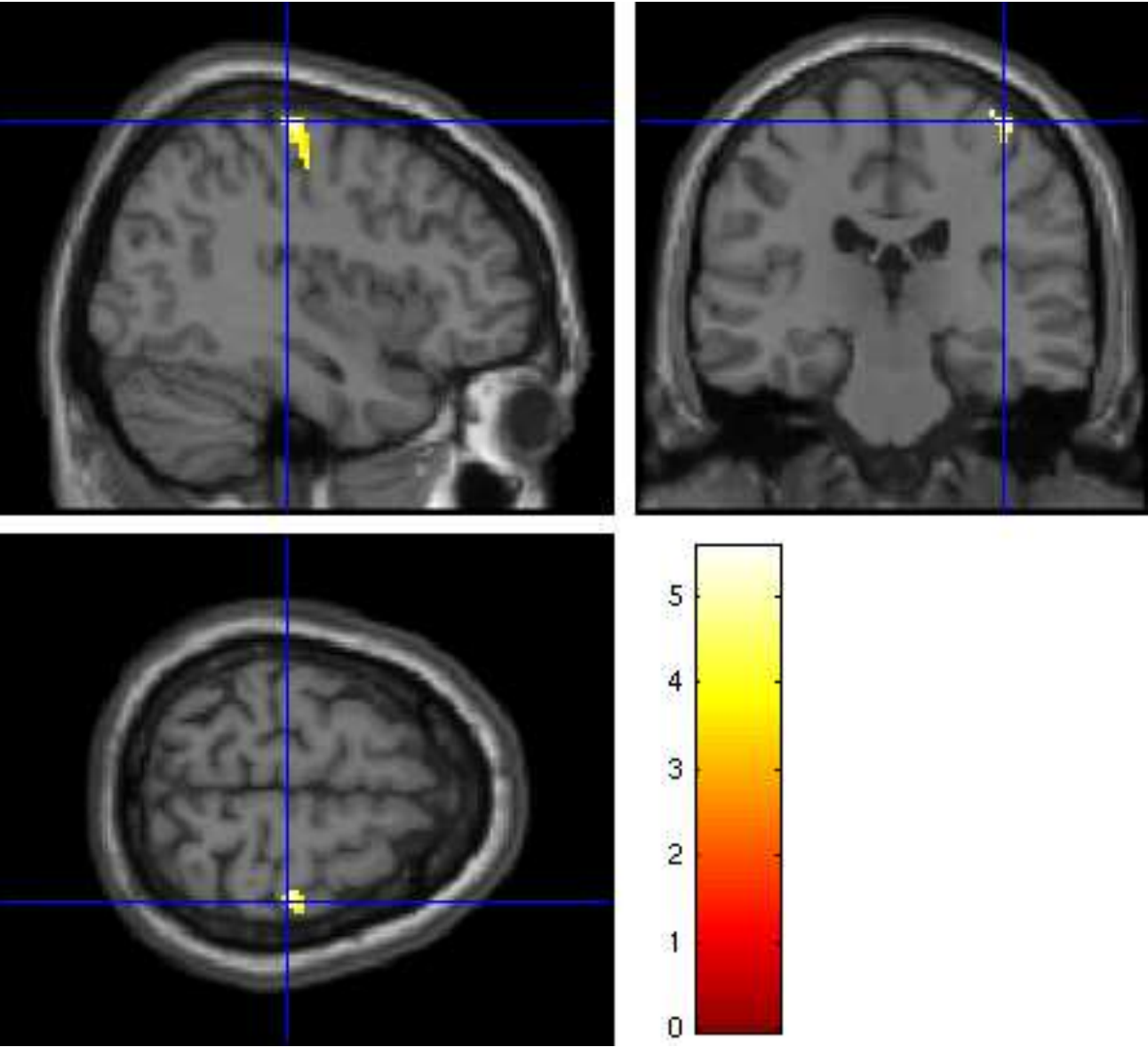}&
\hspace*{-0.1cm}\includegraphics[width=2.5cm, height=2.5cm]{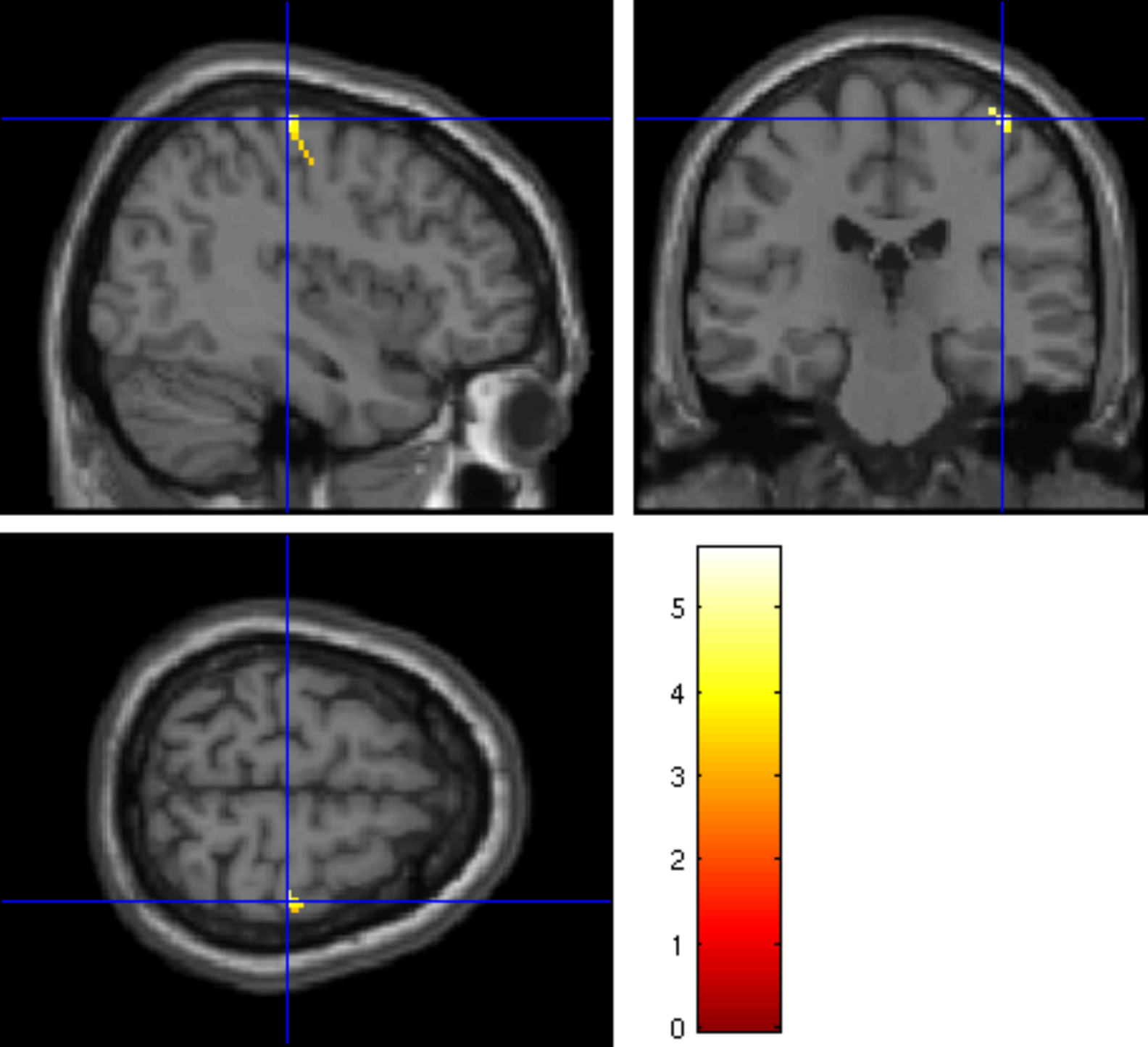}
\end{tabular}\vspace*{-.3cm}
\caption{Subject-level Student's $t$-maps superimposed to anatomical MRI for the \LcRc contrast. Data have been 
reconstructed using the \mSENSE, UWR-SENSE and 4D-UWR-SENSE  (with and without spatial smoothing for the latter), 
respectively. Neurological convention. The blue 
cross shows the global maximum activation peak. \label{fig:res_T_Lc-Rc}}
\end{figure}

\begin{table}[!ht]
\centering 
\caption{Significant statistical results at the subject-level for the \LcRc contrast (corrected for 
multiple comparisons at $p=0.05$). Images were reconstructed using the \mSENSE, UWR-SENSE and 
4D-UWR-SENSE  (with and without spatial smoothing for the latter) algorithms for $R=2$ and $R=4$.}
\begin{tabular}{|c|c|c|c|c|c|c|}
\cline{3-7}
\cline{3-7}
\multicolumn{2}{c}{}&\multicolumn{2}{|c|}{cluster-level}&\multicolumn{3}{|c|}{voxel-level}\\
\cline{3-7}
\multicolumn{2}{c|}{}&p-value&Size&p-value&T-score& Position\\
\hline
\multirow{9}{*}{$R=2$}&
\multicolumn{1}{|c|}{\mSENSE} &$ <10^{-3}$ &79&$ <10^{-3}$ & 6.49&38 -26 66 \\
\cline{2-7}
\cline{2-7}
&\multirow{2}{*}{UWR-SENSE} &$ <10^{-3}$ &144&0.004& 5.82& 40 -22 63 \\
\cline{3-7}
& &$0.03$ &21&0.064& 4.19& 24 -8 63 \\
\cline{2-7}
\cline{2-7}
&\multirow{3}{*}{4D-UWR-SENSE} &$<10^{-3}$ &\textbf{189}&0.001&7.03& 34 -24 69  \\
\cline{3-7}
& &$<10^{-3}$ &53&0.001&4.98&50 -18 42\\
\cline{3-7}
& &$<10^{-3}$ &47&0.001&5.14&32 -6 66 \\
\cline{2-7}
\cline{2-7}
&\multirow{3}{*}{Unsmoothed 4D-UWR-SENSE} &$<10^{-3}$ &112&0.001&\textbf{7.26}& 34 -24 69  \\
\cline{3-7}
& &$<10^{-3}$ &21&0.001&4.77&32 -6 66\\
\cline{3-7}
& &$<10^{-3}$ &19&0.001&4.98&50 -18 42\\
\hline
\hline
\multirow{4}{*}{$R=4$}&\multicolumn{1}{|c|}{\mSENSE} &0.006 & 21 & 0.295&4.82&34 -28 63\\
\cline{2-7}
\cline{2-7}
&\multicolumn{1}{|c|}{UWR-SENSE} &$< 10^{-3}$ & 33& 0.120& 5.06&40 -24 66\\
\cline{2-7}
\cline{2-7}
&\multicolumn{1}{|c|}{4D-UWR-SENSE} &$< 10^{-3}$&\textbf{51}& 0.006&5.57&40 -24 66\\
\cline{2-7}
\cline{2-7}
&\multirow{1}{*}{Unsmoothed 4D-UWR-SENSE} &$<10^{-3}$ &25&0.001&\textbf{5.7}& 40 -24 66  \\
\hline
\end{tabular}
\label{tab:StatRes2allRcLc}
\end{table}

Fig.~\ref{fig:LcRc_variability} reports on the robustness of the proposed pMRI pipeline to the between-subject variability for this motor contrast. Since sensory functions are expected to generate larger BOLD effects~(higher SNR) and appear more stable, our comparison takes place at $R=4$. 
Two subject-level Student's $t$-maps reconstructed using the different pMRI algorithms are compared in Fig.~\ref{fig:LcRc_variability}. For the second subject, one can observe that the \mSENSE~algorithm fails to detect any activation cluster in the right motor cortex. By contrast, our 4D-UWR-SENSE method retrieves more coherent activity for this second subject in the expected region.

\begin{figure}[!ht]
\centering
\begin{tabular}{cc c c}
&\mSENSE&UWR-SENSE&4D-UWR-SENSE\\
\hspace*{-0.1cm}\raisebox{1.8cm}{\footnotesize \texttt{Subj.~1}}&
\hspace*{-0.3cm}\includegraphics[width=3.2cm, height=3.2cm]{Lc-Rc_MF_acqR4_mSENSE}&
\hspace*{-0.35cm}\includegraphics[width=3.2cm, height=3.2cm]{Lc-Rc_MF_acqR4_UWRSENSE_2D}&
\hspace*{-0.35cm}\includegraphics[width=3.2cm, height=3.2cm]{Lc-Rc_MF_acqR4_UWRSENSE_4D}\\
\hspace*{-0.1cm}\raisebox{1.8cm}{\footnotesize \texttt{Subj.~5}}&
\hspace*{-0.3cm}\includegraphics[width=3.2cm, height=3.2cm]{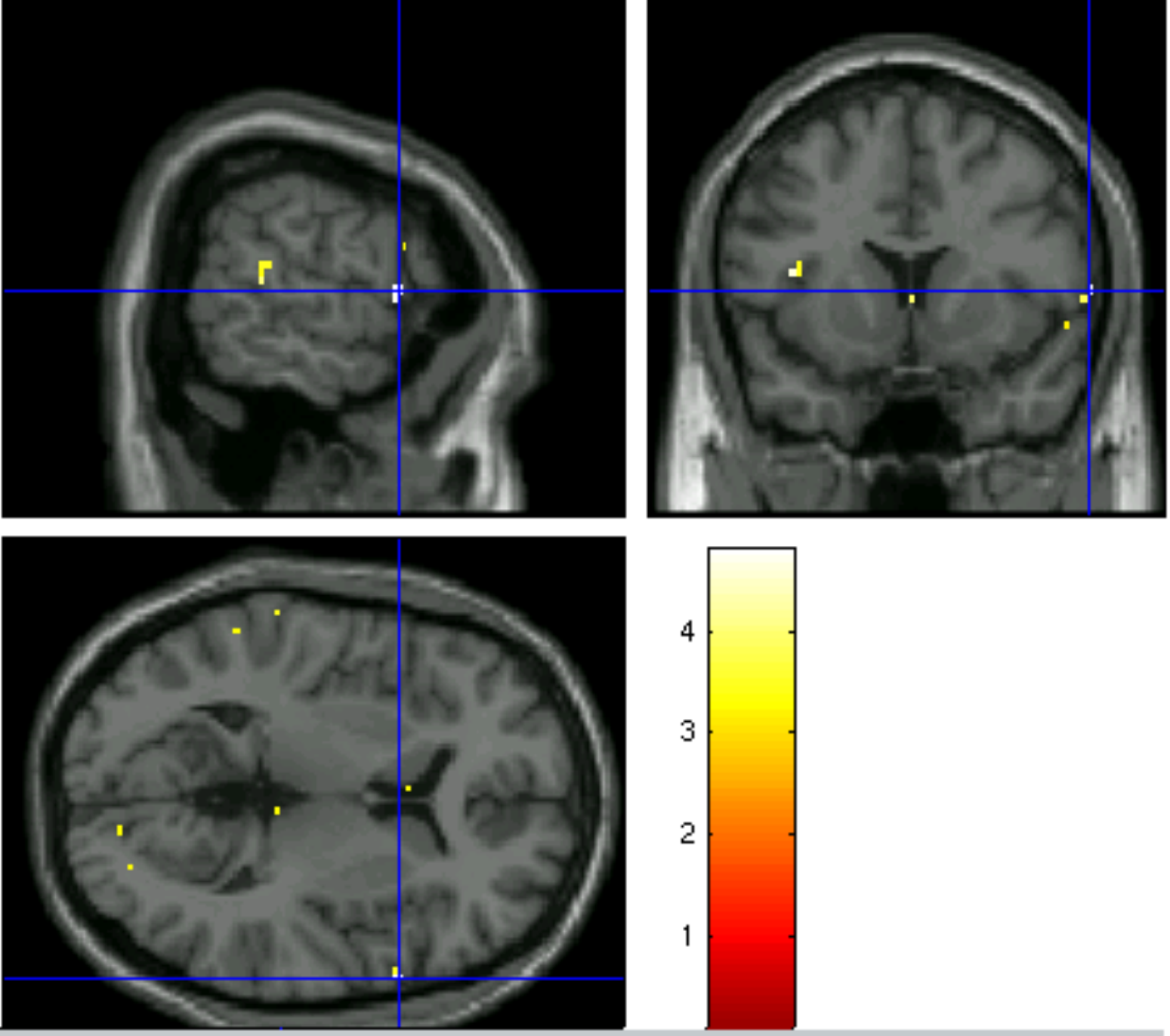}&
\hspace*{-0.35cm}\includegraphics[width=3.2cm, height=3.2cm]{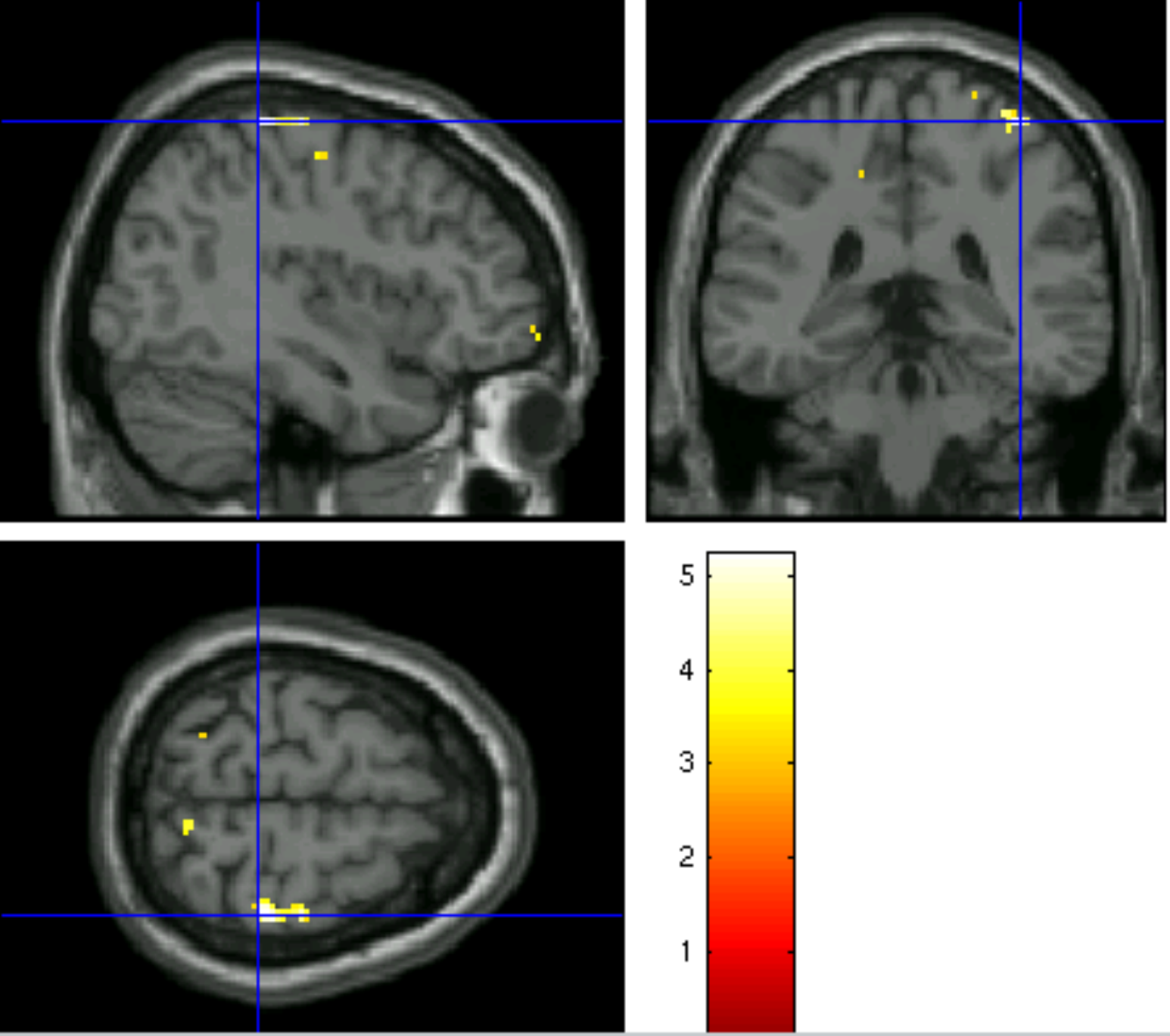}&
\hspace*{-0.35cm}\includegraphics[width=3.2cm, height=3.2cm]{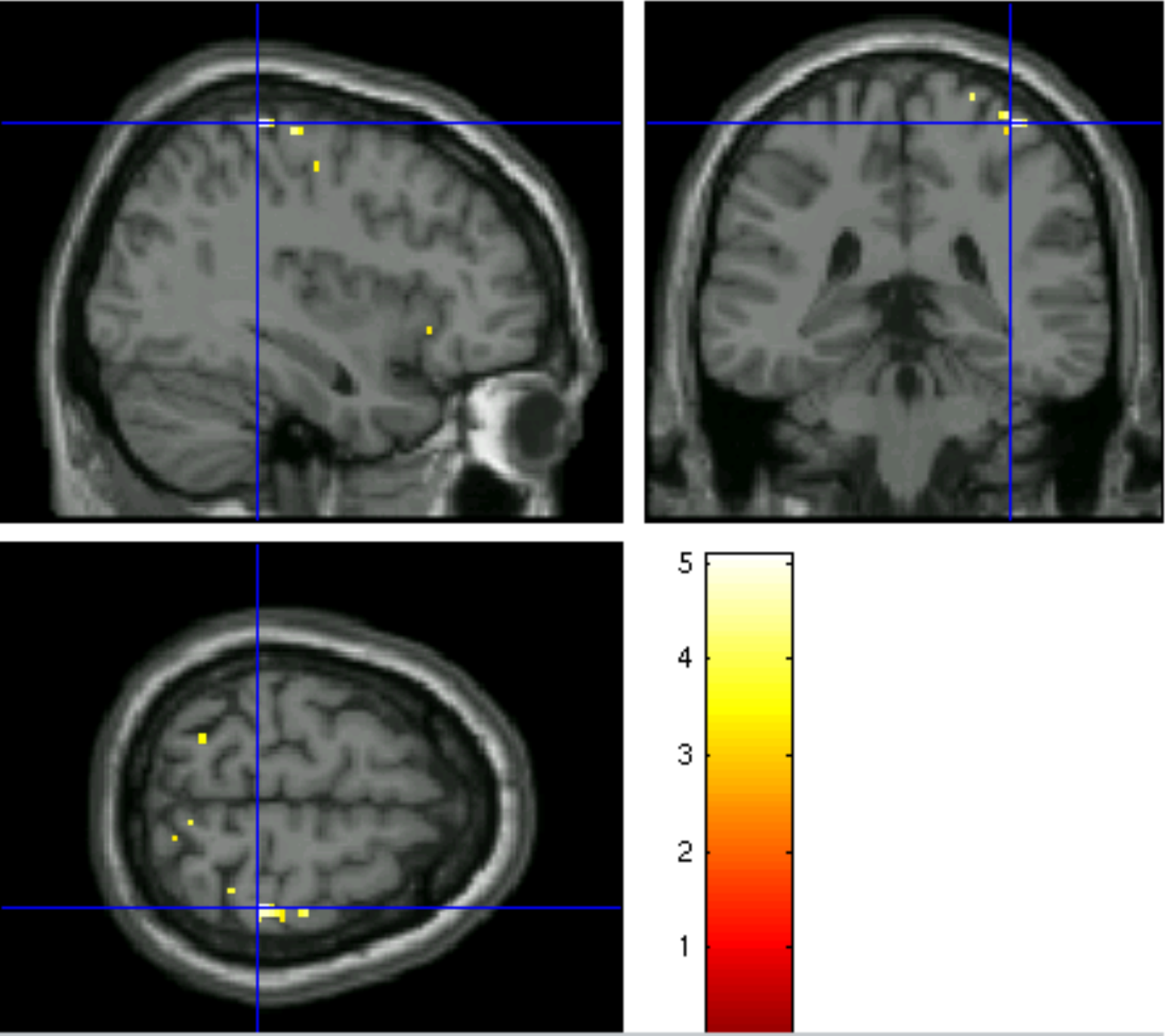}
\end{tabular}\vspace*{-.3cm}
\caption{Between-subject variability of detected activation for the \LcRc contrast at $R=4$. 
Neurological convention. The blue cross shows the global maximum activation peak.\label{fig:LcRc_variability}}
\end{figure}


For the \ACAS contrast, Fig.~\ref{fig:res_T_A-V}~[top] shows, for the most significant slice and $R=2$, that all pMRI reconstruction algorithms succeed in finding evoked activity in the left parietal and frontal cortices, more precisely in the inferior parietal lobule and middle frontal gyrus according to the AAL template\footnote{available in the \texttt{xjView} toolbox of SPM5.}. 
Tab.~\ref{tab:StatRes2all} also confirms a bilateral activity pattern in parietal regions for $R=2$. Moreover, for $R=4$, Fig.~\ref{fig:res_T_A-V}~[bottom] illustrates that our pipeline~(UWR-SENSE and 4D-UWR-SENSE) 
and especially the proposed 4D-UWR-SENSE scheme enables to retrieve reliable frontal activity elicited by mental calculation, which is lost by the the \mSENSE~algorithm. 
From a quantitative viewpoint, the proposed 4D-UWR-SENSE algorithm finds larger clusters whose local maxima are more significant than the ones obtained using \mSENSE~and UWR-SENSE, as reported in 
Tab.~\ref{tab:StatRes2all}. Concerning the most significant cluster for $R=2$, the peak positions remain stable whatever the reconstruction algorithm. However, examining their significance level, one can first measure the benefits of wavelet-based regularization when comparing UWR-SENSE with \mSENSE~results and then additional positive effects of temporal regularization and 3D wavelet decomposition when looking at the 4D-UWR-SENSE results. These benefits are also demonstrated for $R=4$.

\begin{figure}[!htp]
\centering
\begin{tabular}{c c c c}
&\mSENSE&UWR-SENSE&4D-UWR-SENSE\\
\hspace*{-0.4cm}\raisebox{2cm}{$R=2$}&\includegraphics[width=3.2cm, height=3.2cm]{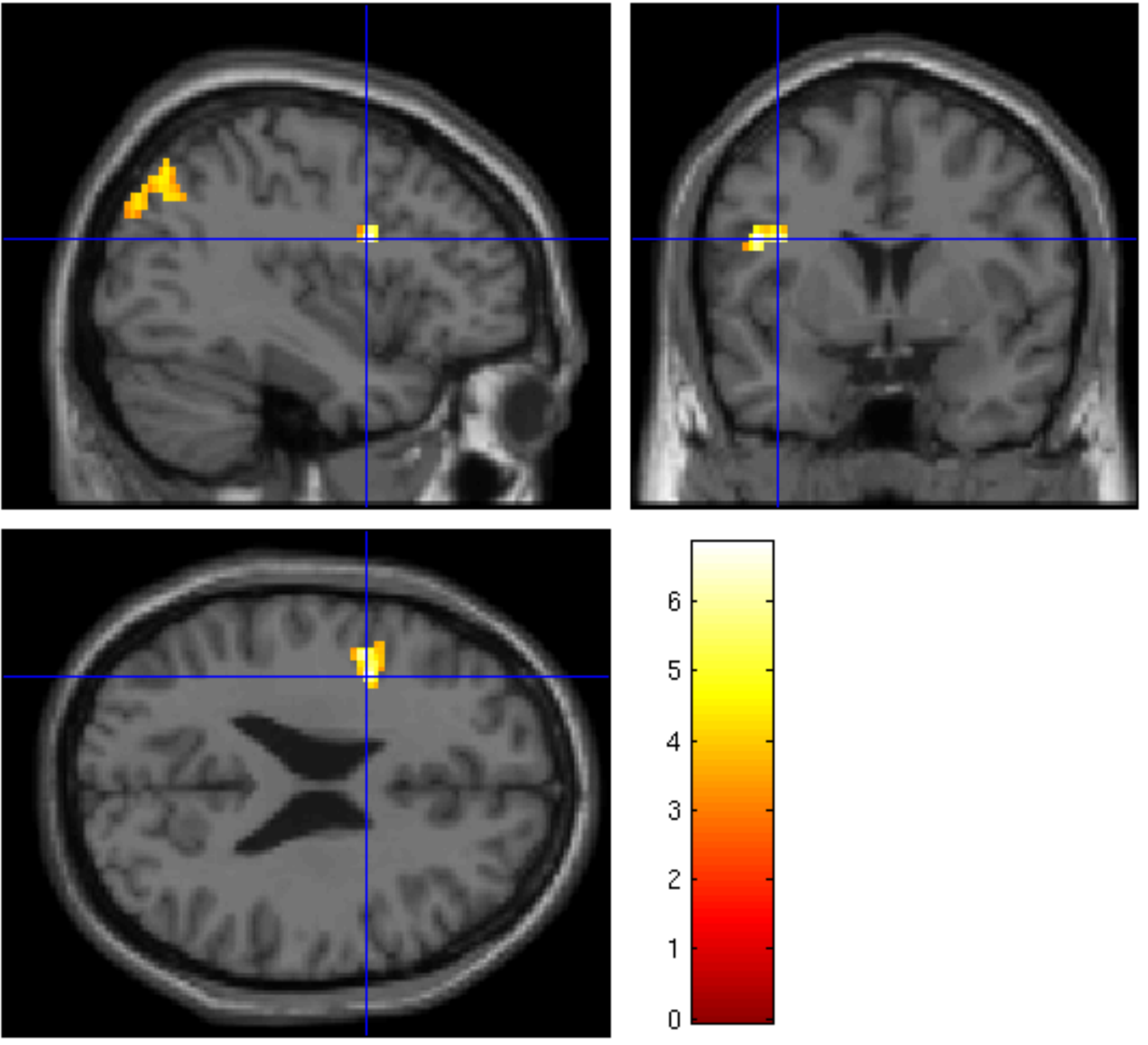}&
\hspace*{-0.3cm}\includegraphics[width=3.2cm, height=3.2cm]{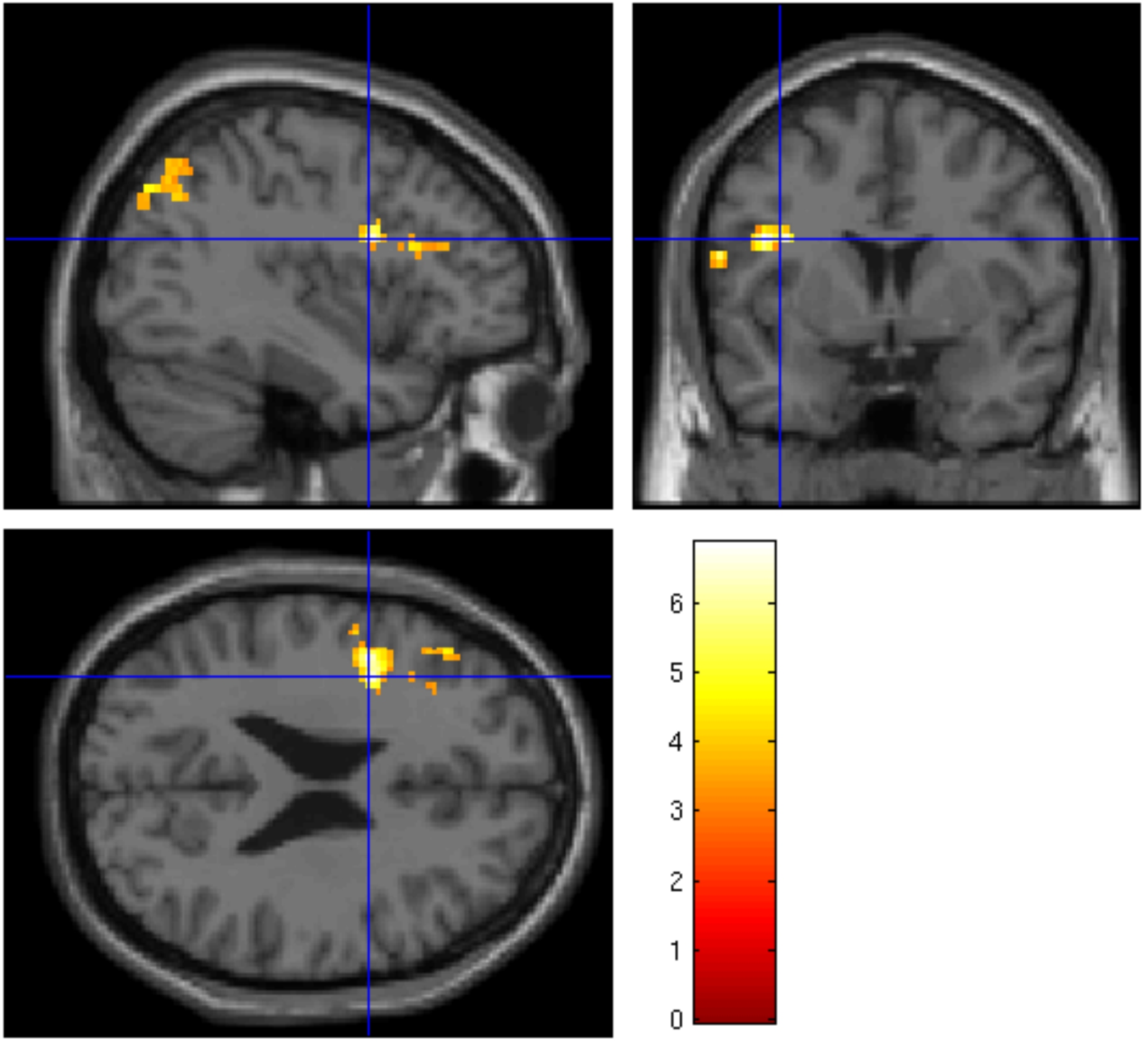}&
\hspace*{-0.3cm}\includegraphics[width=3.2cm, height=3.2cm]{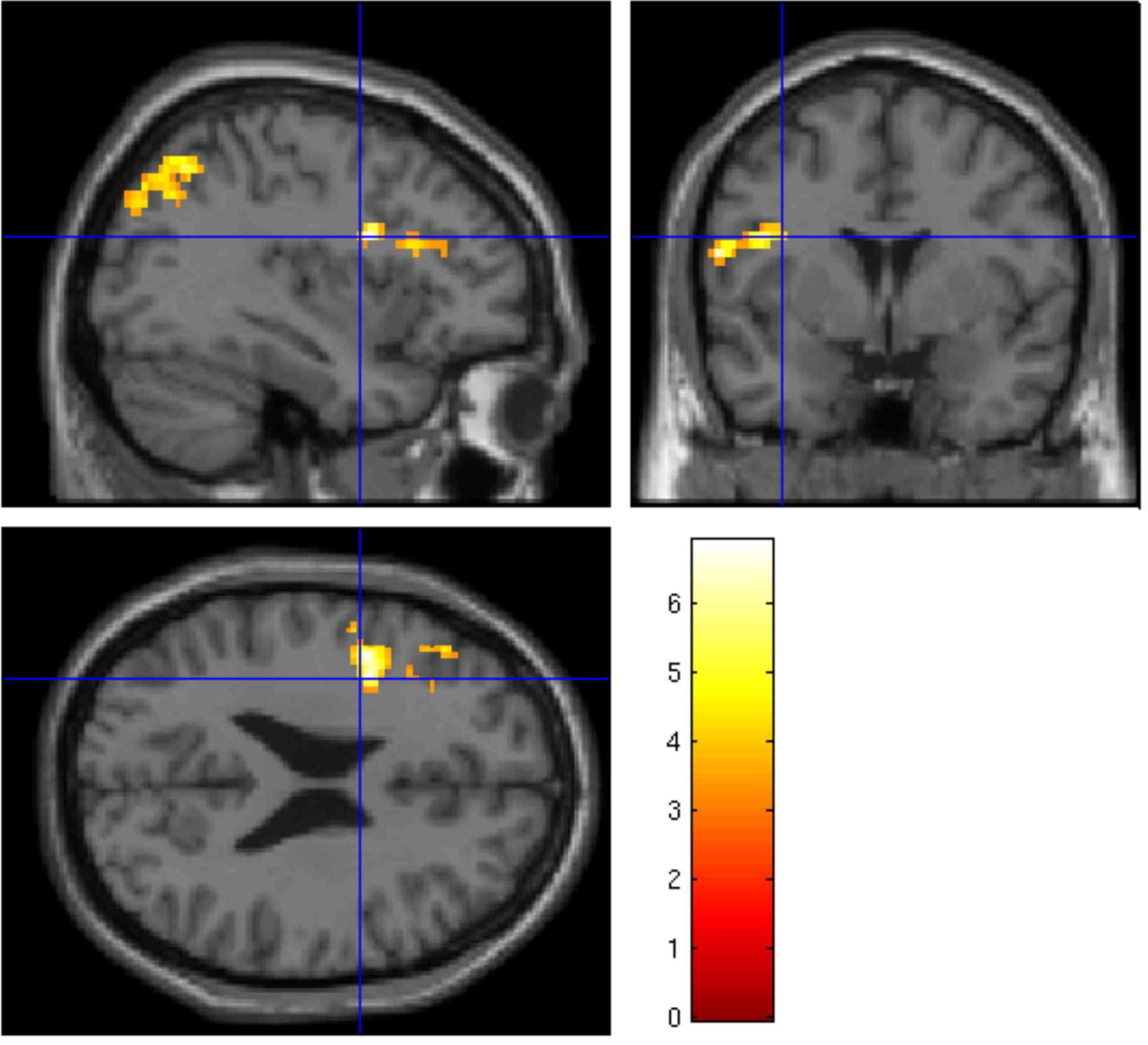}\\
\hspace*{-0.4cm}\raisebox{2cm}{$R=4$}&\includegraphics[width=3.2cm, height=3.2cm]{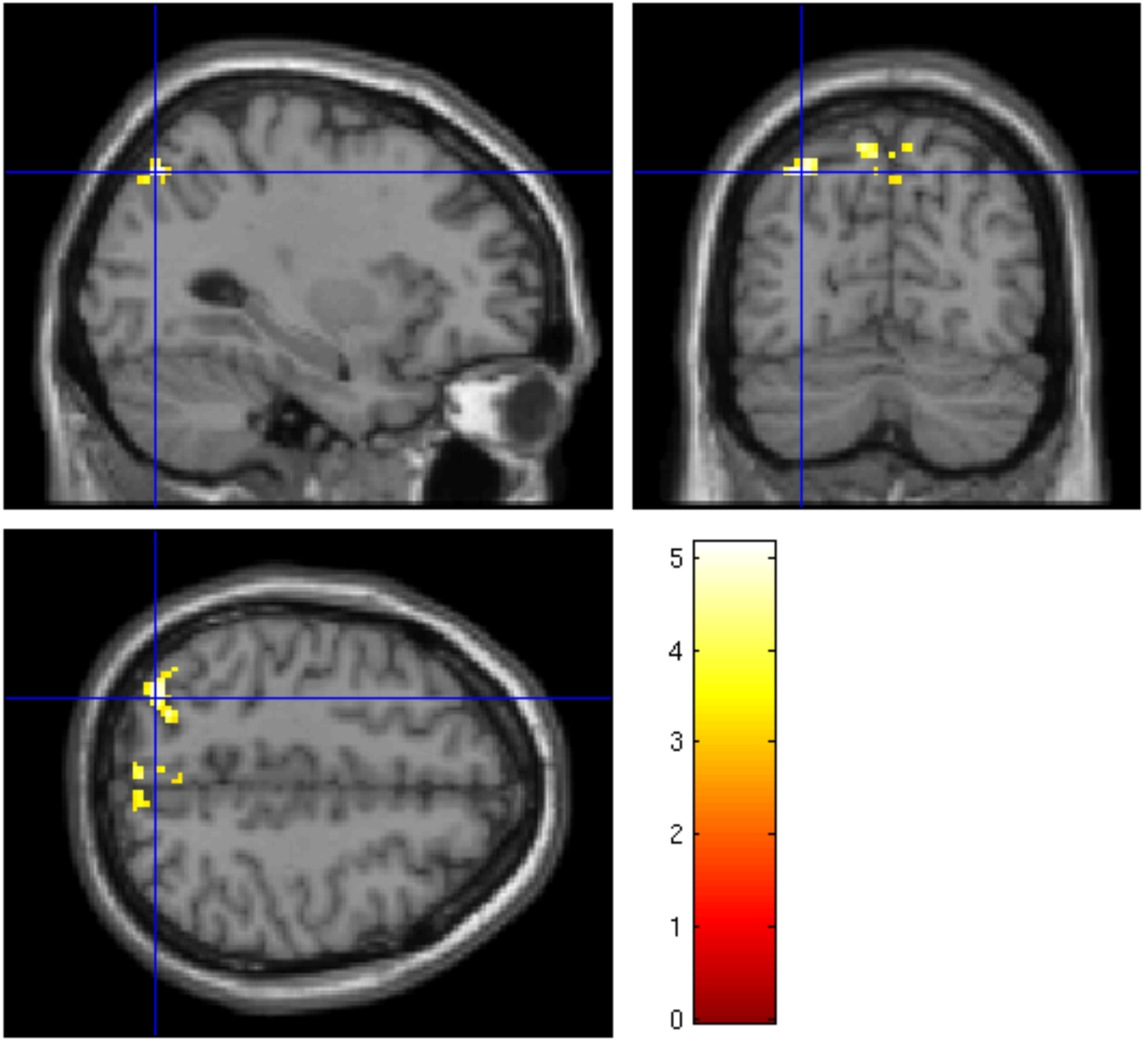}&
\hspace*{-0.3cm}\includegraphics[width=3.2cm, height=3.2cm]{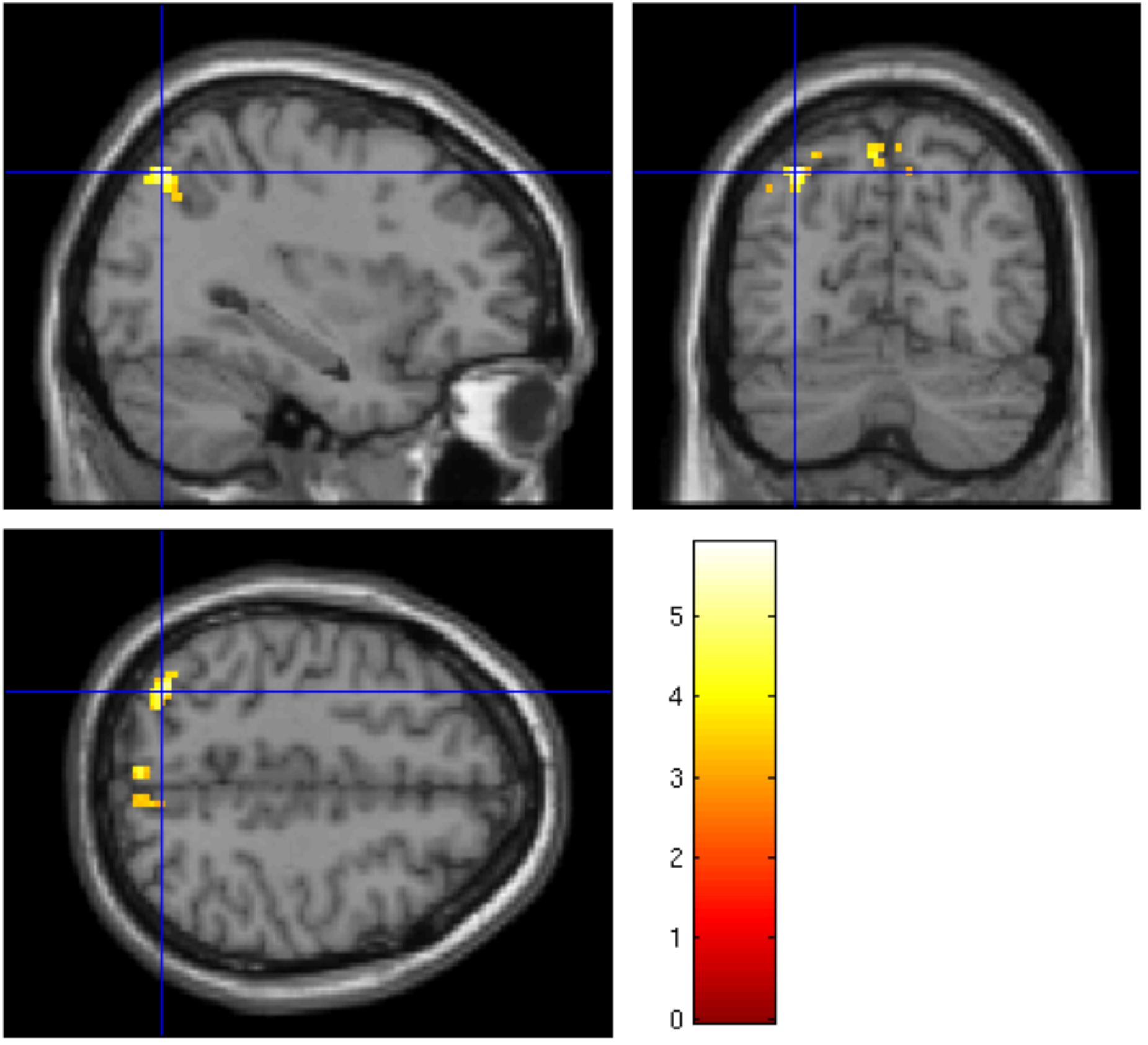}&
\hspace*{-0.3cm}\includegraphics[width=3.2cm, height=3.2cm]{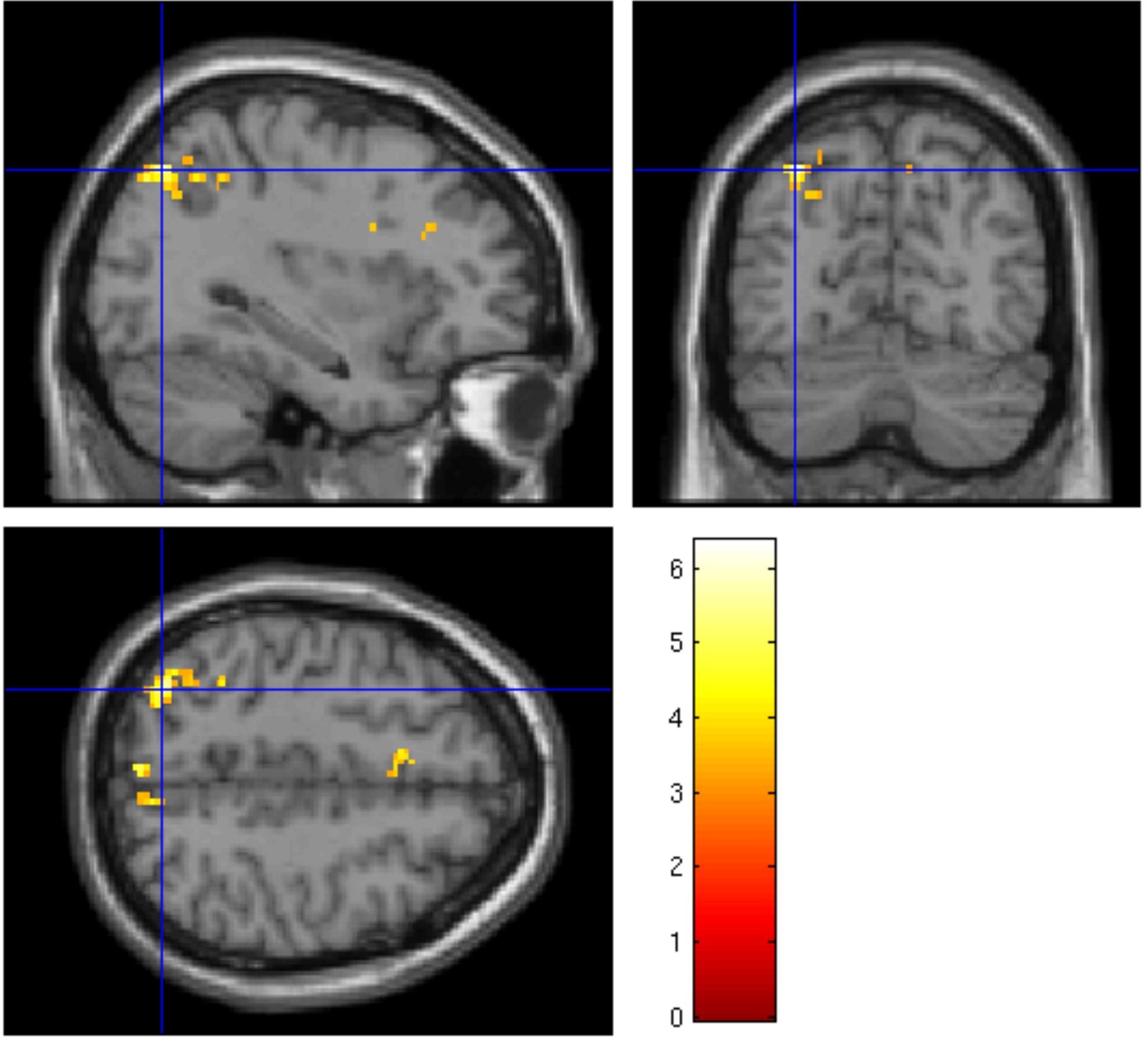}
\end{tabular}\vspace*{-.3cm}
\caption{Subject-level Student's $t$-maps superimposed to anatomical MRI for the \ACAS contrast. 
Data have been reconstructed using the~\mSENSE, UWR-SENSE and 4D-UWR-SENSE, respectively. 
Neurological convention: {\bf left is left}. The blue cross shows the global maximum activation peak. \label{fig:res_T_A-V}}
\end{figure}

\begin{table}[!htp]
\centering 
\caption{Significant statistical results at the subject-level for the \ACAS contrast (corrected 
for multiple comparisons at $p=0.05$). Images were reconstructed using the~\mSENSE, UWR-SENSE 
and 4D-UWR-SENSE algorithm for $R=2$ and $R=4$.}
\begin{tabular}{|c|c|c|c|c|c|c|}
\cline{3-7}
\cline{3-7}
\multicolumn{2}{c}{}&\multicolumn{2}{|c|}{cluster-level}&\multicolumn{3}{|c|}{voxel-level}\\
\cline{3-7}
\multicolumn{2}{c|}{}&p-value&Size&p-value&T-score& Position\\
\hline
&\multirow{4}{*}{\mSENSE} & $< 10^{-3}$ &320 &$< 10^{-3}$ & 6.40&-32 -76 45 \\
\cline{3-7}
\multirow{12}{*}{$R=2$} & & $< 10^{-3}$  &163 &$< 10^{-3}$ & 5.96&-4 -70 54 \\
\cline{3-7}
& &$< 10^{-3}$  &121 &$< 10^{-3}$ & 6.34&34 -74 39 \\
\cline{3-7}
&&$< 10^{-3}$  &94 &$< 10^{-3}$ & 6.83&-38 4 24 \\
\cline{2-7}
\cline{2-7}
&\multirow{4}{*}{UWR-SENSE} & $< 10^{-3}$  &407 &$< 10^{-3}$& 6.59&-32 -76 45 \\
\cline{3-7}
& & $< 10^{-3}$  &164 &$< 10^{-3}$& 5.69&-6 -70 54 \\
\cline{3-7}
& & $< 10^{-3}$  &159 &$< 10^{-3}$& 5.84&32 -70 39 \\
\cline{3-7}
& & $< 10^{-3}$  &155 &$< 10^{-3}$& 6.87&-44 4 24 \\
\cline{2-7}
\cline{2-7}
&\multirow{4}{*}{4D-UWR-SENSE} &$< 10^{-3}$ &\textbf{454} &$< 10^{-3}$ & 6.54& -32 -76 45 \\
\cline{3-7}
& & $< 10^{-3}$ &199& $< 10^{-3}$ & 5.43& -6 26 21 \\
\cline{3-7}
& &$< 10^{-3}$ &183 & $< 10^{-3}$ & 5.89& 32 -70 39 \\
\cline{3-7}
& &$< 10^{-3}$ &170 & $< 10^{-3}$ &\textbf{ 6.90}& -44 4 24 \\
\hline
\hline
\multirow{6}{*}{$R=4$}&\multicolumn{1}{|c|}{\mSENSE} &$< 10^{-3}$ & 58& 0.028& 5.16&-30 -72 48\\
\cline{2-7}
\cline{2-7}
&\multirow{2}{*}{UWR-SENSE} &$< 10^{-3}$ & 94& 0.003& 5.91&-32 -70 48\\
\cline{3-7}
& &$< 10^{-3}$ & 60& 0.044& 4.42&-6 -72 54\\
\cline{2-7}
\cline{2-7}
&\multirow{3}{*}{4D-UWR-SENSE} &$< 10^{-3}$ & \textbf{152} &$< 10^{-3}$&\textbf{6.36}&-32 -70 48 \\
\cline{3-7}
& &$< 10^{-3}$ & 36 &0.009&5.01&-4 -78 48 \\
\cline{3-7}
& &$< 10^{-3}$ & 29 &0.004&5.30&-34 6 27 \\
\hline
\end{tabular}
\label{tab:StatRes2all}
\end{table}

Fig.~\ref{fig:ACAS_variability} illustrates another property of the proposed pMRI pipeline, i.e. its robustness to 
the between-subject variability. Indeed, when comparing subject-level Student's $t$-maps reconstructed using the 
different pipelines~($R=2$), it can
be observed that the \mSENSE~algorithm fails to detect any activation cluster in the expected regions for the 
second subject~(see Fig.~\ref{fig:ACAS_variability}~[bottom]). By contrast, our 4D-UWR-SENSE method retrieves 
more coherent activity while not exactly at the same position as for the first subject.

\begin{figure}[!htp]
\centering
\begin{tabular}{cc c c}
&\mSENSE&UWR-SENSE&4D-UWR-SENSE\\
\hspace*{-0.1cm}\raisebox{1.8cm}{\footnotesize \texttt{Subj.~1}}&\hspace*{-0.3cm}\includegraphics[width=3.2cm, height=3.2cm]{mSENSE_R2_CS_LE.pdf}&
\hspace*{-0.35cm}\includegraphics[width=3.2cm, height=3.2cm]{2D_R2_CS_LE}&
\hspace*{-0.35cm}\includegraphics[width=3.2cm, height=3.2cm]{4D_R2_CS_LE}\\
\hspace*{-0.1cm}\raisebox{1.8cm}{\footnotesize \texttt{Subj.~2}}&
\hspace*{-0.3cm}\includegraphics[width=3.2cm, height=3.2cm]{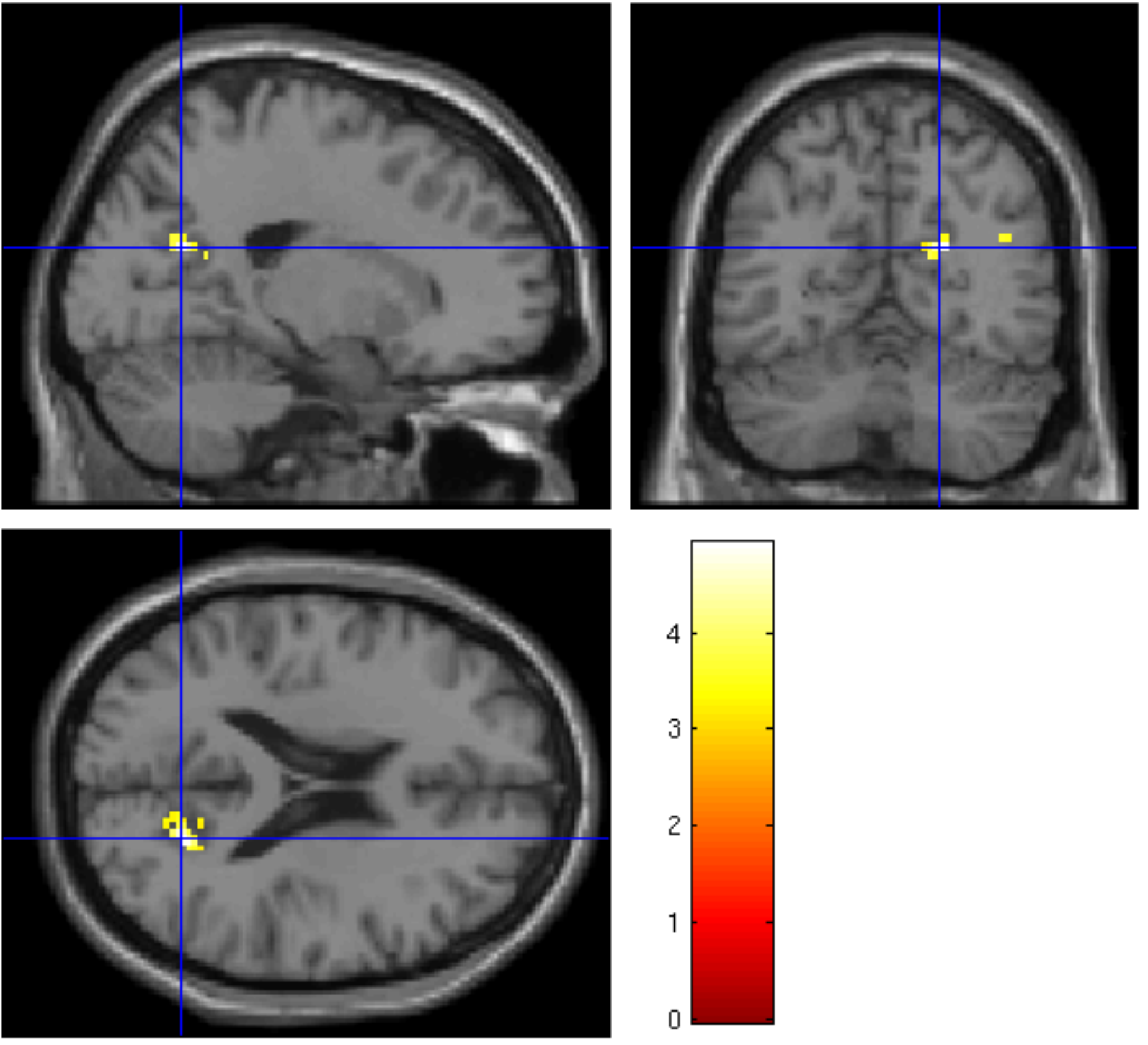}&
\hspace*{-0.35cm}\includegraphics[width=3.2cm, height=3.2cm]{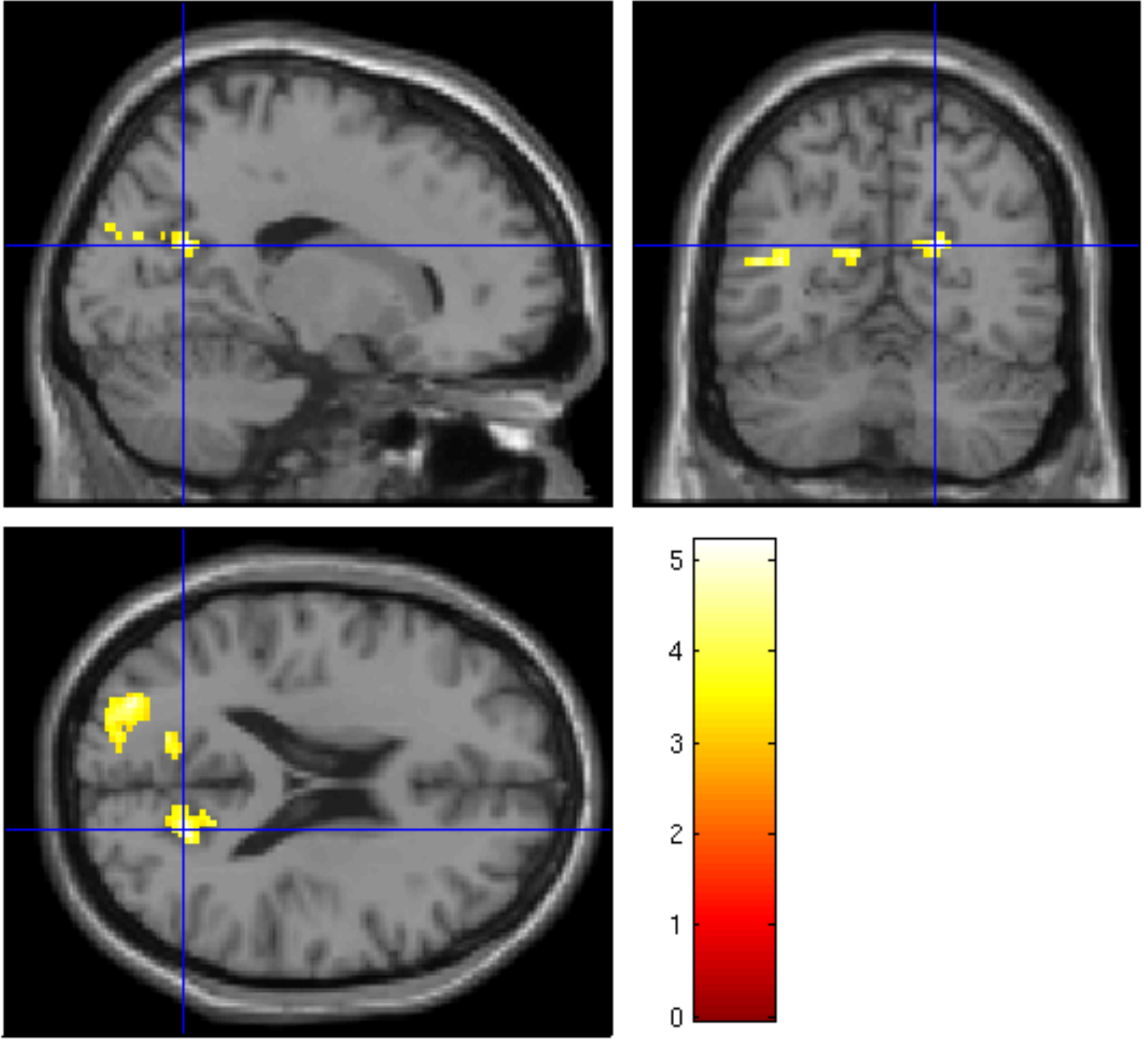}&
\hspace*{-0.35cm}\includegraphics[width=3.2cm, height=3.2cm]{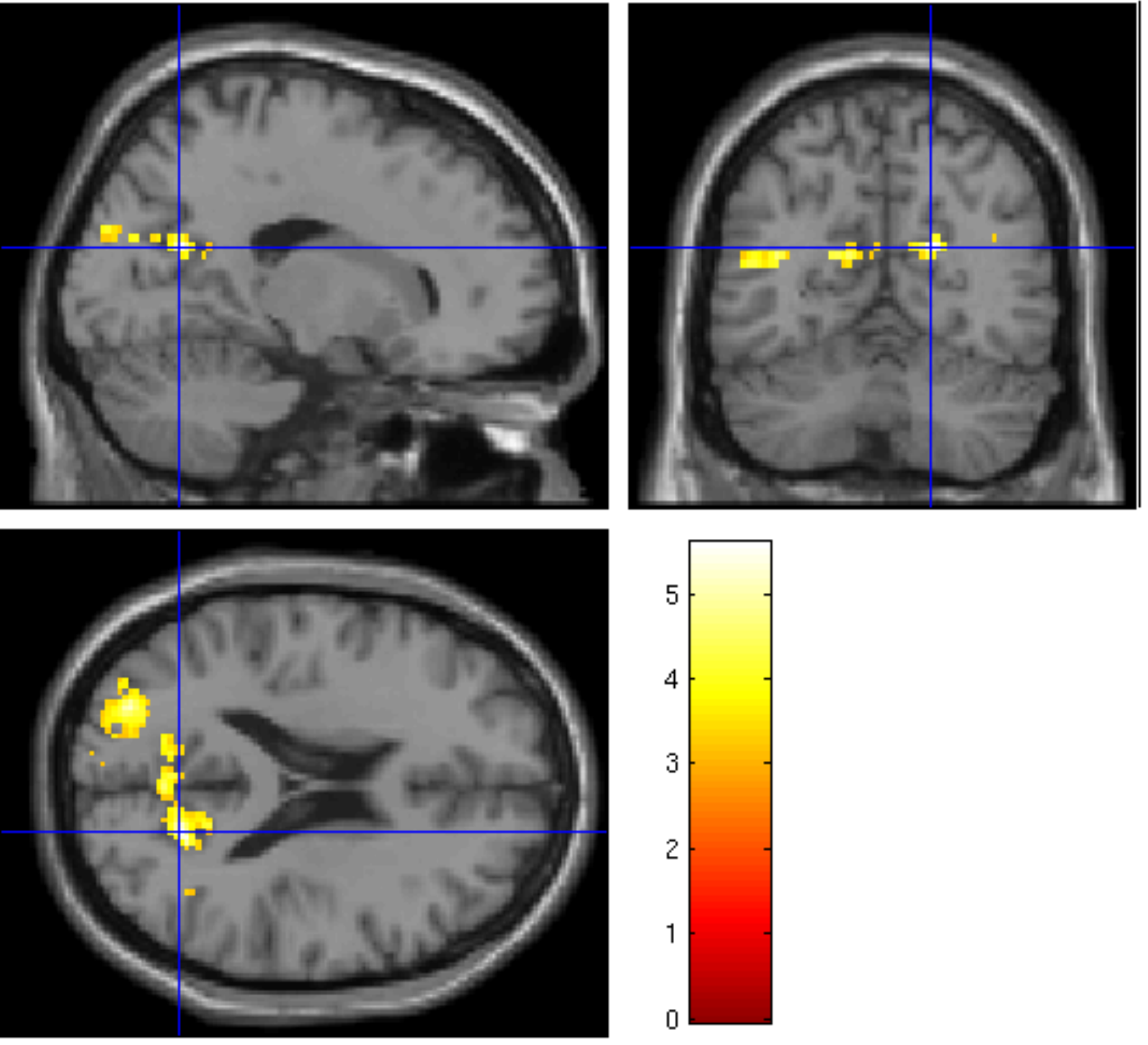}
\end{tabular}\vspace*{-.3cm}
\caption{Between-subject variability of detected activation for the \ACAS contrast at $R=2$. 
Neurological convention. The blue cross shows the global maximum activation peak.\label{fig:ACAS_variability}}
\end{figure}

To summarize, for these two contrasts our 4D-UWR-SENSE algorithm always outperforms the alternative reconstruction methods 
used in this paper
in terms of statistical 
significance~(number of clusters, cluster extent, peak values,...) but also in terms of robustness.

\subsubsection{Intrinsic smoothing characterization} 

 To characterize the intrinsic smoothing effect of our reconstruction method, we vary the 
 FWHM parameter of the spatial smoothing we apply to \mSENSE~data and derive the correspondence between the 
 two approaches. To investigate the spatial smoothing effect, Tab.~\ref{tab:comp_smooth_r2} shows statistical results 
obtained for the \LcRc contrast at $R=2$ using \mSENSE~ and 3D-UWR-SENSE. 

\begin{table}[!h]
\centering 
\caption{Statistical results for $R=2$ (\LcRc contrast) at the cluster and voxel levels using \mSENSE~and 3D-UWR-SENSE with different 
FWHMs of the Gaussian spatial filtering (pre-processing).}
\begin{tabular}{|c|c|c|c|c|c|c|}
\hline
  \multirow{2}{*}{Method} &  \multirow{2}{*}{Smoothing level}&\multicolumn{2}{|c|}{Cluster level} &\multicolumn{3}{|c|}{Voxel level} \\
  \cline{3-7}
& & p-value&Size&p-value&T-score&Position\\
 \hline
 \multirow{6}{*}{\mSENSE}&None&$< 10^{-3}$&37&0.002&5.92&32 -30 63\\
 \cline{2-7}
 &  \cellcolor[gray]{0.7}$3 \times 3 \times 3$ mm$^3$&\cellcolor[gray]{0.7}$< 10^{-3}$&\cellcolor[gray]{0.7}
 87&\cellcolor[gray]{0.7}$< 10^{-3}$&\cellcolor[gray]{0.7}6.29&\cellcolor[gray]{0.7}32 -30 63\\
  \cline{2-7}
  &\cellcolor[gray]{0.7}$4 \times 4 \times 4$ mm$^3$&\cellcolor[gray]{0.7}$< 10^{-3}$&\cellcolor[gray]{0.7}123 &
  \cellcolor[gray]{0.7}$< 10^{-3}$&\cellcolor[gray]{0.7}6.38&\cellcolor[gray]{0.7}32 -30 63\\
   \cline{2-7}
   & $5 \times 5 \times 5$ mm$^3$&$< 10^{-3}$&161 &$< 10^{-3}$&6.39&32 -28 63\\
      \cline{2-7}
   & $6 \times 6 \times 6$ mm$^3$&$< 10^{-3}$&207 &$< 10^{-3}$&6.43&32 -28 63\\
      \cline{2-7}
   & $8 \times 8 \times 8$ mm$^3$&$< 10^{-3}$&261 &0.001&6.11&32 -28 63\\
 \hline
  \multirow{6}{*}{3D-UWR-SENSE}&\cellcolor[gray]{0.7}None&\cellcolor[gray]{0.7}$< 10^{-3}$&\cellcolor[gray]{0.7} 95&
  \cellcolor[gray]{0.7}0.017&\cellcolor[gray]{0.7}5.51&\cellcolor[gray]{0.7}40 -22 63\\
 \cline{2-7}
  &$3 \times 3 \times 3$ mm$^3$&$< 10^{-3}$&201&0.005&5.77&40 -22 63\\
  \cline{2-7}
  &$4 \times 4 \times 4$ mm$^3$&$< 10^{-3}$&276 &0.002&6.38&36 -22 66\\
    \cline{2-7}
  &$4.8 \times 4.8 \times 4.8$ mm$^3$&$< 10^{-3}$&346 &$< 10^{-3}$&6.22&36 -22 66\\
   \cline{2-7}
   & $5 \times 5 \times 5$ mm$^3$&$< 10^{-3}$&333 &0.001&6.19&36 -22 66\\
     \cline{2-7}
  &$6 \times 6 \times 6$ mm$^3$&$< 10^{-3}$&384 &0.001&6.22&36 -22 63\\
   \cline{2-7}
   & $8 \times 8 \times 8$ mm$^3$&$< 10^{-3}$&435 &$< 10^{-3}$&6.19&38 -22 66\\
 \hline 
   \multirow{6}{*}{4D-UWR-SENSE}&None&$< 10^{-3}$&112&$< 10^{-3}$&7.26&34 -24 69\\
 \cline{2-7}
  &$3 \times 3 \times 3$ mm$^3$&$< 10^{-3}$&189&$< 10^{-3}$&7.05&34 -24 69\\
  \cline{2-7}
  &$4 \times 4 \times 4$ mm$^3$&$< 10^{-3}$&368 &$< 10^{-3}$&6.94&34 -24 69\\
    \cline{2-7}
  &$4.8 \times 4.8 \times 4.8$ mm&$< 10^{-3}$&507 &$< 10^{-3}$&6.94&34 -22 69\\
   \cline{2-7}
   & $5 \times 5 \times 5$ mm$^3$&$< 10^{-3}$&464 &$< 10^{-3}$&6.90&34 -22 69\\
     \cline{2-7}
  &$6 \times 6 \times 6$ mm$^3$&$< 10^{-3}$&560 &$< 10^{-3}$&6.88&34 -22 69\\
   \cline{2-7}
   & $8 \times 8 \times 8$ mm$^3$&$< 10^{-3}$&789 &$< 10^{-3}$&6.78&36 -22 69\\
 \hline
\end{tabular}
\label{tab:comp_smooth_r2}
\end{table}


When comparing statistical results corresponding to the most significant peak 
(see Tab.~\ref{tab:comp_smooth_r2}), we can notice that~\mSENSE~reaches the performance of 3D-UWR-SENSE only with a spatial 
smoothing which lies between FWHM=$3 \times 3 \times 3$~mm$^3$ and FWHM=$4 \times 4 \times 4$~mm$^3$ 
(see gray lines in Tab.~\ref{tab:comp_smooth_r2}). 
We therefore 
can conclude that the intrinsic spatial smoothing of the proposed method can be estimated as a Gaussian 
smoothing with FWHM$_{\mathrm{3D-UWR-SENSE}} \approx 3.5 \times 3.5 \times 3.5$~mm$^3$. 
Based on this conclusion, one can for instance compare \mSENSE~results smoothed at FWHM=$6 \times 6 \times 6$~mm$^3$ 
to the same effective smoothing with 3D-UWR-SENSE. 
For doing so, we have calculate the additional spatial smoothing to apply to 3D-UWR-SENSE images.  
Straightforward calculations based on the relation between the FWHMs of the reconstruction 
methods\footnote{FWHM=$2\sigma \sqrt{2\log 2}$, where we have the following relation between standard 
deviations of the absolute, pre-processing and reconstruction method smoothing: 
$\sigma_{\mathrm{absolute}} = \sqrt{\sigma_{\mathrm{preprocessing}}^2 + \sigma_{\mathrm{method}}^2}$} 
(FWHM$_{\mathrm{\mSENSE}}=1 \times 1 \times 1$mm$^3$ and 
FWHM$_{\mathrm{3D-UWR-SENSE}}\approx3.5 \times 3.5 \times 3.5$~mm$^3$) and the pre-processing smoothing 
FWHM show that 
3D-UWR-SENSE images have to be smoothed with a FWHM$\approx 4.8 \times 4.8 \times 4.8$~mm$^3$ Gaussian filter. Results 
corresponding to this smoothing level are illustrated in Tab.~\ref{tab:comp_smooth_r2}. 
Comparisons with those of 
\mSENSE~filtered at FWHM=$6 \times 6 \times 6$~mm$^3$ show that 3D-UWR-SENSE clearly outperforms \mSENSE~especially in 
terms of spatial extent of the most significant cluster while giving close T-score maxima.\\
As regards temporal smoothing, Tab.~\ref{tab:comp_smooth_r2} also shows statistical results 
obtained for the \LcRc contrast at $R=2$ using 4D-UWR-SENSE. When compared to those obtained with 3D-UWR-SENSE, 
one can clearly notice the high impact of the temporal regularization both in terms of 
cluster extent and T-score maxima. This conclusion holds for different spatial smoothing levels.

%

\subsubsection{Group-level analysis} 
Due to between-subject anatomical and functional variability, group-level analysis is necessary in order to derive robust and reproducible conclusions at the population level. For this validation, random effect analyses~(RFX) involving fifteen healthy subjects have been conducted on the contrast maps we previously investigated at the subject level. More precisely, 
one-sample Student's $t$ test was performed on the subject-level contrast images~(eg, \LcRc, \ACAS,... images) using SPM5.

\begin{figure}[!ht]
\centering
\begin{tabular}{c c c c}
&\mSENSE&UWR-SENSE&4D-UWR-SENSE\\
\hspace*{-0.4cm}\raisebox{2cm}{$R=2$}&
\hspace*{-0.3cm}\includegraphics[width=3.2cm, height=3cm]{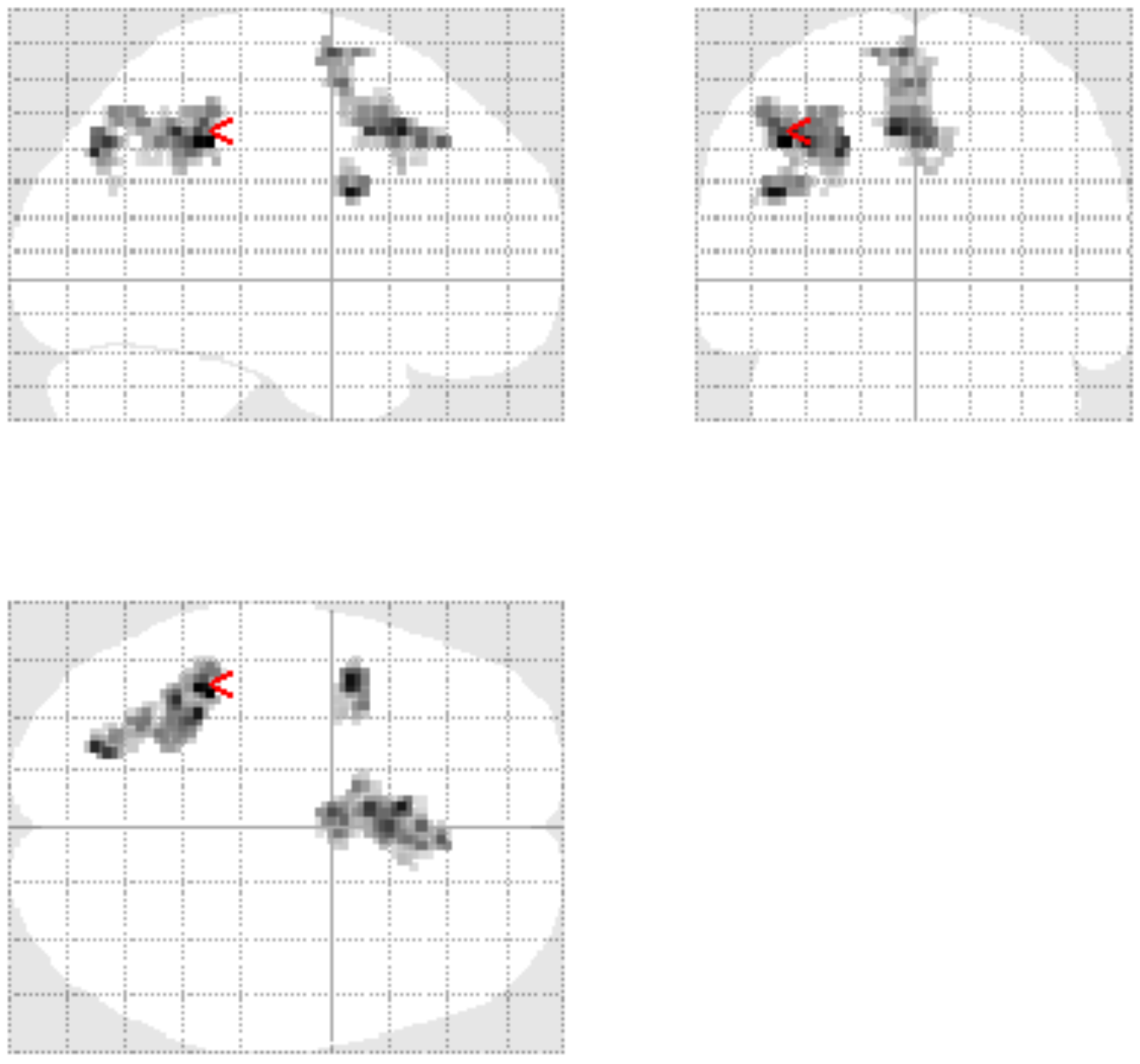}&
\hspace*{-0.3cm}\includegraphics[width=3.2cm, height=3cm]{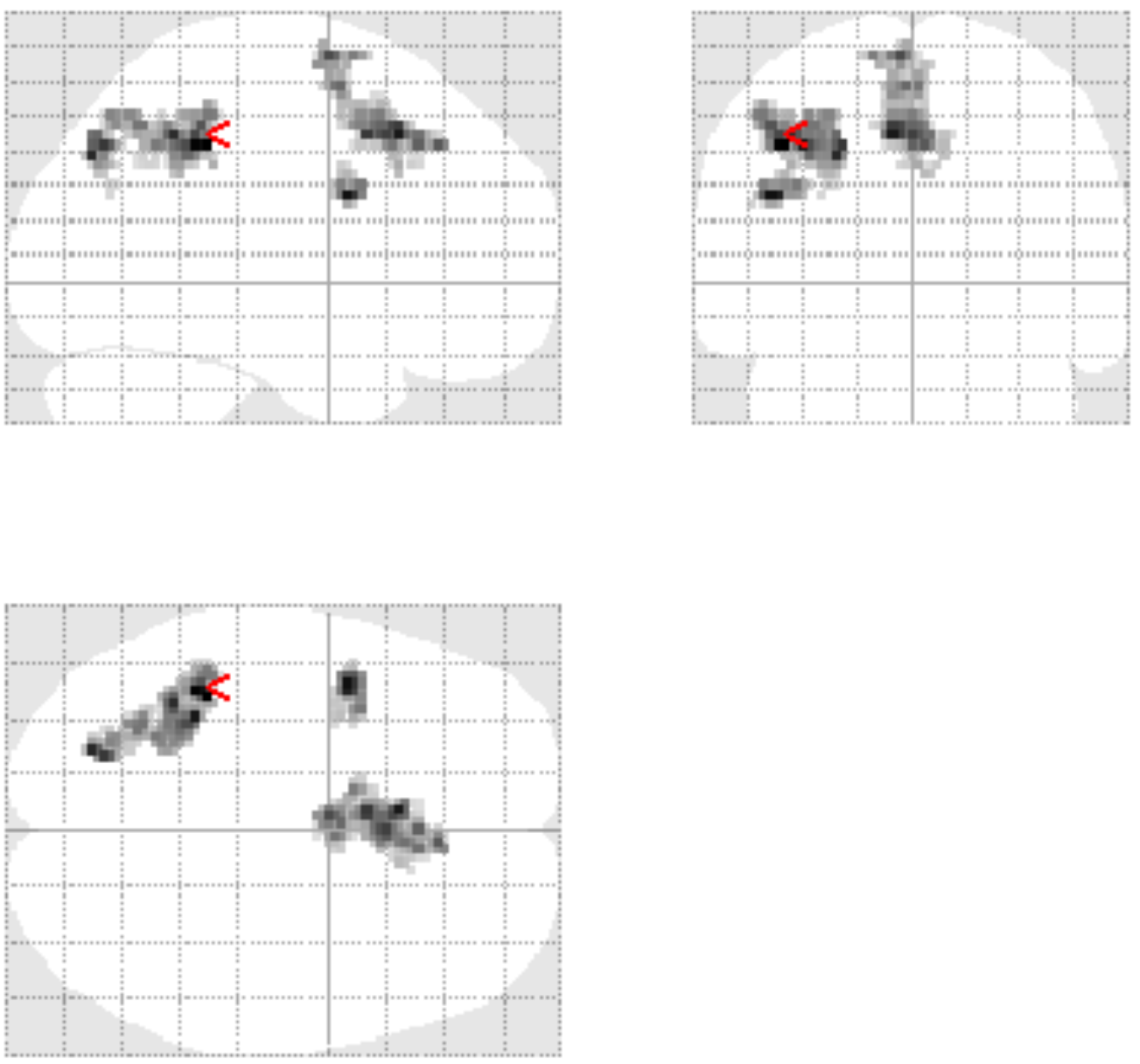}&
\hspace*{-0.3cm}\includegraphics[width=3.2cm, height=3cm]{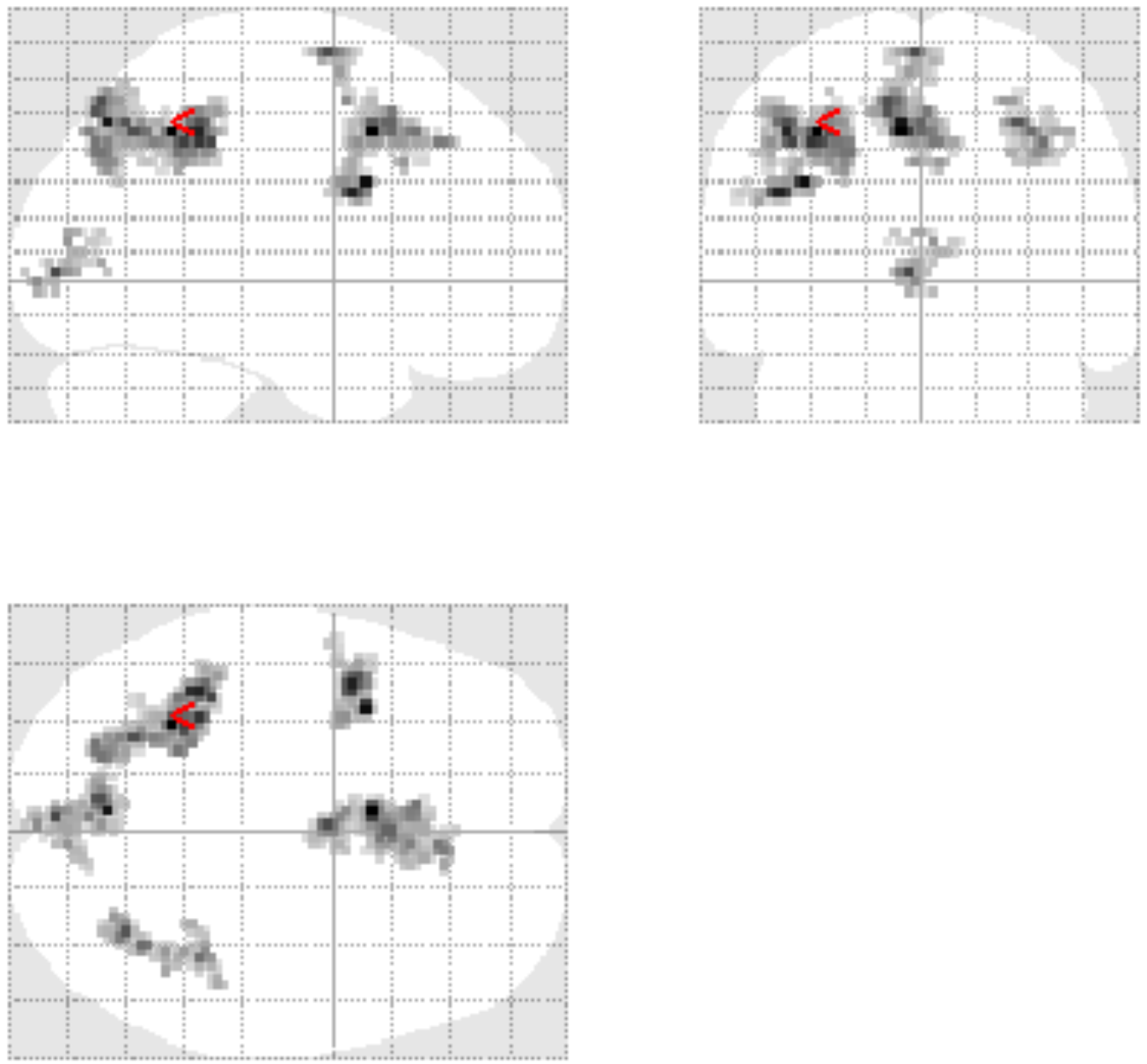}\\
\hspace*{-0.4cm}\raisebox{2cm}{$R=4$}&
\includegraphics[width=3.2cm, height=3cm]{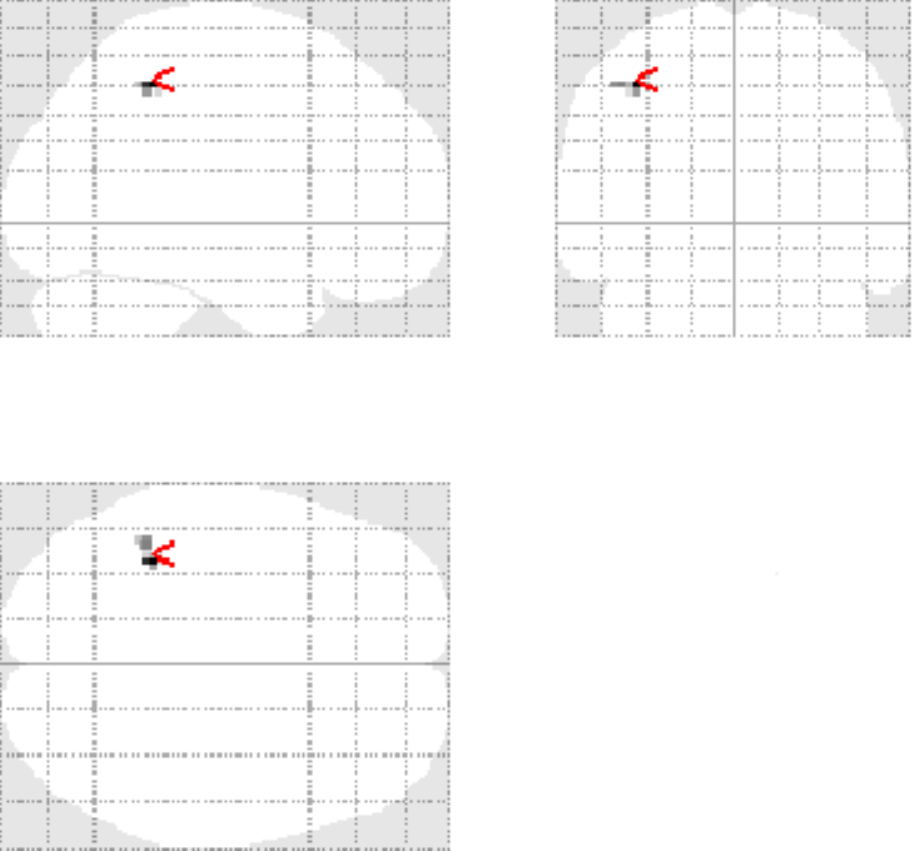}&
\hspace*{-0.3cm}\includegraphics[width=3.2cm, height=3cm]{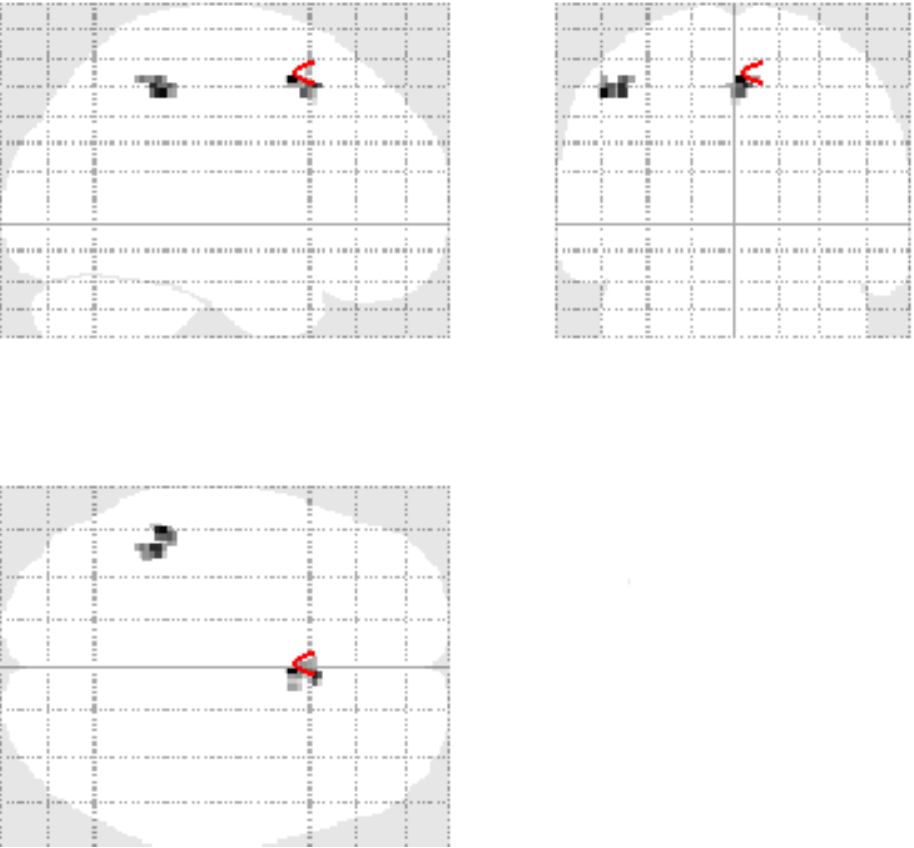}&
\hspace*{-0.3cm}\includegraphics[width=3.2cm, height=3cm]{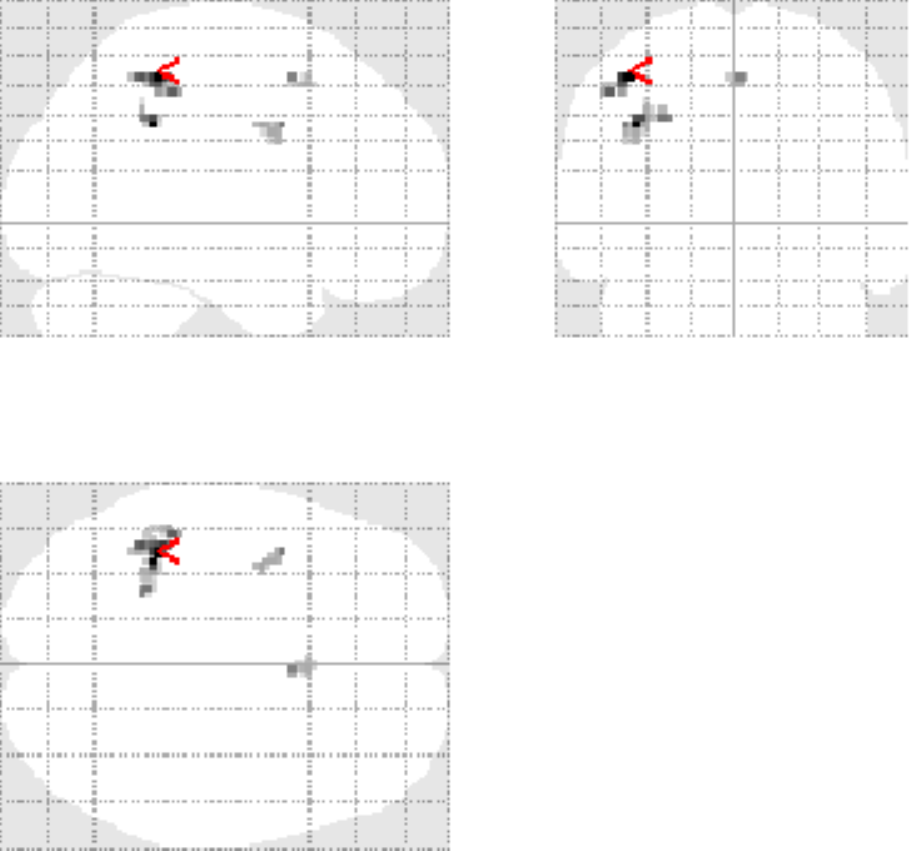}
\end{tabular}\vspace*{-.3cm}
 \caption{Group-level Student's $t$-maps for the \ACAS contrast where data have been reconstructed using the \mSENSE, UWR-SENSE and 4D-UWR-SENSE for $R=2$ and $R=4$. Neurological convention. 
 Red arrows indicate the global maximum activation peak.\label{fig:res_G_A-V}}
\end{figure}

For the \ACAS contrast, Maximum Intensity Projection~(MIP) Student's $t$-maps are shown in Fig.~\ref{fig:res_G_A-V}. First, they illustrate
that irrespective of the reconstruction method larger and more
significant activations are found on datasets acquired with $R=2$
providing the better SNR. Second, for $R=2$, visual inspection of
Fig.~\ref{fig:res_G_A-V}~[top] confirms that only the 4D-UWR-SENSE
algorithm allows us to retrieve significant bilateral activations in
the parietal cortices~(see axial MIP slices) in addition to larger
cluster extent and a gain in significance level for the stable
clusters across the different reconstructors. Similar conclusions can
be drawn when looking at Fig.~\ref{fig:res_G_A-V}~[bottom] for
$R=4$. Complementary results are available in
Tab.~\ref{tab:StatRes2allGrA-V} for $R=2$ and $R=4$.

\begin{table}[!ht]
\centering 
\caption{Significant statistical results at the group-level for the \ACAS contrast (corrected for multiple comparisons at $p=0.05$). Images were reconstructed using the \mSENSE, UWR-SENSE and 4D-UWR-SENSE algorithms for $R=2$ and $R=4$.}
\begin{tabular}{|c|c|c|c|c|c|c|}
\cline{3-7}
\cline{3-7}
\multicolumn{2}{c}{}&\multicolumn{2}{|c|}{cluster-level}&\multicolumn{3}{|c|}{voxel-level}\\
\cline{3-7}
\multicolumn{2}{c|}{}&p-value&Size&p-value&T-score& Position\\
\hline
\multirow{9}{*}{$R=2$}&\multirow{3}{*}{\mSENSE} &$< 10^{-3}$ & 361 & 0.014&7.68&-6 -22 45\\
\cline{3-7}
& &$< 10^{-3}$ &331 & 0.014&8.23&-40 -38 42\\
\cline{3-7}
& &$< 10^{-3}$ &70 & 0.014&7.84&-44 6 27\\
\cline{2-7}
\cline{2-7}
&\multirow{3}{*}{UWR-SENSE} &$< 10^{-3}$ & 361&0.014& 7.68&-6 22 45\\
\cline{3-7}
& &$< 10^{-3}$&331&0.014& 7.68 &-44 -38 42\\
\cline{3-7}
& &$< 10^{-3}$ & 70&0.014& 7.84 &-44 6 27\\
\cline{2-7}
\cline{2-7}
&\multirow{3}{*}{4D-UWR-SENSE} &$< 10^{-3}$& \textbf{441} & $< 10^{-3}$&\textbf{9.45}&-32 -50 45\\
\cline{3-7}
& &$< 10^{-3}$ & 338 &$< 10^{-3}$&9.37&-6 12 45 \\
\cline{3-7}
& &$< 10^{-3}$ & 152 & 0.010&7.19&30 -64 48 \\
\hline
\hline
\multirow{6}{*}{$R=4$}&\multicolumn{1}{|c|}{\mSENSE} &0.003& 14& 0.737&5.13&-38 -42 51\\
\cline{2-7}
\cline{2-7}
&\multirow{2}{*}{UWR-SENSE} &$< 10^{-3}$ & \textbf{41} & 0.274&5.78&-50 -38 -48 \\
\cline{3-7}
& &$< 10^{-3}$ & 32 & 0.274&5.91&2 12 54 \\
\cline{2-7}
\cline{2-7}
&\multirow{3}{*}{4D-UWR-SENSE} &$< 10^{-3}$& 37 & 0.268&\textbf{6.46}&-40 -40 54\\
\cline{3-7}
& &$< 10^{-3}$& 25& 0.268&6.37 &-38 -42 36\\\cline{3-7}
 & &$< 10^{-3}$ & 18 & 0.273 & 5 & -42 8 36 \\
\hline
\end{tabular}
\label{tab:StatRes2allGrA-V}
\end{table}

These results allow us to 
numerically validate this visual comparison:
\begin{itemize}
\item  Whatever the reconstruction method in use, the statistical performance is much more significant using $R=2$, especially at the cluster level since the cluster extent decreases by one order of magnitude.

\item Voxel and cluster-level results are enhanced using the 4D-UWR-SENSE approach instead of the \mSENSE~reconstruction or its early UWR-SENSE version.
\end{itemize}
Fig.~\ref{fig:res_G_Lc-Rc} reports similar group-level MIP results for $R=2$ and $R=4$ concerning the \LcRc contrast. 

\begin{figure}[!htp]
\centering
\begin{tabular}{c c c c}
&\mSENSE&UWR-SENSE&4D-UWR-SENSE\\
\hspace*{-0.4cm}\raisebox{2cm}{$R=2$}&\hspace*{-0.3cm}
\includegraphics[width=3.2cm, height=3cm]{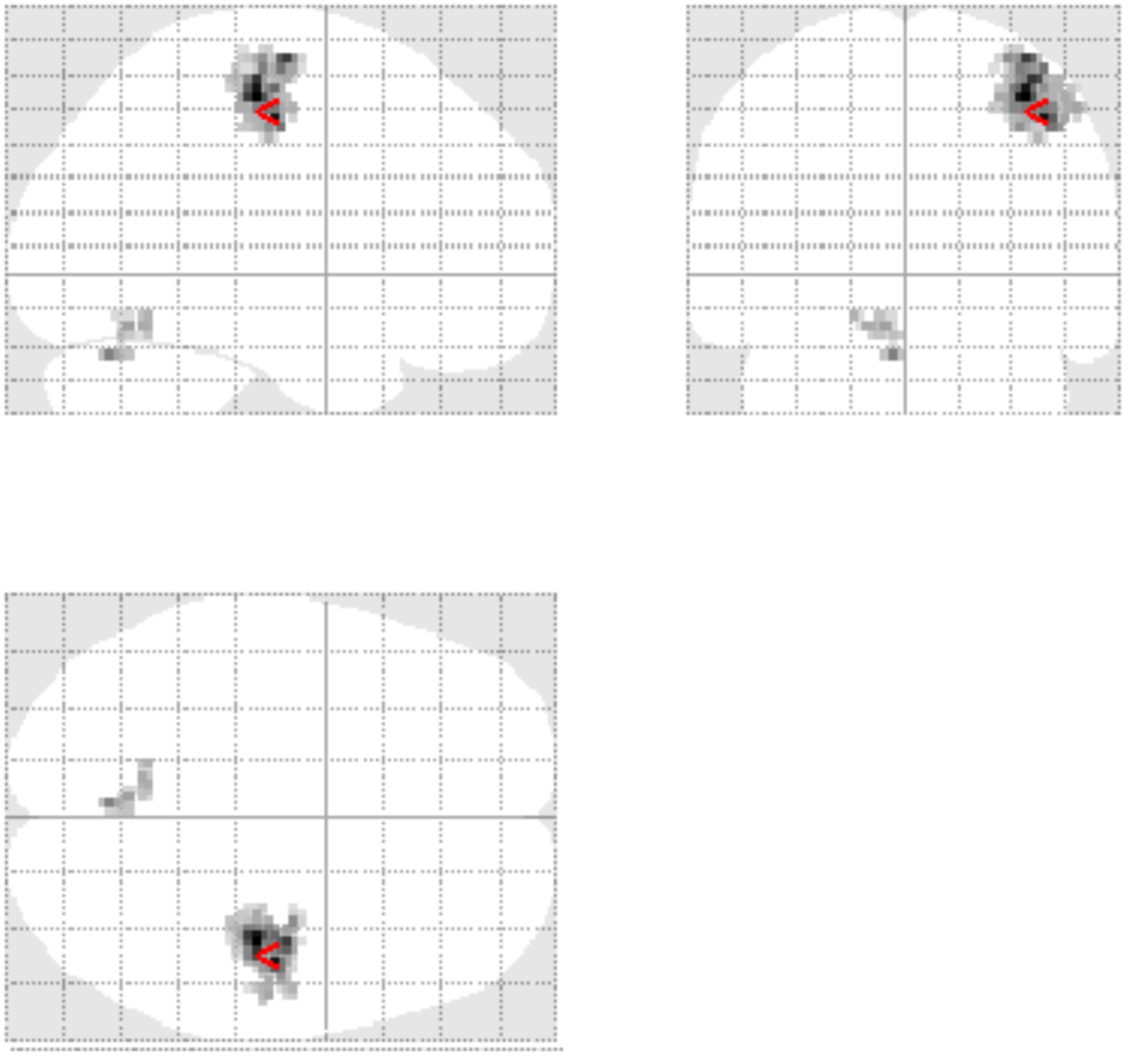}&
\hspace*{-0.3cm}\includegraphics[width=3.2cm, height=3cm]{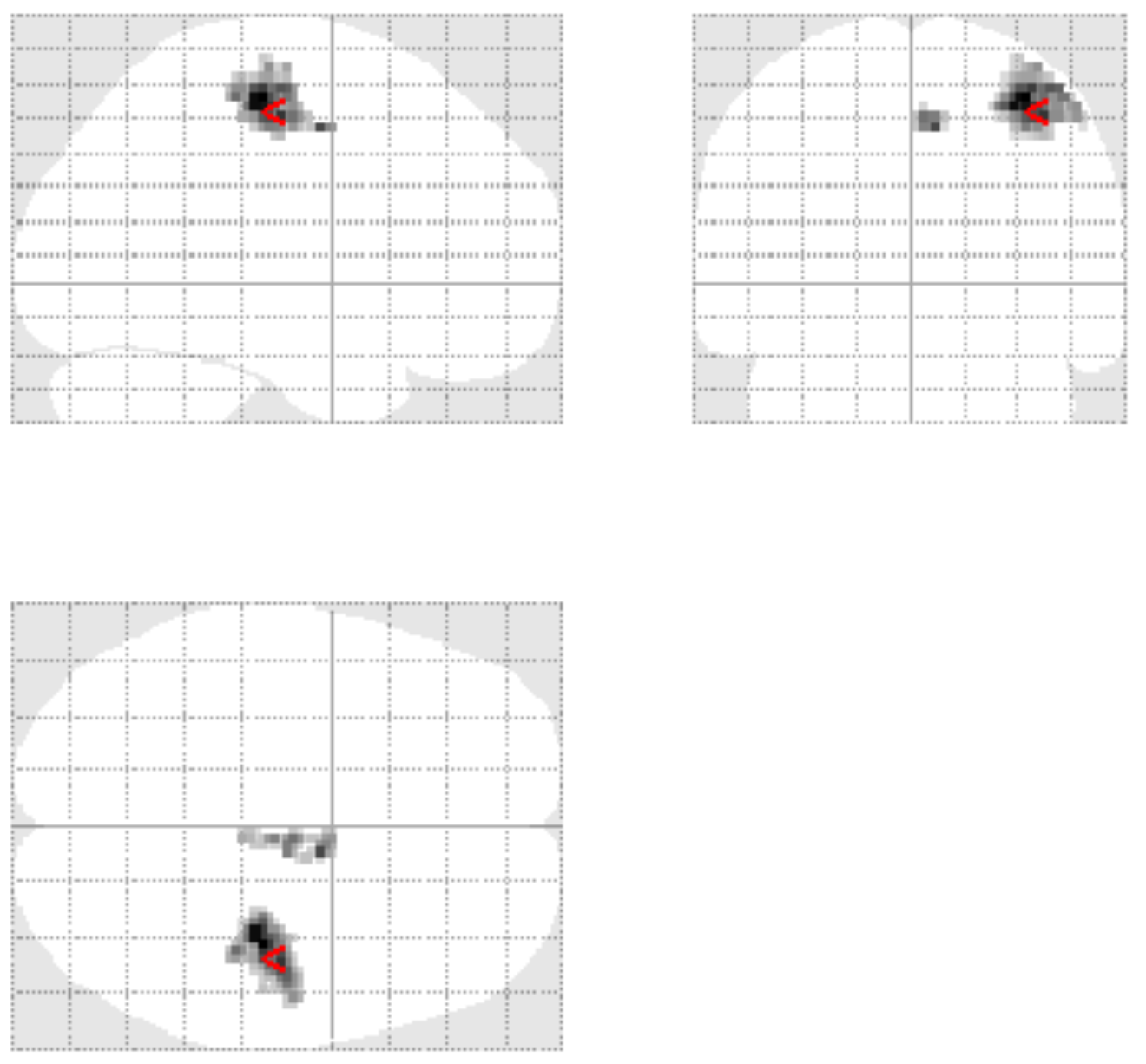}&
\hspace*{-0.3cm}\includegraphics[width=3.2cm, height=3cm]{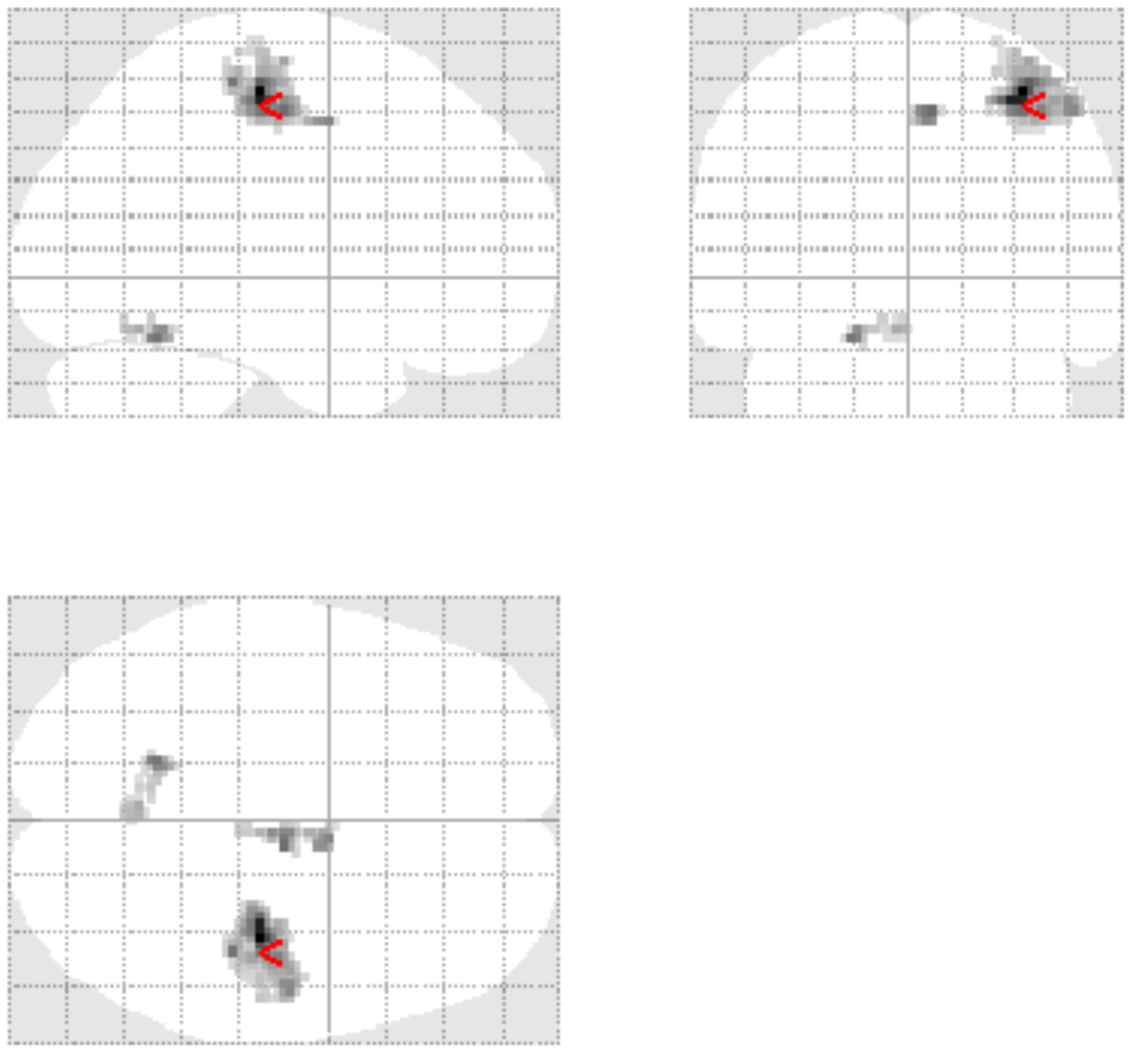}\\
\hspace*{-0.4cm}\raisebox{2cm}{$R=4$}&\hspace*{-0.3cm}
\includegraphics[width=3.2cm, height=3cm]{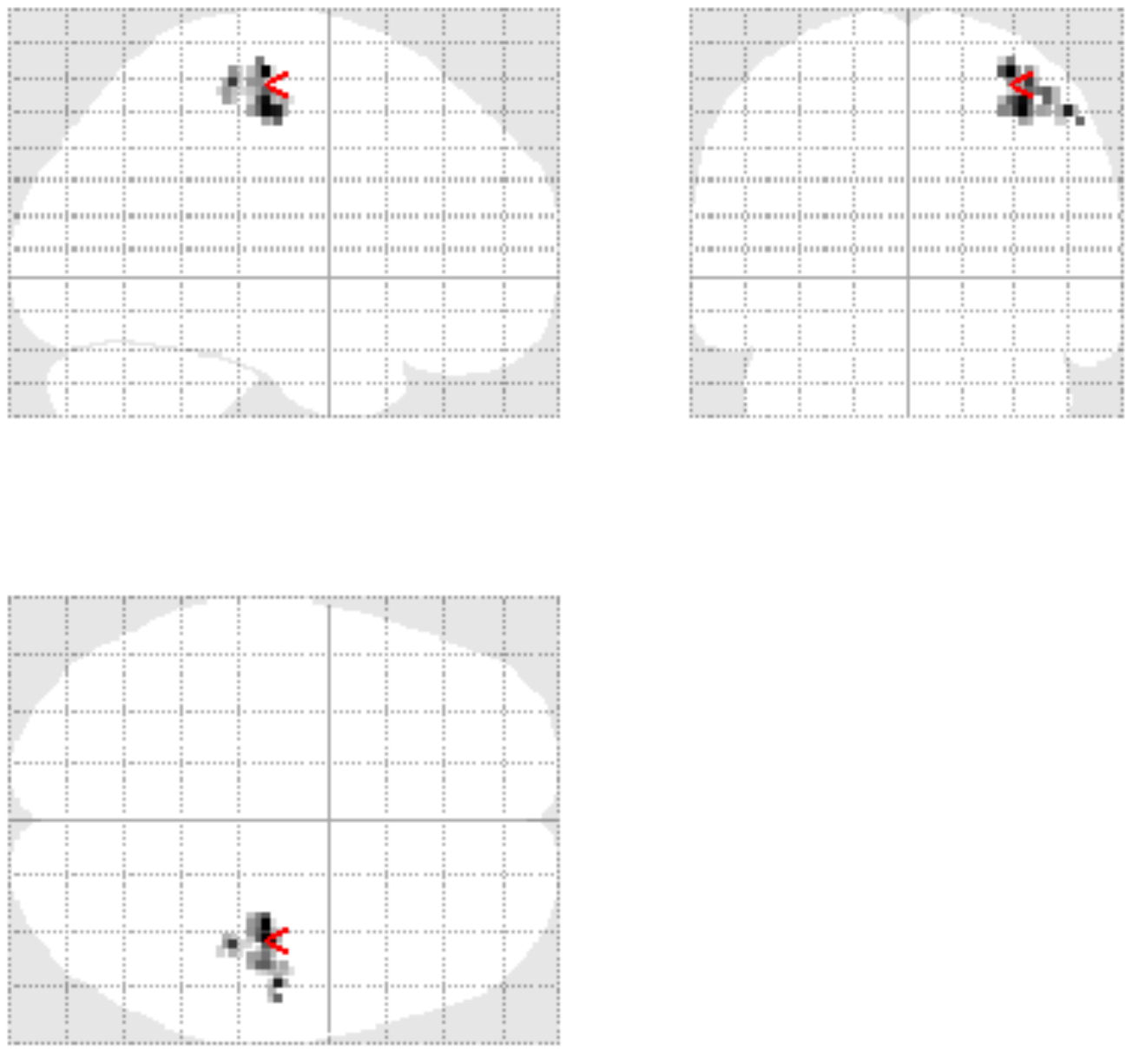}&
\hspace*{-0.3cm}\includegraphics[width=3.2cm, height=3cm]{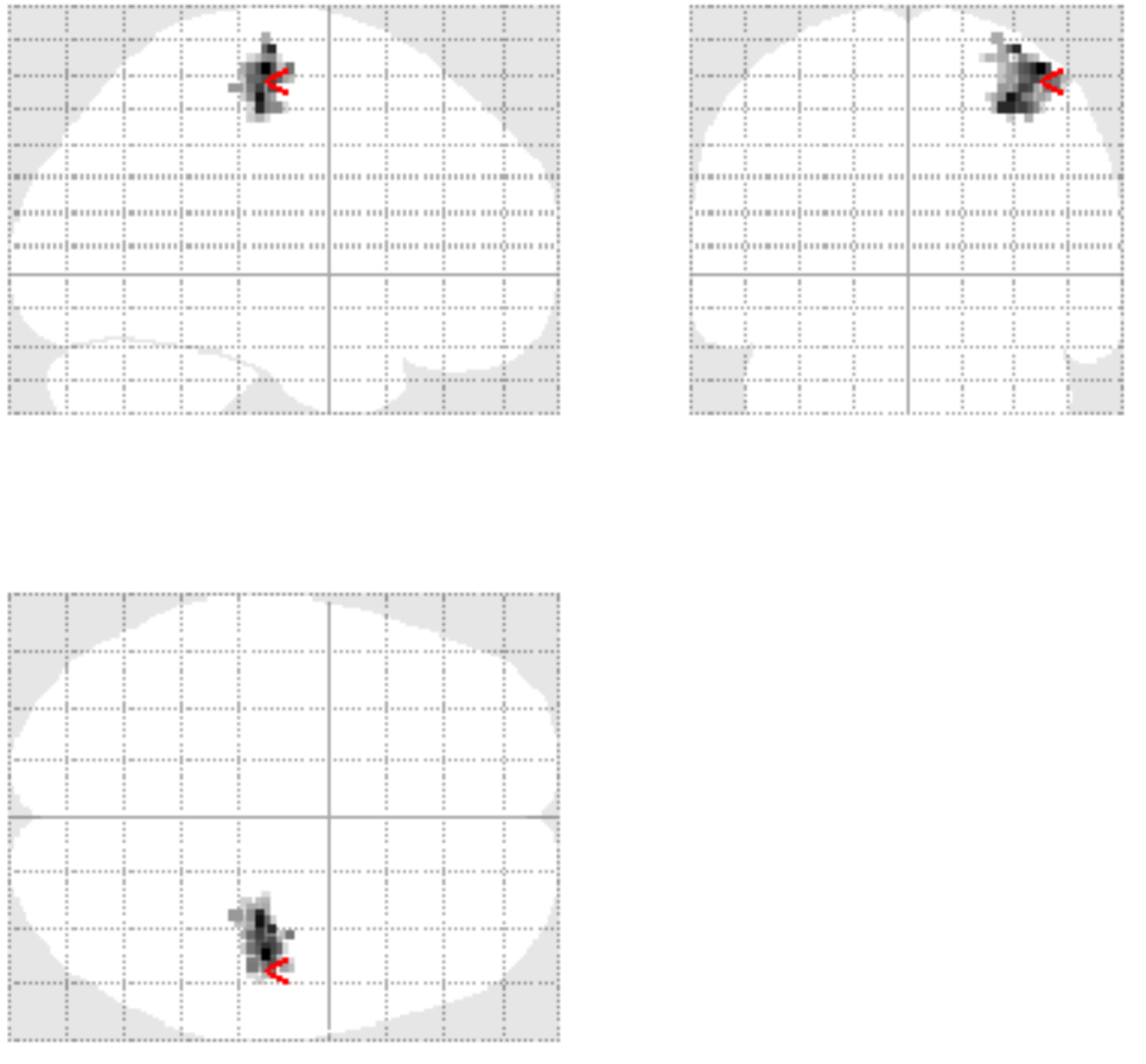}&
\hspace*{-0.3cm}\includegraphics[width=3.2cm, height=3cm]{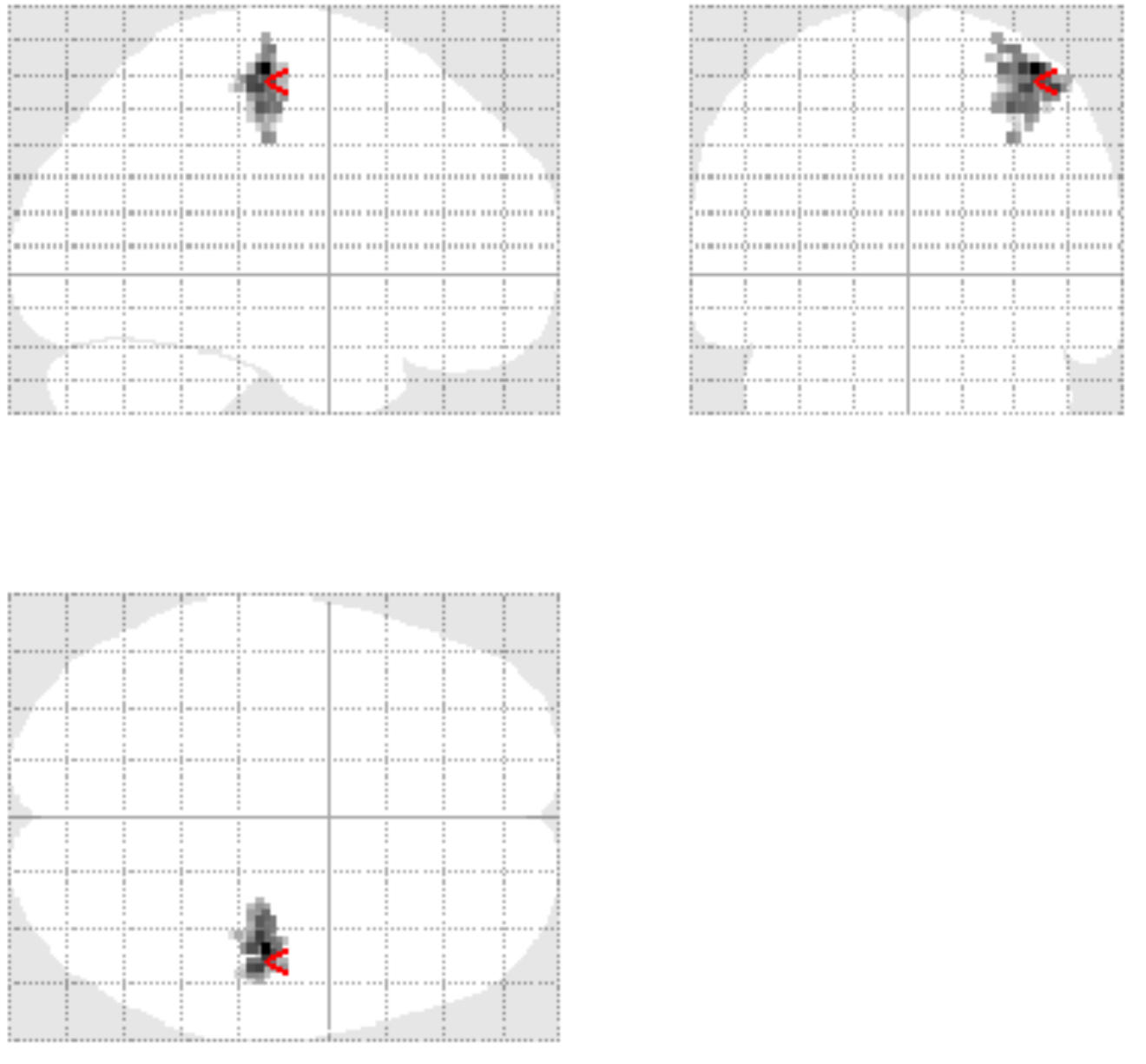}
\end{tabular}\vspace*{-.3cm}
\caption{Group-level Student's $t$-maps for the \LcRc contrast where data have been reconstructed using the \mSENSE, 
UWR-SENSE and 4D-UWR-SENSE for $R=2$ and $R=4$. Neurological convention. Red arrows indicate the global maximum activation peak.
\label{fig:res_G_Lc-Rc}}
\end{figure}

It is shown that whatever the acceleration factor $R$ in use, our pipeline enables to detect a much more spatially extended 
activation area in the motor cortex. This visual inspection is quantitatively confirmed in 
Tab.~\ref{tab:StatRes2allGrLc-Rc} when comparing the detected clusters using our 4D-UWR-SENSE approach with those found by
\mSENSE, again irrespective of $R$. Finally, the 4D-UWR-SENSE algorithm outperforms the UWR-SENSE one, which corroborates the
benefits of the proposed spatio-temporal regularization scheme.
\begin{table}[!htp]
\centering 
\caption{Significant statistical results at the group-level for the \LcRc contrast (corrected for multiple comparisons at $p=0.05$). Images were reconstructed using the \mSENSE, UWR-SENSE and 4D-UWR-SENSE algorithms for $R=2$ and $R=4$.}
\begin{tabular}{|c|c|c|c|c|c|c|}
\cline{3-7}
\cline{3-7}
\multicolumn{2}{c}{}&\multicolumn{2}{|c|}{cluster-level}&\multicolumn{3}{|c|}{voxel-level}\\
\cline{3-7}
\multicolumn{2}{c|}{}&p-value&Size&p-value&T-score& Position\\
\hline
\multirow{7}{*}{$R=2$}&
\multirow{2}{*}{\mSENSE} &$< 10^{-3}$ & 354 &$< 10^{-3}$&9.48&38 -22 54\\
\cline{3-7}
& &0.001 &44 & 0.665&6.09&-4 -68 -24\\
\cline{2-7}
&\multirow{2}{*}{UWR-SENSE} &$< 10^{-3}$ &350& 0.005& 9.83&36 -22 57\\
\cline{3-7}
& &$< 10^{-3}$ & 35&0.286& 7.02&4 -12 51\\
\cline{2-7}
&\multirow{3}{*}{4D-UWR-SENSE} &$< 10^{-3}$& \textbf{377} & 0.001&\textbf{11.34}&36 -22 57\\
\cline{3-7}
& &$< 10^{-3}$ & 53& $< 10^{-3}$&7.50&8 -14 51 \\
\cline{3-7}
&\multicolumn{1}{|c|}{} &$< 10^{-3}$ & 47& $< 10^{-3}$&7.24&-18 -54 -18 \\
\hline
\hline
\multirow{3}{*}{$R=4$}&\multicolumn{1}{|c|}{\mSENSE} & $< 10^{-3}$& 38& 0.990&5.97&32 -20 45\\
\cline{2-7}
&\multicolumn{1}{|c|}{UWR-SENSE}  &$< 10^{-3}$ &163 &0.128&7.51&46 -18 60\\
\cline{2-7}
&\multicolumn{1}{|c|}{4D-UWR-SENSE}&$< 10^{-3}$ & \textbf{180}& 0.111& \textbf{7.61}&46 -18 60\\
\hline
\end{tabular}
\label{tab:StatRes2allGrLc-Rc}
\end{table}



\section{Discussion}\label{sec:discussion}

Through illustrated results, we showed that whole brain acquisition can be routinely used at a spatial in-plane resolution
of $2 \times 2 \mathrm{mm}^2$ in a 
short and constant repetition time~($\text{TR} = 2.4$s) provided that a reliable pMRI reconstruction pipeline is chosen. 
In this paper, we demonstrated that our 4D-UWR-SENSE reconstruction algorithm meets this goal. To draw this conclusion, qualitative comparisons have been made directly on reconstructed images using our pipeline involving the 3D and 4D-UWR-SENSE algorithms or \mSENSE. On anatomical data where the acquisition scheme is fully 3D, our results confirm the usefulness of the 3D wavelet regularization for attenuating 3D spatially propagating artifacts. On the other hand, our results on functional data show that, even when the acquisition scheme is 2D sequential, reconstruction artifacts are attenuated by resorting simultaneously to the 3D wavelet and temporal regularizations. In the case of interleaved 2D acquisition scheme where contiguous slices are acquired every $\text{TR}/2$, motion artifacts may dramatically alter the reconstruction quality using the \mSENSE~method. Although the actual version of the proposed algorithm does not account for such artifacts, a trade-off between the two regularizers may be found to cope with this issue.

Quantitatively speaking, our comparison took place at the statistical analysis 
level and relied on quantitative criteria~(voxel- and cluster-level corrected p-values, $t$-scores, peak positions) both at the subject and group levels. In particular, we showed that our 4D-UWR-SENSE approach outperforms both its UWR-SENSE 
ancestor~\cite{Chaari_MEDIA_2011} and the Siemens \mSENSE~reconstruction in terms of statistical significance and robustness. This emphasized the benefits of combining temporal and 3D regularization in the wavelet domain. The usefulness of 3D regularization in reconstructing 3D anatomical images was also shown, especially in more degraded situations~($R=4$) where regularization plays a prominent role. 
The validity of our conclusions lies in the reasonable size of our datasets since the same participants were scanned using two different pMRI acceleration factors~($R=2$ and $R=4$).

At the considered spatio-temporal compromise~($2\times2\times3$mm$^3$ and
$\text{TR}=2.4$~s), we also illustrated the impact of increasing the
acceleration factor~(passing from $R=2$ to $R=4$) on the statistical
sensitivity at the subject and group levels for a given reconstruction
algorithm. We performed this comparison to anticipate what could be
the statistical performance for detecting evoked brain activity on
data requiring this acceleration factor, such as high spatial
resolution EPI images~(e.g., $1.5 \times 1.5\mathrm{mm}^2$ in-plane
resolution) acquired in the same short $\text{TR}$. Our conclusions
were balanced depending on the contrast of interest: when looking at
the \ACAS contrast involving the fronto-parietal circuit, it turned
out that $R=4$ was not reliable enough to recover significant group-level activity at 3~Tesla: the SNR 
loss was too important and should be compensated by an increase of the static magnetic field~(e.g. passing from 3 to 7~Tesla). 
However, the situation becomes acceptable for the \LcRc motor contrast, which elicits activation in motor regions: our results 
brought evidence that the 4D-UWR-SENSE approach enables the use of $R=4$ for this contrast.

%

\section{Conclusion}\label{sec:conclusion}

The contribution of the present paper was twofold. First, we proposed  a novel reconstruction method that relies on  a 3D wavelet transform and accounts for temporal dependencies in successive fMRI volumes. As a particular
case, the proposed method allows us to deal with 3D acquired anatomical data
when a single volume is acquired. 
Second, when artifacts were superimposed to brain activation, 
we showed that the choice of the pMRI reconstruction algorithm has a significant
influence on the statistical sensitivity at the subject and group-levels in fMRI and may enable whole brain neuroscience studies at high spatial resolution. 
Our results brought evidence that the compromise between acceleration factor and
spatial in-plane resolution should be selected with care depending on
the regions involved in the fMRI paradigm. As a consequence, high
resolution fMRI studies can be conducted using high speed
acquisition~(short $\text{TR}$ and large $R$ value) provided that the
expected BOLD effect is strong, as experienced in primary motor,
visual and auditory cortices. Of course, the use of an efficient
reconstruction method such as the one proposed is a pre-requisite to
shift this compromise towards larger $R$ values and higher spatial
resolution and it could be optimally combined with ultra high magnetic fields~($\geqslant 7$~T).

A direct extension of the present work, which is actually in progress,
consists of studying the impact of tight frames instead of wavelet
bases to define more suitable 3D transforms. However, unsupervised reconstruction becomes more challenging in this framework since the estimation of hyper-parameters becomes 
cumbersome~(see~\cite{Chaari_TSP_2010} for details). 
Integrating some pre-processing steps in the reconstruction model may also be of great interest to account for motion artifacts in the regularization step, especially for interleaved 2D acquisition schemes. Such an extension deserves integration of recent works on joint correction of motion and slice-timing such 
as~\cite{Roche_2011}.

Ongoing work will also concern the combination of the present contribution
with the joint detection estimation approach of evoked
activity~\cite{Makni08,Vincent10,Badillo13} to go beyond the GLM framework and
to evaluate how the pMRI reconstruction algorithm also impacts HRF estimation. 
Another extension of our work would concern the combination of our wavelet-regularized reconstruction with the WSPM approach~\cite{VanDeVille_2007} 
in which statistical analysis is directly performed in the wavelet transform domain.\\

\begin{center}
 {\LARGE \textbf{Appendix}}
\end{center}

\appendix 

\section{Optimization procedure for the 4D reconstruction}\label{append:a1}

Based on the formulation hereabove, the criterion to be minimized can be written as follows:
\begin{align}
\label{eq:Reg_4D_3}
\mathcal{J}_{\rm ST} (\zeta) &= \mathcal{J}_{\rm TWLS}(\zeta) + g(\zeta) + h(\zeta)
\end{align}
where $\mathcal{J}_{\rm TWLS}$ is defined as 
\begin{align}
 \mathcal{J}_{\rm TWLS}(\zeta) &= \sum\limits_{t = 1}^{N_r} \mathcal{J}_{\rm WLS}(\zeta^t) \nonumber \\
 &=\sum\limits_{t = 1}^{N_r} \sum\limits_{\mathbf{r}\in \{1,\ldots,X\} \times \{1,\ldots,Y/R\}\times  \{1,\ldots,Z\}} 
\Vert \vect{d}^t(\vect{r}) - \vect{S}(\vect{r})(T^*\zeta^t)(\vect{r}) \Vert^2_{\vect{\Psi}^{-1}}.
\end{align}

The minimization of $\mathcal{J}_{\rm ST}$ is performed by resorting to the concept of proximity operators \cite{Moreau_65}, which was found to be fruitful in a number of recent works in convex 
optimization \cite{Chaux_C_07,Combettes_PL_2005_mms_Signal_rbpfbs,Combettes_09bis}. In what follows, we recall the definition of a proximity operator:

\begin{definition} {\rm \cite{Moreau_65}} \label{def:prox}
Let $\Gamma_0(\chi)$ be the class of proper lower semicontinuous convex functions from a separable real Hilbert 
 space $\chi$ to $]-\infty,+\infty]$ and let $\varphi \in \Gamma_0(\chi)$. For every $\mathsf{x} \in \chi$, 
 the function $\varphi+\Vert \cdot-\mathsf{x} \Vert^2/2$ achieves its infimum at a unique point denoted by 
 $\mathrm{prox}_{\varphi}\mathsf{x}$. The operator $\mathrm{prox}_{\varphi}\; : \; \chi \rightarrow \chi$ 
 is the proximity operator of $\varphi$.
\end{definition}

In this work, as the observed data are complex-valued, the definition of proximity operators is extended to a class of convex functions defined for complex-valued variables. For the function
\begin{align}
\Phi \colon \mathbb{C}^K &\to ]-\infty,+\infty]\\ \nonumber
x &\mapsto \phi^{\mathrm{Re}}(\mathrm{Re}(x))+ \phi^{\mathrm{Im}}(\mathrm{Im}(x)),
\end{align} 
where $\phi^{\mathrm{Re}}$ and $\phi^{\mathrm{Im}}$ are functions in $\Gamma_0(\RR^K)$ and $\mathrm{Re}(x)$~(respectively $\mathrm{Im}(x)$) is the vector of the real parts~(respectively imaginary parts) of the components of $x\in \mathbb{C}^K$, the proximity operator is defined as
\begin{align}
\mathrm{prox}_{\Phi} \colon \mathbb{C}^K & \to \mathbb{C}^K \\ \nonumber
x &\mapsto \mathrm{prox}_{\phi^{\mathrm{Re}}}(\mathrm{Re}(x))+\imath
\mathrm{prox}_{\phi^{\mathrm{Im}}}(\mathrm{Im}(x)).
\label{eq:defproxc}
\end{align}

Let us now provide the expressions of proximity operators involved in our reconstruction problem.

\subsection{Proximity operator of the data fidelity term}

According to standard rules on the calculation of proximity operators \cite[Table 1.1]{Combettes_09bis}, the proximity operator of the data fidelity term $\mathcal{J}_{\rm WLS}$ is given for every vector of coefficients $\zeta^t$ (with $t\in\{1,\ldots,N_r\}$) by $\mathrm{prox}_{\mathcal{J}_{\rm WLS}}(\zeta^t) = T u^t$, where the image $u^t$ is such that $\forall \mathbf{r}\in \{1,\ldots,X\}\times \{1,\ldots,Y/R\}\times \{1,\ldots,Z\}$,
\begin{equation}
\vect{u}^t(\mathbf{r})= \big(\vect{I}_R + 2\vect{S}^{\hermit}(\mathbf{r})\vect{\Psi}^{-1}\vect{S}(\mathbf{r}) \big)^{-1} 
\big({\boldsymbol{\rho}^{t}}(\mathbf{r}) + 2\vect{S}^{\hermit}(\mathbf{r})\vect{\Psi}^{-1}\vect{d}^t(\mathbf{r})\big),
\end{equation} 
where ${\rho^{t}} =  T^*\zeta^{t}$.

\subsection{Proximity operator of the spatial regularization function}

According to \cite{Chaari_MEDIA_2011}, for every resolution level $j$ and orientation $o$, the proximity operator of the spatial regularization function $\Phi_{o,j}$ is given by
\begin{multline}
\forall \xi \in \CC,\qquad
\mathrm{prox}_{\Phi_{o,j}} \xi= 
\dfrac{\mathrm{sign}(\mathrm{Re}(\xi-\mu_{o,j}))}{\beta_{o,j}^{\mathrm{Re}}+1}\max\{|\mathrm{Re}(\xi-\mu_{o,j})|-
\alpha_{o,j}^{\mathrm{Re}},0\}\\ +
\imath \dfrac{\mathrm{sign}(\mathrm{Im}(\xi-\mu_{o,j}))}{\beta_{o,j}^{\mathrm{Im}}+1}\max\{|\mathrm{Im}(\xi-\mu_{o,j})|-
\alpha_{o,j}^{\mathrm{Im}},0\}+\mu_{o,j}
\end{multline}
where the $\mathrm{sign}$ function is defined as follows:
\begin{equation}
\forall \xi\in\mathbb{R},\qquad  \mathrm{sign}(\xi)= \begin{cases} +1 & \text{if} \; \xi \geq 0\\
-1 & \text{otherwise.} \end{cases} \nonumber
\end{equation}

\subsection{Proximity operator of the temporal regularization function}

A simple expression of the proximity operator of function $h$ is not available. We thus propose to split this regularization term as a sum of two more tractable functions $h_1$ and $h_2$:
\begin{equation}
\label{eq:JT1}
 h_1(\zeta) =  \kappa \sum_{t = 1}^{N_r/2} \Vert T^*\zeta^{2t} - T^*\zeta^{2t-1} \Vert_p^p
\end{equation}
and 
\begin{equation}
\label{eq:JT2}
h_2(\zeta) = \kappa \sum_{t = 1}^{N_r/2-1} \Vert T^*\zeta^{2t+1} - T^*\zeta^{2t} \Vert_p^p.
\end{equation}

Since $h_1$ (respectively $h_2$) is separable w.r.t the time variable $t$, its proximity operator can easily be calculated based on the proximity operator of each of the involved terms in the sum of Eq.~\eqref{eq:JT1} (respectively Eq.~\eqref{eq:JT2}).

Indeed, let us consider the following function 
\begin{align}
 \Psi: \CC^K \times \CC^K &\longrightarrow \RR \\ \nonumber
  (\zeta^t,\zeta^{t-1}) &\mapsto \kappa \Vert T^*\zeta^t - T^*\zeta^{t-1} \Vert_p^p = \psi \circ H (\zeta^t,\zeta^{t-1}),
\end{align}
where $\psi = \kappa\Vert T^*\cdot \Vert_p^p$ and $H$ is the linear operator defined as
\begin{align}
  H: \CC^K \times \CC^K &\longrightarrow \CC^K \\ \nonumber
  (a,b) &\mapsto a-b.
\end{align}

Its associated adjoint operator $H^*$ is therefore given by
\begin{align}
  H^*: \CC^K &\longrightarrow \CC^K \times \CC^K \\ \nonumber
  a &\mapsto (a,-a).
\end{align}

Since we have $H H^* = 2\rm{Id}$, the proximity operator of $\Psi$ can easily be calculated using \cite[Prop.~11]{Combettes_PL_2007_istsp_Douglas_rsatncvsr}:
\begin{equation}
\prox_{\Psi} = \prox_{\psi \circ H} = \mathrm{Id} + \dfrac{1}{2}H^*\circ ( \prox_{2\psi} - \mathrm{Id}) \circ H.
\end{equation}
The calculation of $\prox_{2\psi}$ is discussed in \cite{Chaux_C_07}.

\subsection{Parallel Proximal Algorithm (PPXA)}

The function to be minimized has been reexpressed as
\begin{align}
\label{eq:Reg_4Dbis}
\mathcal{J}_{\rm ST} (\zeta) = & \sum_{t = 1}^{N_r} \sum_{\mathbf{r}\in \{1,\ldots,X\} \times \{1,\ldots,Y/R\}\times  \{1,\ldots,Z\}} 
\Vert \vect{d}^t(\vect{r}) - \vect{S}(\vect{r})(T^*\zeta^t)(\vect{r}) \Vert^2_{\vect{\Psi}^{-1}}\nonumber \\
&+ g(\zeta) + h_1(\zeta) +  h_2(\zeta).
\end{align}

Since $\mathcal{J}_{\rm ST}$ is made up of more than two
non-necessarily differentiable terms, an appropriate solution for
minimizing such an optimality criterion is
PPXA~\cite{Combettes_PL_08}. In particular, it is important to note
that this algorithm does not require subiterations as was the case for
the constrained optimization algorithm proposed in~\cite{Chaari_MEDIA_2011}. In addition, the computations in this algorithm can be performed in a parallel manner and the convergence of the algorithm to an optimal solution to the minimization problem is guaranteed.

The resulting algorithm for the minimization of the optimality
criterion in Eq.~\eqref{eq:Reg_4Dbis} is given in
Algorithm~\ref{algo:4D}. In this algorithm, the weights $\omega_i$
have been fixed to $1/4$ for every $i\in \{1,\ldots,4\}$. The
parameter $\gamma$ has been set to 200 since this value was observed
to lead to the fastest convergence in practice. The stopping parameter $\varepsilon$ has been set to $10^{-4}$. Using these parameters, the algorithm typically converges in less than 50 iterations.

\begin{algorithm}[!ht]
\caption{\small {\bf 4D-UWR-SENSE}: spatio-temporal regularized reconstruction.}
\small Set $\gamma \in ]0,+\infty[$, $\varepsilon \in ]0,1[$, $(\omega_i)_{1 \leq i \leq 4} \in ]0,1[^4$ such that $\sum_{i=1}^4 \omega_i = 1$, $n=0$,  
$(\zeta_i^{(n)})_{1 \leq i \leq 4} \in (\CC^{K\times N_r})^4$ 
where $\zeta_i^{(n)} = (\zeta_i^{1,(n)},\zeta_i^{2,(n)},\ldots,\zeta_i^{N_r,(n)})$, and 
$\zeta_i^{t,(n)} = \big((\zetab^{t,(n)}_{i,a}), ((\zetab^{t,(n)}_{i,o,j}))_{o\in \mathbb{O},1 \le j \le j_\mathrm{max}}\big)$ for every 
$i \in \{1,\ldots,4\}$ and 
$t \in \{1,\ldots,N_r\}$. Set also $\zeta^{(n)} = \sum_{i=1}^4 \omega_i \zeta_i^{(n)}$ and $\mathcal{J}_{\rm ST}^{(n)} = 0$.
\begin{algorithmic}[1]
\REPEAT
	\STATE Set $p_4^{1,(n)} = \zeta_4^{1,(n)}$. 
	\FOR {$t=1$ to $N_r$ }
	\STATE Compute $p_1^{t,(n)} = \mathrm{prox}_{\gamma \mathcal{J}_{\rm WLS} / \omega_1}(\zeta_1^{t,(n)})$.	  
	\STATE Compute $p_2^{t,(n)} = \big(\prox_{\gamma \Phi_a/\omega_2}(\zetab^{t,(n)}_{2,a}), (\prox_{\gamma \Phi_{o,j}/\omega_2}(\zetab^{t,(n)}_{2,o,j}))_{o\in \mathbb{O},1 \le j \le j_\mathrm{max}}\big)$.
\IF{$t$ is even}
	\STATE Compute $(p_3^{t,(n)},p_3^{t-1,(n)}) = \prox_{\gamma \Psi/\omega_3}(\zeta^{t,(n)}_3,\zeta^{t-1,(n)}_3)$
\ELSIF{$t$ is odd and $t>1$}
	\STATE Compute $(p_4^{t,(n)},p_4^{t-1,(n)}) = \prox_{\gamma \Psi/\omega_4}(\zeta^{t,(n)}_4,\zeta^{t-1,(n)}_4)$.
\ENDIF
		  \IF{$t>1$} \STATE Set $P^{t-1,(n)} = \sum_{i=1}^4 \omega_i p_{i}^{t-1,(n)}$. \ENDIF
	\ENDFOR
	\STATE Set $p_4^{N_r,(n)} = \zeta_4^{N_r,(n)}$. 
	\STATE Compute $P^{N_r,(n)} = \sum_{i=1}^4 \omega_i  p_{i}^{N_r,(n)}$.
	\STATE Set $P^{(n)} = (P^{1,(n)},P^{2,(n)},\ldots,P^{N_r,(n)})$.
	\STATE Set $\lambda_n \in [0,2]$.
		\FOR {$i=1$ to $4$ }
		\STATE Set $p_i^{(n)} = (p_i^{1,(n)},p_i^{2,(n)},\ldots,p_i^{N_r,(n)})$.
		\STATE Compute $\zeta_i^{(n)} =\zeta_i^{(n)} + \lambda_n(2P^{(n)} - \zeta^{(n)} - p_i^{(n)}) $.
		\ENDFOR
	\STATE Compute $\zeta^{(n+1)} =\zeta^{(n)} + \lambda_n(P^{(n)} - \zeta^{(n)}) $.
	\STATE \label{s:a2-f} $n \leftarrow n+1$.
\UNTIL {$| \mathcal{J}_{\rm ST}(\zeta^{(n)})-\mathcal{J}_{\rm ST}(\zeta^{(n-1)})| \le \varepsilon\mathcal{J}_{\rm ST}(\zeta^{(n-1)})$}.
\STATE Set $\hat{\zeta} = \zeta^{(n)}$.
\RETURN $\hat{\rho}^t=T^*\hat{\zeta}^{t}$ for every $t \in \{1,\ldots,N_r\}$.
\end{algorithmic}\label{algo:4D}
\end{algorithm}

\section{Maximum likelihood estimation of regularization parameters}\label{append:a2}

A rigorous way of addressing the regularization parameter choice would be to consider that the sum of the regularization functions $g$ and $h$ corresponds to the minus-log-likelihood of a prior distribution 
$f(\cdot;\Thetab)$ where
$$\Thetab=\bigpth{\mub_{a,j_{\rm max}},\vect{\alpha}_{a,j_{\rm max}},\vect{\beta}_{a,j_{\rm max}},\bigpth{\mub_{o,j},\vect{\alpha}_{o,j}, \vect{\beta}_{o,j}}_{o\in \mathbb{O},1 \leq j \leq j_{\rm max}},\kappa,p },$$
and to maximize the \emph{integrated} likelihood of the data. This would however entail two main difficulties. On the one hand, this would require to integrate out the sought image decomposition $\zeta$ and to iterate between image reconstruction and hyper-parameter estimation. Methods allowing us to perform this task are computationally 
intensive~\cite{Dempster77}. On the second hand, the partition function of the distribution $f(\cdot;\Thetab)$ does not take a closed form and we would thus need to resort to numerical
methods \cite{Vieth_95,Risser09b,Risser11} to 
compute it. To alleviate the computational burden, akin to~\cite{Jalobeanu02} we shall proceed differently by assuming that a reference full FOV image $\widetilde{\rho}$ is available, and so is its wavelet decomposition $\widetilde{\zeta}=T\widetilde{\rho}$. 
In practice, our reference image $\widetilde{\rho}$ is obtained using 1D-SENSE reconstruction at the same $R$ value. We then apply an approximate ML procedure which consists of estimating separately the spatial and temporal parameters. Although this approach is not optimal from a theoretical standpoint, it is quite simple and it was observed to provide satisfactory results in practice. Alternative solutions based on Monte Carlo 
methods \cite{Chaari_TSP_2010} or Stein's principle \cite{chaux_08} can also be thought of, at the expense of an additional computational complexity.

\subsection{Spatial regularization parameters}

For the spatial hyper-parameter estimation task, we will assume that the real and imaginary parts of the wavelet coefficients\footnote{A similar approach is adopted for the approximation coefficients.} are modelled by the following \emph{Generalized Gauss-Laplace}~(GGL) distribution:
\begin{equation}
\forall \xi \in \mathbb{R}, \quad f(\xi;\mu,\alpha,\beta)= \sqrt{\frac{\beta}{2\pi}}
\dfrac{e^{-(\alpha|\xi-\mu|+\frac{\beta}{2} (\xi-\mu)^2+
\frac{\alpha^2}{2\beta})}}{\mathrm{erfc}(\frac{\alpha}{\sqrt{2\beta}})}.
\end{equation} 
For each resolution level $j$ and orientation $o$, $\widehat{\mu}_{o,j}^{\rm Re}$, $\widehat{\alpha}_{o,j}^{\rm Re}$ and $\widehat{\beta}_{o,j}^{\rm Re}$ are estimated from $\widetilde{\zetab}_{o,j}$ as follows (we proceed similarly to estimate $\widehat{\mu}_{o,j}^{\rm Im}$, $\widehat{\alpha}_{o,j}^{\rm Im}$ and $\widehat{\beta}_{o,j}^{\rm Im}$ by replacing ${\rm Re}(\cdot)$ by ${\rm Im}(\cdot)$):

\begin{align}
(\widehat{\mu}_{o,j}^{\rm Re},\widehat{\alpha}_{o,j}^{\rm Re},\widehat{\beta}_{o,j}^{\rm Re})&= \argmax_{(\mu,\alpha,\beta)\in \RR \times \RR_+\times
\RR_+^*} f({\rm Re}(\widetilde{\zetab}_{o,j});\mu,\alpha,\beta) \nonumber \\
&= \argmax_{(\mu,\alpha,\beta)\in \RR \times \RR_+\times
\RR_+^*} \sum_{k=1}^{K_j} \log f({\rm Re}(\widetilde{\zeta}_{o,j,k});\mu,\alpha,\beta) \nonumber\\
 & =\argmin_{(\mu,\alpha,\beta)\in \RR \times \RR_+\times
 \RR_+^*}
\Bigl\{
 \alpha \sum_{k=1}^{K_j} |{\rm Re}(\widetilde{\zeta}_{o,j,k} -\mu)| \!\!+\!\! \frac{\beta}{2}\sum_{k=1}^{K_j} |{\rm Re}(\widetilde{\zeta}_{o,j,k}-\mu)|^2 \nonumber \\
 &+\frac{K_j\alpha^2}{2\beta} - \frac{K_j}{2} \log\beta + K_j \log \bigpth{ \mathrm{erfc}(\frac{\alpha}{\sqrt{2\beta}})  }
\Bigr \}.
\end{align} 
This three-dimensional minimization problem does not admit a closed form solution. Hence, we can compute the ML estimated parameters using the zero-order Powell 
optimization method~\cite{Bertsekas02}.

\subsection{Temporal regularization parameter}

For the temporal hyper-parameter estimation task, we will assume that, at a given voxel, the temporal noise is distributed according to the following generalized Gaussian (GG) distribution:
\begin{equation}
 \forall \epsilon\in \RR, \quad f(\epsilon;\kappa,p)= \frac{p\kappa^{1/p}e^{-\kappa |\epsilon|^p}}{2\Gamma(1/p)}.
\end{equation}
Akin to the spatial hyper-parameter estimation, reference images $(\widetilde{\rho}^t)_{1\leq t \leq N_r}$ are made available based on a 1D-SENSE reconstruction, where $\forall t\in \{1,\ldots,N_r\}$, $\widetilde{\rho}^t = T^*\widetilde{\zeta}^t$. We consider that at spatial position $\vect{r}$, the temporal noise vector $\vect{\epsilon}_{\vect{r}}=[\widetilde{\rho}^2(\vect{r})-\widetilde{\rho}^1(\vect{r}), \widetilde{\rho}^3(\vect{r})-\widetilde{\rho}^2(\vect{r}),\ldots,\widetilde{\rho}^{N_r}(\vect{r})-\widetilde{\rho}^{N_r-1}(\vect{r})]^{\trans}$ is a realization of a full independent GG prior distribution and we adjust the temporal hyper-parameter vector $(\kappa,p)$ directly from it.  It should be noted here that the considered model for the temporal noise accounts for correlations between successive observations usually  considered in the fMRI literature. It also presents more flexibility than the Gaussian model, which corresponds to the particular case when $p=2$. Estimates $\widehat{\kappa}$ and $\
widehat{p}$ of the parameters are then obtained as follows:
\begin{align}
(\widehat{\kappa},\widehat{p}) &= \argmax_{(\kappa,p) \in \RR_+ \times [1,+\infty[} f(\vect{\epsilon}_{\vect{r}};\kappa,p) \nonumber \\
&= \argmax_{(\kappa,p) \in \RR_+ \times [1,+\infty[} \log f(\vect{\epsilon}_{\vect{r}};\kappa,p) \nonumber \\
&= \argmin_{(\kappa,p) \in \RR_+ \times [1,+\infty[} \kappa   \sum_{t=1}^{N_r-1} |\widetilde{\rho}^{t+1}(\vect{r})-\widetilde{\rho}^t(\vect{r})|^p-(N_r-1)\log\Big(\frac{p\kappa^{1/p}}{2\Gamma(1/p)}\Big). 
\end{align}
Note that in the above minimization, for a given value of $p$, the optimal value of $\kappa$ admits the following closed form:
\begin{equation}
 \widehat{\kappa} = \frac{N_r-1}{p\sum_{t=1}^{N_r-1} |\widetilde{\rho}^{t+1}(\vect{r})-\widetilde{\rho}^t(\vect{r})|^p}.
\end{equation}
A zero-order Powell optimization method can then be used to solve the resulting one-variable minimization problem. To reduce the computational complexity of this estimation, it is only performed on the brain mask, and the temporal regularization parameter $\kappa$ is set to zero for voxels belonging to the image background.

{\ifthenelse{\boolean{publ}}{\footnotesize}{\small}
 \bibliographystyle{bmc_article}  
  \bibliography{revuedef,NeuroImage_v5} }     

\ifthenelse{\boolean{publ}}{\end{multicols}}{}

\end{bmcformat}

\end{document}

%% file: Definitions.tex
\newcommand{\CC}{{\ensuremath{\mathbb C}}}

\def\trans{\mbox{\tiny\textsf{T}}}
\def\hermit{\mbox{\tiny\textsf{H}}}
\newcommand{\eqdef}{\mbox{$\buildrel\triangle\over =$}}

\newcommand{\scal}[2]{\left\langle{{#1}|{#2}}\right\rangle}

\newcommand{\RR}{\ensuremath{\mathbb R}}

\newcommand{\prox}{\ensuremath{\mathrm{prox}}}

\newcommand{\vect}[1]{{\boldsymbol #1}}    

\def\qed{\ifmmode\hbox{\hfill\sqb}\else{\ifhmode\unskip\fi%
\nobreak\hfil
\penalty50\hskip1em\null\nobreak\hfil$\blacksquare$
\parfillskip=0pt\finalhyphendemerits=0\endgraf}\fi}

\def\argmax{\mathop{\mathrm{arg\,max}}} 
\def\argmin{\mathop{\mathrm{arg\,min}}} 
\RequirePackage{amsmath}
\RequirePackage{xspace}
\RequirePackage{bbm}

\def\XS{\xspace}
\DeclareMathAlphabet{\mathb}{OML}{cmm}{b}{it}
\def\sbm#1{\ensuremath{\mathb{#1}}}                
\def\sbmm#1{\ensuremath{\boldsymbol{#1}}}          

\def\pth#1{\left(#1\right)}

\def\bigpth#1{\bigl(#1\bigr)}

\def\Sb{{\sbm{S}}\XS}

\def\MAP{^{\kern1pt{\rm MAP}\kern-1pt}}

\usepackage{color,soul}

      \def\Thetab    {{\sbmm{\Theta}}\XS}

\def\mub         {{\sbmm{\mu}}\XS}

\def\zetab       {{\sbmm{\zeta}}\XS}

\def\PM{\kern0pt^{\textrm{{\scriptsize PM}}}\kern0pt}
\def\MMAP{\kern1pt^{\textrm{{\tiny MMAP}}}\kern-1pt} 

\def\rem#1{}                    
\def\mSENSE{\texttt{mSENSE}\XS}

\def\LcRc{\texttt{Lc-Rc}\XS}

\def\ACAS{\texttt{aC-aS}\XS}

%% file: JASP_CHAARI_revision2.bbl

\begin{thebibliography}{10}
\providecommand{\url}[1]{[#1]}
\providecommand{\urlprefix}{}

\bibitem{Kochunov05b}
Kochunov P, Rivi\`ere D, Lancaster JL, Mangin JF, Cointepas Y, Glahn D, Fox P,
  Rogers J: \textbf{Development of high-resolution {MRI} imaging and image
  processing for live and post-mortem primates}. In \emph{{{H}uman {B}rain
  {M}apping}}, \emph{Volume 26 (1)}, Toronto, Canada 2005.

\bibitem{Rabrait07}
Rabrait C, Ciuciu P, Rib{\`e}s A, Poupon C, Leroux P, Lebon V, Dehaene-Lambertz
  G, Bihan DL, Lethimonnier F: \textbf{High temporal resolution functional
  {MRI} using parallel echo volume imaging}. \emph{{M}agnetic {R}esonance
  {I}maging} 2008, \textbf{27}(4):744--753.

\bibitem{Sodickson_D_97}
Sodickson DK, Manning WJ: \textbf{Simultaneous acquisition of spatial harmonics
  ({SMASH}): fast imaging with radiofrequency coil arrays}. \emph{{M}agnetic
  {R}esonance in {M}edicine} 1997, \textbf{38}(4):591--603.

\bibitem{pruessmann_99}
Pruessmann KP, Weiger M, Scheidegger MB, Boesiger P: \textbf{{SENSE}:
  sensitivity encoding for fast {MRI}}. \emph{Magnetic Resonance in Medicine}
  1999, \textbf{42}(5):952--962.

\bibitem{griswold_02}
Griswold MA, Jakob PM, Heidemann RM, Nittka M, Jellus V, Wang J, Kiefer B,
  Haase A: \textbf{Generalized autocalibrating partially parallel acquisitions
  {GRAPPA}}. \emph{{M}agnetic {R}esonance in {M}edicine} 2002,
  \textbf{47}(6):1202--1210.

\bibitem{Candes_06}
Cand{\`e}s E, Romberg J, Tao T: \textbf{Robust uncertainty principles: exact
  signal reconstruction from highly incomplete frequency information}.
  \emph{{{IEEE} {T}ransactions on {I}nformation {T}heory}} 2006,
  \textbf{52}(2):489--509.

\bibitem{Lustig07}
Lustig M, Donoho D, Pauly JM: \textbf{Sparse {MRI}: The Application of
  Compressed Sensing for Rapid {MR} Imaging}. \emph{Magnetic Resonance in
  Medicine} 2007, \textbf{58}:1182--1195.

\bibitem{Liang09}
Liang D, Liu B, Wang J, Ying L: \textbf{{Accelerating SENSE using compressed
  sensing}}. \emph{{{M}agnetic {R}esonance in {M}edicine}} 2009,
  \textbf{62}(6):1574--84.

\bibitem{Boyer12}
Boyer C, Ciuciu P, Weiss P, M\'eriaux S: \textbf{{HYR$^2$PICS}: Hybrid
  Regularized Reconstruction for combined Parallel Imaging and Compressive
  Sensing in {MRI}}. In \emph{9th {{I}nternational {S}ymposium on {B}iomedical
  {I}maging ({ISBI})}}, Barcelona, Spain 2012:66--69.

\bibitem{Madore99}
Madore B, Glover GH, Pelc NJ: \textbf{{Unaliasing by {F}ourier-encoding the
  overlaps using the temporal dimension (UNFOLD), applied to cardiac imaging
  and f{MRI}}}. \emph{{{M}agnetic {R}esonance in {M}edicine}} 1999,
  \textbf{42}(5):813--28.

\bibitem{Tsao03}
Tsao J, Boesiger P, Pruessmann KP: \textbf{{k-t BLAST and k-t SENSE: dynamic
  MRI with high frame rate exploiting spatiotemporal correlations}}.
  \emph{{{M}agnetic {R}esonance in {M}edicine}} 2003, \textbf{50}(5):1031--42.

\bibitem{Tsao05}
Tsao J, Kozerke S, Boesiger P, Pruessmann KP: \textbf{{Optimizing
  spatiotemporal sampling for k-t BLAST and k-t SENSE: application to
  high-resolution real-time cardiac steady-state free precession}}.
  \emph{Magnetic resonance in medicine} 2005, \textbf{53}(6):1372--82.

\bibitem{Huang05}
Huang F, Akao J, Vijayakumar S, Duensing GR, Limkeman M: \textbf{{k-t GRAPPA: a
  k-space implementation for dynamic MRI with high reduction factor}}.
  \emph{{{M}agnetic {R}esonance in {M}edicine}} 2005, \textbf{54}(5):1172--84.

\bibitem{Jung07}
Jung H, Ye JC, Kim EY: \textbf{{Improved k-t BLAST and k-t SENSE using
  FOCUSS}}. \emph{Physics in medicine and biology} 2007,
  \textbf{52}(11):3201--26.

\bibitem{Jung09}
Jung H, Sung K, Nayak KS, Kim EY, Ye JC: \textbf{{k-t FOCUSS: a general
  compressed sensing framework for high resolution dynamic MRI}}.
  \emph{{{M}agnetic {R}esonance in {M}edicine}} 2009, \textbf{61}:103--16.

\bibitem{Damoiseaux06}
Damoiseaux JS, Rombouts SA, Barkhof F, Scheltens P, Stam CJ, Smith SM, Beckmann
  CF: \textbf{Consistent resting-state networks across healthy subjects}.
  \emph{{{P}roceedings of the {N}ational {A}cademy of {Sciences} of the
  {U}nited {S}tates of {A}merica}} 2006, \textbf{103}(37):13848--13853.

\bibitem{Dale99}
Dale AM: \textbf{Optimal experimental design for event-related f{MRI}}.
  \emph{{{H}uman {B}rain {M}apping}} 1999, \textbf{8}:109--114.

\bibitem{Varoquaux10}
Varoquaux G, Sadaghiani S, Pinel P, Kleinschmidt A, Poline JB, Thirion B:
  \textbf{A group model for stable multi-subject {ICA} on {fMRI} datasets}.
  \emph{{{N}euroimage}} 2010, \textbf{51}:288--299.

\bibitem{Ciuciu12}
Ciuciu P, Varoquaux G, Abry P, Sadaghiani S, Kleinschmidt A: \textbf{Scale-Free
  and Multifractal Time Dynamics of f{MRI} Signals during Rest and Task}.
  \emph{Frontiers in physiology} 2012, \textbf{3}(Article 186):1--18.

\bibitem{Birn02}
Birn R, Cox R, Bandettini PA: \textbf{Detection versus estimation in
  event-related f{MRI}: choosing the optimal stimulus timing}.
  \emph{{{N}euroimage}} 2002, \textbf{15}:252--264.

\bibitem{Logothetis08}
Logothetis NK: \textbf{What we can do and what we cannot do with f{MRI}}.
  \emph{Nature} 2008, \textbf{453}(7197):869--878.

\bibitem{deZwart02}
de~Zwart J, Gelderen PV, Kellman P, Duyn JH: \textbf{{Application of
  sensitivity-encoded echo-planar imaging for blood oxygen level-dependent
  functional brain imaging}}. \emph{{{M}agnetic {R}esonance in {M}edicine}}
  2002, \textbf{48}(6):1011--20.

\bibitem{Preibisch03}
Preibisch C: \textbf{{Functional MRI using sensitivity-encoded echo planar
  imaging (SENSE-EPI)}}. \emph{Neuroimage} 2003, \textbf{19}(2):412--421.

\bibitem{deZwart06}
de~Zwart J, Gelderen PV, Golay X, Ikonomidou VN, Duyn JH: \textbf{{Accelerated
  parallel imaging for functional imaging of the human brain}}. \emph{NMR
  Biomed} 2006, \textbf{19}(3):342--51.

\bibitem{Utting10}
Utting JF, Kozerke S, Schnitker R, Niendorf T: \textbf{{Comparison of k-t
  SENSE/k-t BLAST with conventional SENSE applied to BOLD fMRI}}.
  \emph{{{J}ournal of {M}agnetic {R}esonance {I}maging}} 2010,
  \textbf{32}:235--41.

\bibitem{Liang_02}
Liang ZP, Bammer R, Ji J, Pelc NJ, Glover GH: \textbf{Making better {SENSE}:
  wavelet denoising, {T}ikhonov regularization, and total least squares}. In
  \emph{{I}nternational {S}ociety for {M}agnetic {R}esonance in {M}edicine},
  Hawa\"{\i}, USA 2002:2388.

\bibitem{Ying_L_04}
Ying L, Xu D, Liang ZP: \textbf{{O}n {T}ikhonov Regularization for image
  reconstruction in parallel {MRI}}. In \emph{{IEEE E}ngineering in {M}edicine
  and {B}iology {S}ociety}, San Francisco, USA 2004:1056--1059.

\bibitem{Liu_08_1}
Zou YM, Ying L, Liu B: \textbf{Sparse{SENSE}: application of compressed sensing
  in parallel {MRI}}. In \emph{{IEEE I}nternational Conference on Technology
  and Applications in Biomedicine}, Shenzhen, China 2008:127--130.

\bibitem{chaari_08}
Chaari L, Pesquet JC, Benazza-Benyahia A, Ciuciu P: \textbf{{A}utocalibrated
  Parallel {MRI} Reconstruction in the Wavelet Domain}. In \emph{{IEEE
  I}nternational {S}ymposium on {B}iomedical {I}maging (ISBI)}, Paris, France
  2008:756--759.

\bibitem{Liu_08_2}
Liu B, Abdelsalam E, Sheng J, , Ying L: \textbf{Improved spiral {SENSE}
  reconstruction using a multiscale wavelet model}. In \emph{{IEEE Int. Symp.
  on Biomed. Imag.}}, Paris, France 2008:1505--1508.

\bibitem{Chaari_MEDIA_2011}
Chaari L, Pesquet JC, Benazza-Benyahia A, Ciuciu P: \textbf{A wavelet-based
  regularized reconstruction algorithm for {SENSE} parallel {MRI} with
  applications to neuroimaging}. \emph{Medical Image Analysis} 2011,
  \textbf{15}(2):185--201.

\bibitem{Chaari10e}
Chaari L, M\'eriaux S, Pesquet JC, Ciuciu P: \textbf{Impact of the parallel
  imaging reconstruction algorithm on brain activity detection in f{MRI}}. In
  \emph{International Symposium on Applied Sciences in Biomedical and
  Communication Technologies (ISABEL)}, Rome, Italy 2010:1--5.

\bibitem{Jakob_06}
Jakob P, Griswold M, Breuer F, Blaimer M, Seiberlich N: \textbf{A {3D GRAPPA}
  algorithm for volumetric parallel imaging}. In \emph{{Scientific Meeting
  International Society for Magnetic Resonance in Medicine}}, Seattle, USA
  2006:286.

\bibitem{Aguirre97}
Aguirre GK, Zarahn E, D'Esposito M: \textbf{Empirical analysis of {BOLD} f{MRI}
  statistics. {II}. {S}patially Smoothed Data Collected under Null-Hypothesis
  and Experimental Conditions}. \emph{Neuroimage} 1997, \textbf{5}(3):199--212.

\bibitem{Zarahn97}
Zarahn E, Aguirre GK, D'Esposito M: \textbf{Empirical analysis of {BOLD} f{MRI}
  statistics. {I}. {S}patially unsmoothed data collected under null-hypothesis
  conditions}. \emph{Neuroimage} 1997, \textbf{5}(3):179--197.

\bibitem{Purdon98}
Purdon PL, Weisskoff RM: \textbf{Effect of temporal autocorrelation due to
  physiological noise and stimulus paradigm on voxel-level false-positive rates
  in f{MRI}.} \emph{{Human Brain Mapping}} 1998, \textbf{6}(4):239--249.

\bibitem{Woolrich01}
Woolrich M, Ripley B, Brady M, Smith S: \textbf{Temporal autocorrelation in
  univariate linear modelling of f{MRI} data}. \emph{Neuroimage} 2001,
  \textbf{14}(6):1370--1386.

\bibitem{Worsley02}
Worsley KJ, Liao CH, Aston J, Petre V, Duncan GH, Morales F, Evans AC:
  \textbf{A general statistical analysis for f{MRI} data}. \emph{Neuroimage}
  2002, \textbf{15}:1--15.

\bibitem{Penny03}
Penny WD, Kiebel S, Friston KJ: \textbf{Variational {B}ayesian inference for
  f{MRI} time series}. \emph{Neuroimage} 2003, \textbf{19}(3):727--741.

\bibitem{Chaari_TMI_2012}
Chaari L, Vincent T, Forbes F, Dojat M, Ciuciu P: \textbf{Fast joint
  detection-estimation of evoked brain activity in event-related f{MRI} using a
  variational approach}. \emph{IEEE Transactions on Medical Imaging} 2013,
  \textbf{32}(5):821--837.

\bibitem{Combettes_PL_08}
Combettes PL, Pesquet JC: \textbf{A proximal decomposition method for solving
  convex variational inverse problems}. \emph{Inverse Problems} 2008,
  \textbf{24}(6):27.

\bibitem{Unser_TMI_2011}
Guerquin-Kern M, Haberlin M, Pruessmann KP, Unser M: \textbf{A Fast
  Wavelet-Based Reconstruction Method for Magnetic Resonance Imaging}.
  \emph{IEEE Transactions on Medical Imaging} 2011, \textbf{30}(9):1649--1660.

\bibitem{Sodickson_D_00}
Sodickson DK: \textbf{Tailored SMASH Image Reconstructions for Robust In Vivo
  Parallel {MR} Imaging}. \emph{{M}agnetic {R}esonance in {M}edicine} 2000,
  \textbf{44}(2):243–251.

\bibitem{keeling_03}
Keeling SL: \textbf{Total variation based convex filters for medical imaging}.
  \emph{{A}pplied {M}athematics and {C}omputation} 2003, \textbf{139}:101--119.

\bibitem{Liu_08}
Liu B, King K, Steckner M, Xie J, Sheng J, Ying L: \textbf{Regularized
  sensitivity encoding ({SENSE}) reconstruction using {B}regman iterations}.
  \emph{Magnetic Resonance in Medicine} 2008, \textbf{61}:145 -- 152.

\bibitem{Sumbul_2009}
S{\"u}mb{\"u}l U, Santos JM, Pauly JM: \textbf{Improved Time Series
  Reconstruction for Dynamic Magnetic Resonance Imaging}. \emph{IEEE
  Transactions on Medical Imaging} 2009, \textbf{28}(7):1093--1104.

\bibitem{Pinel_07}
Pinel P, Thirion B, M{\'e}riaux S, Jobert A, Serres J, {Le Bihan} D, Poline JB,
  Dehaene S: \textbf{Fast reproducible identification and large-scale
  databasing of individual functional cognitive networks}. \emph{BMC
  Neuroscience} 2007, \textbf{8}:91.

\bibitem{daubechies_92}
Daubechies I: \emph{{T}en {L}ectures on {W}avelets}. Philadelphia: {S}ociety
  for {I}ndustrial and {A}pplied {M}athematics 1992.

\bibitem{Dehaene99}
Dehaene S: \textbf{Cerebral bases of number processing and calculation}. In
  \emph{The New Cognitive Neurosciences}. Edited by Gazzaniga M, Cambridge,:
  MIT Press 1999:987--998.

\bibitem{Nichols03}
Nichols TE, Hayasaka S: \textbf{{Controlling the Familywise Error Rate in
  Functional Neuroimaging: A Comparative Review}}. \emph{Statistical Methods in
  Medical Research} 2003, \textbf{12}(5):419--446.

\bibitem{Brett_04}
Brett M, Penny W, Kiebel S: \textbf{Introduction to Random Field Theory}. In
  \emph{Human Brain Function}, 2nd edition. Edited by Frackowiak RSJ, Friston
  KJ, Fritch CD, Dolan RJ, Price CJ, Penny WD, Academic Press 2004:867--880.

\bibitem{Chaari_TSP_2010}
Chaari L, Pesquet JC, Tourneret JY, Ciuciu P, Benazza-Benyahia A: \textbf{A
  Hierarchical {B}ayesian Model For Frame Representation}. \emph{IEEE
  Transactions on Signal Processing} 2010, \textbf{58}(11):5560--5571.

\bibitem{Roche_2011}
Roche A: \textbf{A Four-Dimensional Registration Algorithm With Application to
  Joint Correction of Motion and Slice Timing in f{MRI}}. \emph{IEEE
  Transactions on Medical Imaging} 2011, \textbf{30}(8):1546--1554.

\bibitem{Makni08}
Makni S, Idier J, Vincent T, Thirion B, Dehaene-Lambertz G, Ciuciu P: \textbf{A
  fully {B}ayesian approach to the parcel-based detection-estimation of brain
  activity in {fMRI}}. \emph{{{N}euroimage}} 2008, \textbf{41}(3):941--969.

\bibitem{Vincent10}
Vincent T, Risser L, Ciuciu P: \textbf{Spatially adaptive mixture modeling for
  analysis of {within-subject fMRI} time series}. \emph{IEEE Transactions on
  Medical Imaging} 2010, \textbf{29}(4):1059--1074.

\bibitem{Badillo13}
Badillo S, Vincent T, Ciuciu P: \textbf{Group-level impacts of within- and
  between-subject hemodynamic variability in {fMRI}}. \emph{NeuroImage} 2013,
  \textbf{82}:433--448.

\bibitem{VanDeVille_2007}
Van De~Ville D, Seghier M, Lazeyras F, Blu T, Unser M: \textbf{{WSPM}:
  Wavelet-based statistical parametric mapping}. \emph{Neuroimage} 2007,
  \textbf{37}(4):1205--1217.

\bibitem{Moreau_65}
Moreau JJ: \textbf{Proximit{\'e} et dualit{\'e} dans un espace hilbertien}.
  \emph{{B}ulletin de la {S}oci{\'e}t{\'e} {M}ath{\'e}matique de {F}rance}
  1965, \textbf{93}:273--299.

\bibitem{Chaux_C_07}
Chaux C, Combettes P, Pesquet JC, Wajs VR: \textbf{A variational formulation
  for frame-based inverse problems}. \emph{{I}nverse {P}roblems} 2007,
  \textbf{23}(4):1495--1518.

\bibitem{Combettes_PL_2005_mms_Signal_rbpfbs}
Combettes PL, Wajs VR: \textbf{Signal Recovery by proximal forward-backward
  splitting}. \emph{{M}ultiscale {M}odeling and {S}imulation} 2005,
  \textbf{4}:1168--1200.

\bibitem{Combettes_09bis}
Combettes PL, Pesquet JC: \textbf{Proximal splitting methods in signal
  processing}. In \emph{Fixed-Point Algorithms for Inverse Problems in Science
  and Engineering}. Edited by Bauschke HH, Burachik R, Combettes PL, Elser V,
  Luke DR, Wolkowicz H, New York: {Springer Verlag} 2010:185--212.

\bibitem{Combettes_PL_2007_istsp_Douglas_rsatncvsr}
Combettes PL, Pesquet JC: \textbf{A {D}ouglas-{R}achford Splitting Approach to
  Nonsmooth Convex Variational Signal Recovery}. \emph{IEEE Journal of Selected
  Topics in Signal Processing} 2007, \textbf{1}(4):564--574.

\bibitem{Dempster77}
Dempster AP, Laird AP, Rubin DB: \textbf{Maximum likelihood from incomplete
  data via the {EM} algorithm (with discussion)}. \emph{Journal of the Royal
  Statistical Society, Series B} 1977, \textbf{39}:1--38.

\bibitem{Vieth_95}
Vieth M, Kolinski A, Skolnick J: \textbf{A simple technique to estimate
  partition functions and equilibrium constants from {M}onte {C}arlo
  simulations}. \emph{Journal of Chemical Physics} 1995,
  \textbf{102}:6189--6193.

\bibitem{Risser09b}
Risser L, Vincent T, Ciuciu P, Idier J: \textbf{Robust extrapolation scheme for
  fast estimation of {3D I}sing field partition functions. Application to
  within-subject f{MRI} data analysis.} In \emph{12th{P}roc. Medical Image
  Computing and Computer Assisted Intervention}, London, UK: Springer Verlag
  Berlin Heidelberg 2009:975--983.

\bibitem{Risser11}
Risser L, Vincent T, Forbes F, Idier J, Ciuciu P: \textbf{{Min-max}
  extrapolation scheme for fast estimation of {3D Potts} field partition
  functions. Application to the joint detection-estimation of brain activity in
  {fMRI}.} \emph{{{J}ournal of {S}ignal {P}rocessing {S}ystems}} 2011,
  \textbf{65}(3):325--338.

\bibitem{Jalobeanu02}
Jalobeanu A, Blanc-F\'eraud L, Zerubia J: \textbf{Hyperparameter estimation for
  satellite image restoration using a {MCMC} maximum likelihood method}.
  \emph{Pattern Recognition} 2002, \textbf{35}(2).

\bibitem{chaux_08}
Chaux C, Duval L, Benazza-Benyahia A, Pesquet JC: \textbf{A nonlinear {S}tein
  based estimator for multichannel image denoising}. \emph{{IEEE T}ransactions
  on {S}ignal {P}rocessing} 2008, \textbf{56}(8):3855--3870.

\bibitem{Bertsekas02}
Bertsekas DP: \emph{Nonlinear programming, Second Edition}. Belmont, {USA}:
  Athena Scientific 1995.

\end{thebibliography}

\newcommand{\BMCxmlcomment}[1]{}

\BMCxmlcomment{

<refgrp>

<bibl id="B1">
  <title><p>Development of high-resolution {MRI} imaging and image processing
  for live and post-mortem primates</p></title>
  <aug>
    <au><snm>Kochunov</snm><fnm>P.</fnm></au>
    <au><snm>Rivi\`ere</snm><fnm>D.</fnm></au>
    <au><snm>Lancaster</snm><fnm>J. L.</fnm></au>
    <au><snm>Mangin</snm><fnm>J. F.</fnm></au>
    <au><snm>Cointepas</snm><fnm>Y.</fnm></au>
    <au><snm>Glahn</snm><fnm>D.</fnm></au>
    <au><snm>Fox</snm><fnm>P.</fnm></au>
    <au><snm>Rogers</snm><fnm>J.</fnm></au>
  </aug>
  <source>{{H}uman {B}rain {M}apping}</source>
  <publisher>Toronto, Canada</publisher>
  <pubdate>2005</pubdate>
  <volume>26 (1)</volume>
</bibl>

<bibl id="B2">
  <title><p>High temporal resolution functional {MRI} using parallel echo
  volume imaging</p></title>
  <aug>
    <au><snm>Rabrait</snm><fnm>C.</fnm></au>
    <au><snm>Ciuciu</snm><fnm>P.</fnm></au>
    <au><snm>Rib{\`e}s</snm><fnm>A.</fnm></au>
    <au><snm>Poupon</snm><fnm>C.</fnm></au>
    <au><snm>Leroux</snm><fnm>P.</fnm></au>
    <au><snm>Lebon</snm><fnm>V.</fnm></au>
    <au><snm>Dehaene Lambertz</snm><fnm>G.</fnm></au>
    <au><snm>Bihan</snm><fnm>DL</fnm></au>
    <au><snm>Lethimonnier</snm><fnm>F.</fnm></au>
  </aug>
  <source>{M}agnetic {R}esonance {I}maging</source>
  <pubdate>2008</pubdate>
  <volume>27</volume>
  <issue>4</issue>
  <fpage>744</fpage>
  <lpage>-753</lpage>
</bibl>

<bibl id="B3">
  <title><p>Simultaneous acquisition of spatial harmonics ({SMASH}): fast
  imaging with radiofrequency coil arrays</p></title>
  <aug>
    <au><snm>Sodickson</snm><fnm>D. K.</fnm></au>
    <au><snm>Manning</snm><fnm>W. J.</fnm></au>
  </aug>
  <source>{M}agnetic {R}esonance in {M}edicine</source>
  <pubdate>1997</pubdate>
  <volume>38</volume>
  <issue>4</issue>
  <fpage>591</fpage>
  <lpage>-603</lpage>
</bibl>

<bibl id="B4">
  <title><p>{SENSE}: sensitivity encoding for fast {MRI}</p></title>
  <aug>
    <au><snm>Pruessmann</snm><fnm>K. P.</fnm></au>
    <au><snm>Weiger</snm><fnm>M.</fnm></au>
    <au><snm>Scheidegger</snm><fnm>M. B.</fnm></au>
    <au><snm>Boesiger</snm><fnm>P.</fnm></au>
  </aug>
  <source>Magnetic Resonance in Medicine</source>
  <pubdate>1999</pubdate>
  <volume>42</volume>
  <issue>5</issue>
  <fpage>952</fpage>
  <lpage>-962</lpage>
</bibl>

<bibl id="B5">
  <title><p>Generalized autocalibrating partially parallel acquisitions
  {GRAPPA}</p></title>
  <aug>
    <au><snm>Griswold</snm><fnm>M. A.</fnm></au>
    <au><snm>Jakob</snm><fnm>P. M.</fnm></au>
    <au><snm>Heidemann</snm><fnm>R. M.</fnm></au>
    <au><snm>Nittka</snm><fnm>M.</fnm></au>
    <au><snm>Jellus</snm><fnm>V.</fnm></au>
    <au><snm>Wang</snm><fnm>J.</fnm></au>
    <au><snm>Kiefer</snm><fnm>B.</fnm></au>
    <au><snm>Haase</snm><fnm>A.</fnm></au>
  </aug>
  <source>{M}agnetic {R}esonance in {M}edicine</source>
  <pubdate>2002</pubdate>
  <volume>47</volume>
  <issue>6</issue>
  <fpage>1202</fpage>
  <lpage>-1210</lpage>
</bibl>

<bibl id="B6">
  <title><p>Robust uncertainty principles: exact signal reconstruction from
  highly incomplete frequency information</p></title>
  <aug>
    <au><snm>Cand{\`e}s</snm><fnm>E.</fnm></au>
    <au><snm>Romberg</snm><fnm>J.</fnm></au>
    <au><snm>Tao</snm><fnm>T.</fnm></au>
  </aug>
  <source>{{IEEE} {T}ransactions on {I}nformation {T}heory}</source>
  <pubdate>2006</pubdate>
  <volume>52</volume>
  <issue>2</issue>
  <fpage>489</fpage>
  <lpage>-509</lpage>
</bibl>

<bibl id="B7">
  <title><p>Sparse {MRI}: The Application of Compressed Sensing for Rapid {MR}
  Imaging</p></title>
  <aug>
    <au><snm>Lustig</snm><fnm>M.</fnm></au>
    <au><snm>Donoho</snm><fnm>D.</fnm></au>
    <au><snm>Pauly</snm><fnm>J. M.</fnm></au>
  </aug>
  <source>Magnetic Resonance in Medicine</source>
  <pubdate>2007</pubdate>
  <volume>58</volume>
  <fpage>1182</fpage>
  <lpage>1195</lpage>
</bibl>

<bibl id="B8">
  <title><p>{Accelerating SENSE using compressed sensing}</p></title>
  <aug>
    <au><snm>Liang</snm><fnm>D.</fnm></au>
    <au><snm>Liu</snm><fnm>B.</fnm></au>
    <au><snm>Wang</snm><fnm>J.</fnm></au>
    <au><snm>Ying</snm><fnm>L.</fnm></au>
  </aug>
  <source>{{M}agnetic {R}esonance in {M}edicine}</source>
  <pubdate>2009</pubdate>
  <volume>62</volume>
  <issue>6</issue>
  <fpage>1574</fpage>
  <lpage>-84</lpage>
</bibl>

<bibl id="B9">
  <title><p>{HYR$^2$PICS}: Hybrid Regularized Reconstruction for combined
  Parallel Imaging and Compressive Sensing in {MRI}</p></title>
  <aug>
    <au><snm>Boyer</snm><fnm>C.</fnm></au>
    <au><snm>Ciuciu</snm><fnm>P.</fnm></au>
    <au><snm>Weiss</snm><fnm>P.</fnm></au>
    <au><snm>M\'eriaux</snm><fnm>S.</fnm></au>
  </aug>
  <source>9th {{I}nternational {S}ymposium on {B}iomedical {I}maging
  ({ISBI})}</source>
  <publisher>Barcelona, Spain</publisher>
  <pubdate>2012</pubdate>
  <fpage>66</fpage>
  <lpage>-69</lpage>
</bibl>

<bibl id="B10">
  <title><p>{Unaliasing by {F}ourier-encoding the overlaps using the temporal
  dimension (UNFOLD), applied to cardiac imaging and f{MRI}}</p></title>
  <aug>
    <au><snm>Madore</snm><fnm>B.</fnm></au>
    <au><snm>Glover</snm><fnm>G. H.</fnm></au>
    <au><snm>Pelc</snm><fnm>N. J.</fnm></au>
  </aug>
  <source>{{M}agnetic {R}esonance in {M}edicine}</source>
  <pubdate>1999</pubdate>
  <volume>42</volume>
  <issue>5</issue>
  <fpage>813</fpage>
  <lpage>-28</lpage>
</bibl>

<bibl id="B11">
  <title><p>{k-t BLAST and k-t SENSE: dynamic MRI with high frame rate
  exploiting spatiotemporal correlations}</p></title>
  <aug>
    <au><snm>Tsao</snm><fnm>J.</fnm></au>
    <au><snm>Boesiger</snm><fnm>P.</fnm></au>
    <au><snm>Pruessmann</snm><fnm>K. P.</fnm></au>
  </aug>
  <source>{{M}agnetic {R}esonance in {M}edicine}</source>
  <pubdate>2003</pubdate>
  <volume>50</volume>
  <issue>5</issue>
  <fpage>1031</fpage>
  <lpage>-42</lpage>
</bibl>

<bibl id="B12">
  <title><p>{Optimizing spatiotemporal sampling for k-t BLAST and k-t SENSE:
  application to high-resolution real-time cardiac steady-state free
  precession}</p></title>
  <aug>
    <au><snm>Tsao</snm><fnm>J.</fnm></au>
    <au><snm>Kozerke</snm><fnm>S.</fnm></au>
    <au><snm>Boesiger</snm><fnm>P.</fnm></au>
    <au><snm>Pruessmann</snm><fnm>K. P.</fnm></au>
  </aug>
  <source>Magnetic resonance in medicine</source>
  <pubdate>2005</pubdate>
  <volume>53</volume>
  <issue>6</issue>
  <fpage>1372</fpage>
  <lpage>-82</lpage>
</bibl>

<bibl id="B13">
  <title><p>{k-t GRAPPA: a k-space implementation for dynamic MRI with high
  reduction factor}</p></title>
  <aug>
    <au><snm>Huang</snm><fnm>F.</fnm></au>
    <au><snm>Akao</snm><fnm>J.</fnm></au>
    <au><snm>Vijayakumar</snm><fnm>S.</fnm></au>
    <au><snm>Duensing</snm><fnm>G. R.</fnm></au>
    <au><snm>Limkeman</snm><fnm>M.</fnm></au>
  </aug>
  <source>{{M}agnetic {R}esonance in {M}edicine}</source>
  <pubdate>2005</pubdate>
  <volume>54</volume>
  <issue>5</issue>
  <fpage>1172</fpage>
  <lpage>-84</lpage>
</bibl>

<bibl id="B14">
  <title><p>{Improved k-t BLAST and k-t SENSE using FOCUSS}</p></title>
  <aug>
    <au><snm>Jung</snm><fnm>H.</fnm></au>
    <au><snm>Ye</snm><fnm>J. C.</fnm></au>
    <au><snm>Kim</snm><fnm>E. Y.</fnm></au>
  </aug>
  <source>Physics in medicine and biology</source>
  <pubdate>2007</pubdate>
  <volume>52</volume>
  <issue>11</issue>
  <fpage>3201</fpage>
  <lpage>-26</lpage>
</bibl>

<bibl id="B15">
  <title><p>{k-t FOCUSS: a general compressed sensing framework for high
  resolution dynamic MRI}</p></title>
  <aug>
    <au><snm>Jung</snm><fnm>H.</fnm></au>
    <au><snm>Sung</snm><fnm>K.</fnm></au>
    <au><snm>Nayak</snm><fnm>K. S.</fnm></au>
    <au><snm>Kim</snm><fnm>E. Y.</fnm></au>
    <au><snm>Ye</snm><fnm>J. C.</fnm></au>
  </aug>
  <source>{{M}agnetic {R}esonance in {M}edicine}</source>
  <pubdate>2009</pubdate>
  <volume>61</volume>
  <issue>1</issue>
  <fpage>103</fpage>
  <lpage>-16</lpage>
</bibl>

<bibl id="B16">
  <title><p>Consistent resting-state networks across healthy
  subjects</p></title>
  <aug>
    <au><snm>Damoiseaux</snm><fnm>J. S.</fnm></au>
    <au><snm>Rombouts</snm><fnm>S. A.</fnm></au>
    <au><snm>Barkhof</snm><fnm>F.</fnm></au>
    <au><snm>Scheltens</snm><fnm>P.</fnm></au>
    <au><snm>Stam</snm><fnm>C. J.</fnm></au>
    <au><snm>Smith</snm><fnm>S. M.</fnm></au>
    <au><snm>Beckmann</snm><fnm>C. F.</fnm></au>
  </aug>
  <source>{{P}roceedings of the {N}ational {A}cademy of {Sciences} of the
  {U}nited {S}tates of {A}merica}</source>
  <pubdate>2006</pubdate>
  <volume>103</volume>
  <issue>37</issue>
  <fpage>13848</fpage>
  <lpage>-13853</lpage>
</bibl>

<bibl id="B17">
  <title><p>Optimal experimental design for event-related f{MRI}</p></title>
  <aug>
    <au><snm>Dale</snm><fnm>A. M.</fnm></au>
  </aug>
  <source>{{H}uman {B}rain {M}apping}</source>
  <pubdate>1999</pubdate>
  <volume>8</volume>
  <fpage>109</fpage>
  <lpage>-114</lpage>
</bibl>

<bibl id="B18">
  <title><p>A group model for stable multi-subject {ICA} on {fMRI}
  datasets</p></title>
  <aug>
    <au><snm>Varoquaux</snm><fnm>G.</fnm></au>
    <au><snm>Sadaghiani</snm><fnm>S.</fnm></au>
    <au><snm>Pinel</snm><fnm>P.</fnm></au>
    <au><snm>Kleinschmidt</snm><fnm>A.</fnm></au>
    <au><snm>Poline</snm><fnm>J. B.</fnm></au>
    <au><snm>Thirion</snm><fnm>B.</fnm></au>
  </aug>
  <source>{{N}euroimage}</source>
  <pubdate>2010</pubdate>
  <volume>51</volume>
  <issue>1</issue>
  <fpage>288</fpage>
  <lpage>-299</lpage>
</bibl>

<bibl id="B19">
  <title><p>Scale-Free and Multifractal Time Dynamics of f{MRI} Signals during
  Rest and Task</p></title>
  <aug>
    <au><snm>Ciuciu</snm><fnm>P</fnm></au>
    <au><snm>Varoquaux</snm><fnm>G</fnm></au>
    <au><snm>Abry</snm><fnm>P</fnm></au>
    <au><snm>Sadaghiani</snm><fnm>S</fnm></au>
    <au><snm>Kleinschmidt</snm><fnm>A</fnm></au>
  </aug>
  <source>Frontiers in physiology</source>
  <pubdate>2012</pubdate>
  <volume>3</volume>
  <issue>Article 186</issue>
  <fpage>1</fpage>
  <lpage>-18</lpage>
</bibl>

<bibl id="B20">
  <title><p>Detection versus estimation in event-related f{MRI}: choosing the
  optimal stimulus timing</p></title>
  <aug>
    <au><snm>Birn</snm><fnm>R.</fnm></au>
    <au><snm>Cox</snm><fnm>R.W.</fnm></au>
    <au><snm>Bandettini</snm><fnm>P. A.</fnm></au>
  </aug>
  <source>{{N}euroimage}</source>
  <pubdate>2002</pubdate>
  <volume>15</volume>
  <issue>1</issue>
  <fpage>252</fpage>
  <lpage>-264</lpage>
</bibl>

<bibl id="B21">
  <title><p>What we can do and what we cannot do with f{MRI}</p></title>
  <aug>
    <au><snm>Logothetis</snm><fnm>N. K.</fnm></au>
  </aug>
  <source>Nature</source>
  <pubdate>2008</pubdate>
  <volume>453</volume>
  <issue>7197</issue>
  <fpage>869</fpage>
  <lpage>-878</lpage>
</bibl>

<bibl id="B22">
  <title><p>{Application of sensitivity-encoded echo-planar imaging for blood
  oxygen level-dependent functional brain imaging}</p></title>
  <aug>
    <au><snm>Zwart</snm><fnm>J.</fnm></au>
    <au><snm>Gelderen</snm><fnm>PV</fnm></au>
    <au><snm>Kellman</snm><fnm>P.</fnm></au>
    <au><snm>Duyn</snm><fnm>J. H.</fnm></au>
  </aug>
  <source>{{M}agnetic {R}esonance in {M}edicine}</source>
  <pubdate>2002</pubdate>
  <volume>48</volume>
  <issue>6</issue>
  <fpage>1011</fpage>
  <lpage>-20</lpage>
</bibl>

<bibl id="B23">
  <title><p>{Functional MRI using sensitivity-encoded echo planar imaging
  (SENSE-EPI)}</p></title>
  <aug>
    <au><snm>Preibisch</snm><fnm>C.</fnm></au>
  </aug>
  <source>Neuroimage</source>
  <pubdate>2003</pubdate>
  <volume>19</volume>
  <issue>2</issue>
  <fpage>412</fpage>
  <lpage>-421</lpage>
</bibl>

<bibl id="B24">
  <title><p>{Accelerated parallel imaging for functional imaging of the human
  brain}</p></title>
  <aug>
    <au><snm>Zwart</snm><fnm>J.</fnm></au>
    <au><snm>Gelderen</snm><fnm>PV</fnm></au>
    <au><snm>Golay</snm><fnm>X.</fnm></au>
    <au><snm>Ikonomidou</snm><fnm>V. N.</fnm></au>
    <au><snm>Duyn</snm><fnm>J. H.</fnm></au>
  </aug>
  <source>NMR Biomed</source>
  <pubdate>2006</pubdate>
  <volume>19</volume>
  <issue>3</issue>
  <fpage>342</fpage>
  <lpage>-51</lpage>
</bibl>

<bibl id="B25">
  <title><p>{Comparison of k-t SENSE/k-t BLAST with conventional SENSE applied
  to BOLD fMRI}</p></title>
  <aug>
    <au><snm>Utting</snm><fnm>J. F.</fnm></au>
    <au><snm>Kozerke</snm><fnm>S.</fnm></au>
    <au><snm>Schnitker</snm><fnm>R.</fnm></au>
    <au><snm>Niendorf</snm><fnm>T.</fnm></au>
  </aug>
  <source>{{J}ournal of {M}agnetic {R}esonance {I}maging}</source>
  <pubdate>2010</pubdate>
  <volume>32</volume>
  <issue>1</issue>
  <fpage>235</fpage>
  <lpage>-41</lpage>
</bibl>

<bibl id="B26">
  <title><p>Making better {SENSE}: wavelet denoising, {T}ikhonov
  regularization, and total least squares</p></title>
  <aug>
    <au><snm>Liang</snm><fnm>Z. P.</fnm></au>
    <au><snm>Bammer</snm><fnm>R.</fnm></au>
    <au><snm>Ji</snm><fnm>J.</fnm></au>
    <au><snm>Pelc</snm><fnm>N. J.</fnm></au>
    <au><snm>Glover</snm><fnm>G. H.</fnm></au>
  </aug>
  <source>{I}nternational {S}ociety for {M}agnetic {R}esonance in
  {M}edicine</source>
  <publisher>Hawa\"{\i}, USA</publisher>
  <pubdate>2002</pubdate>
  <fpage>2388</fpage>
</bibl>

<bibl id="B27">
  <title><p>{O}n {T}ikhonov Regularization for image reconstruction in parallel
  {MRI}</p></title>
  <aug>
    <au><snm>Ying</snm><fnm>L.</fnm></au>
    <au><snm>Xu</snm><fnm>D.</fnm></au>
    <au><snm>Liang</snm><fnm>Z. P.</fnm></au>
  </aug>
  <source>{IEEE E}ngineering in {M}edicine and {B}iology {S}ociety</source>
  <publisher>San Francisco, USA</publisher>
  <pubdate>2004</pubdate>
  <fpage>1056</fpage>
  <lpage>-1059</lpage>
</bibl>

<bibl id="B28">
  <title><p>Sparse{SENSE}: application of compressed sensing in parallel
  {MRI}</p></title>
  <aug>
    <au><snm>Zou</snm><fnm>Y. M.</fnm></au>
    <au><snm>Ying</snm><fnm>L.</fnm></au>
    <au><snm>Liu</snm><fnm>B.</fnm></au>
  </aug>
  <source>{IEEE I}nternational Conference on Technology and Applications in
  Biomedicine</source>
  <publisher>Shenzhen, China</publisher>
  <pubdate>2008</pubdate>
  <fpage>127</fpage>
  <lpage>-130</lpage>
</bibl>

<bibl id="B29">
  <title><p>{A}utocalibrated Parallel {MRI} Reconstruction in the Wavelet
  Domain</p></title>
  <aug>
    <au><snm>Chaari</snm><fnm>L.</fnm></au>
    <au><snm>Pesquet</snm><fnm>J. C.</fnm></au>
    <au><snm>Benazza Benyahia</snm><fnm>A.</fnm></au>
    <au><snm>Ciuciu</snm><fnm>P.</fnm></au>
  </aug>
  <source>{IEEE I}nternational {S}ymposium on {B}iomedical {I}maging
  (ISBI)</source>
  <publisher>Paris, France</publisher>
  <pubdate>2008</pubdate>
  <fpage>756</fpage>
  <lpage>-759</lpage>
</bibl>

<bibl id="B30">
  <title><p>Improved spiral {SENSE} reconstruction using a multiscale wavelet
  model</p></title>
  <aug>
    <au><snm>Liu</snm><fnm>B.</fnm></au>
    <au><snm>Abdelsalam</snm><fnm>E.</fnm></au>
    <au><snm>Sheng</snm><fnm>J.</fnm></au>
    <au></au>
    <au><snm>Ying</snm><fnm>L.</fnm></au>
  </aug>
  <source>{IEEE Int. Symp. on Biomed. Imag.}</source>
  <publisher>Paris, France</publisher>
  <pubdate>2008</pubdate>
  <fpage>1505</fpage>
  <lpage>1508</lpage>
</bibl>

<bibl id="B31">
  <title><p>A wavelet-based regularized reconstruction algorithm for {SENSE}
  parallel {MRI} with applications to neuroimaging</p></title>
  <aug>
    <au><snm>Chaari</snm><fnm>L.</fnm></au>
    <au><snm>Pesquet</snm><fnm>J. C.</fnm></au>
    <au><snm>Benazza Benyahia</snm><fnm>A.</fnm></au>
    <au><snm>Ciuciu</snm><fnm>P.</fnm></au>
  </aug>
  <source>Medical Image Analysis</source>
  <pubdate>2011</pubdate>
  <volume>15</volume>
  <issue>2</issue>
  <fpage>185</fpage>
  <lpage>-201</lpage>
</bibl>

<bibl id="B32">
  <title><p>Impact of the parallel imaging reconstruction algorithm on brain
  activity detection in f{MRI}</p></title>
  <aug>
    <au><snm>Chaari</snm><fnm>L.</fnm></au>
    <au><snm>M\'eriaux</snm><fnm>S.</fnm></au>
    <au><snm>Pesquet</snm><fnm>JC</fnm></au>
    <au><snm>Ciuciu</snm><fnm>P.</fnm></au>
  </aug>
  <source>International Symposium on Applied Sciences in Biomedical and
  Communication Technologies (ISABEL)</source>
  <publisher>Rome, Italy</publisher>
  <pubdate>2010</pubdate>
  <fpage>1</fpage>
  <lpage>5</lpage>
</bibl>

<bibl id="B33">
  <title><p>A {3D GRAPPA} algorithm for volumetric parallel imaging</p></title>
  <aug>
    <au><snm>Jakob</snm><fnm>P.</fnm></au>
    <au><snm>Griswold</snm><fnm>M.</fnm></au>
    <au><snm>Breuer</snm><fnm>F.</fnm></au>
    <au><snm>Blaimer</snm><fnm>M.</fnm></au>
    <au><snm>Seiberlich</snm><fnm>N.</fnm></au>
  </aug>
  <source>{Scientific Meeting International Society for Magnetic Resonance in
  Medicine}</source>
  <publisher>Seattle, USA</publisher>
  <pubdate>2006</pubdate>
  <fpage>286</fpage>
</bibl>

<bibl id="B34">
  <title><p>Empirical analysis of {BOLD} f{MRI} statistics. {II}. {S}patially
  Smoothed Data Collected under Null-Hypothesis and Experimental
  Conditions</p></title>
  <aug>
    <au><snm>Aguirre</snm><fnm>G. K.</fnm></au>
    <au><snm>Zarahn</snm><fnm>E.</fnm></au>
    <au><snm>D'Esposito</snm><fnm>M.</fnm></au>
  </aug>
  <source>Neuroimage</source>
  <pubdate>1997</pubdate>
  <volume>5</volume>
  <issue>3</issue>
  <fpage>199</fpage>
  <lpage>-212</lpage>
</bibl>

<bibl id="B35">
  <title><p>Empirical analysis of {BOLD} f{MRI} statistics. {I}. {S}patially
  unsmoothed data collected under null-hypothesis conditions</p></title>
  <aug>
    <au><snm>Zarahn</snm><fnm>E.</fnm></au>
    <au><snm>Aguirre</snm><fnm>G. K.</fnm></au>
    <au><snm>D'Esposito</snm><fnm>M.</fnm></au>
  </aug>
  <source>Neuroimage</source>
  <pubdate>1997</pubdate>
  <volume>5</volume>
  <issue>3</issue>
  <fpage>179</fpage>
  <lpage>-197</lpage>
</bibl>

<bibl id="B36">
  <title><p>Effect of temporal autocorrelation due to physiological noise and
  stimulus paradigm on voxel-level false-positive rates in f{MRI}.</p></title>
  <aug>
    <au><snm>Purdon</snm><fnm>P. L.</fnm></au>
    <au><snm>Weisskoff</snm><fnm>R. M.</fnm></au>
  </aug>
  <source>{Human Brain Mapping}</source>
  <pubdate>1998</pubdate>
  <volume>6</volume>
  <issue>4</issue>
  <fpage>239</fpage>
  <lpage>-249</lpage>
</bibl>

<bibl id="B37">
  <title><p>Temporal autocorrelation in univariate linear modelling of f{MRI}
  data</p></title>
  <aug>
    <au><snm>Woolrich</snm><fnm>M.</fnm></au>
    <au><snm>Ripley</snm><fnm>B.</fnm></au>
    <au><snm>Brady</snm><fnm>M.</fnm></au>
    <au><snm>Smith</snm><fnm>S.</fnm></au>
  </aug>
  <source>Neuroimage</source>
  <pubdate>2001</pubdate>
  <volume>14</volume>
  <issue>6</issue>
  <fpage>1370</fpage>
  <lpage>-1386</lpage>
</bibl>

<bibl id="B38">
  <title><p>A general statistical analysis for f{MRI} data</p></title>
  <aug>
    <au><snm>Worsley</snm><fnm>K. J.</fnm></au>
    <au><snm>Liao</snm><fnm>C. H.</fnm></au>
    <au><snm>Aston</snm><fnm>J.</fnm></au>
    <au><snm>Petre</snm><fnm>V.</fnm></au>
    <au><snm>Duncan</snm><fnm>G. H.</fnm></au>
    <au><snm>Morales</snm><fnm>F.</fnm></au>
    <au><snm>Evans</snm><fnm>A. C.</fnm></au>
  </aug>
  <source>Neuroimage</source>
  <pubdate>2002</pubdate>
  <volume>15</volume>
  <issue>1</issue>
  <fpage>1</fpage>
  <lpage>-15</lpage>
</bibl>

<bibl id="B39">
  <title><p>Variational {B}ayesian inference for f{MRI} time series</p></title>
  <aug>
    <au><snm>Penny</snm><fnm>W. D.</fnm></au>
    <au><snm>Kiebel</snm><fnm>S.</fnm></au>
    <au><snm>Friston</snm><fnm>K. J.</fnm></au>
  </aug>
  <source>Neuroimage</source>
  <pubdate>2003</pubdate>
  <volume>19</volume>
  <issue>3</issue>
  <fpage>727</fpage>
  <lpage>-741</lpage>
</bibl>

<bibl id="B40">
  <title><p>Fast joint detection-estimation of evoked brain activity in
  event-related f{MRI} using a variational approach</p></title>
  <aug>
    <au><snm>Chaari</snm><fnm>L.</fnm></au>
    <au><snm>Vincent</snm><fnm>T.</fnm></au>
    <au><snm>Forbes</snm><fnm>F.</fnm></au>
    <au><snm>Dojat</snm><fnm>M.</fnm></au>
    <au><snm>Ciuciu</snm><fnm>P.</fnm></au>
  </aug>
  <source>IEEE Transactions on Medical Imaging</source>
  <pubdate>2013</pubdate>
  <volume>32</volume>
  <issue>5</issue>
  <fpage>821</fpage>
  <lpage>837</lpage>
</bibl>

<bibl id="B41">
  <title><p>A proximal decomposition method for solving convex variational
  inverse problems</p></title>
  <aug>
    <au><snm>Combettes</snm><fnm>P. L.</fnm></au>
    <au><snm>Pesquet</snm><fnm>J. C.</fnm></au>
  </aug>
  <source>Inverse Problems</source>
  <pubdate>2008</pubdate>
  <volume>24</volume>
  <issue>6</issue>
  <fpage>27</fpage>
</bibl>

<bibl id="B42">
  <title><p>A Fast Wavelet-Based Reconstruction Method for Magnetic Resonance
  Imaging</p></title>
  <aug>
    <au><snm>Guerquin Kern</snm><fnm>M.</fnm></au>
    <au><snm>Haberlin</snm><fnm>M.</fnm></au>
    <au><snm>Pruessmann</snm><fnm>K. P.</fnm></au>
    <au><snm>Unser</snm><fnm>M.</fnm></au>
  </aug>
  <source>IEEE Transactions on Medical Imaging</source>
  <pubdate>2011</pubdate>
  <volume>30</volume>
  <issue>9</issue>
  <fpage>1649</fpage>
  <lpage>1660</lpage>
</bibl>

<bibl id="B43">
  <title><p>Tailored SMASH Image Reconstructions for Robust In Vivo Parallel
  {MR} Imaging</p></title>
  <aug>
    <au><snm>Sodickson</snm><fnm>D. K.</fnm></au>
  </aug>
  <source>{M}agnetic {R}esonance in {M}edicine</source>
  <pubdate>2000</pubdate>
  <volume>44</volume>
  <issue>2</issue>
  <fpage>243–251</fpage>
</bibl>

<bibl id="B44">
  <title><p>Total variation based convex filters for medical
  imaging</p></title>
  <aug>
    <au><snm>Keeling</snm><fnm>S. L.</fnm></au>
  </aug>
  <source>{A}pplied {M}athematics and {C}omputation</source>
  <pubdate>2003</pubdate>
  <volume>139</volume>
  <issue>1</issue>
  <fpage>101</fpage>
  <lpage>-119</lpage>
</bibl>

<bibl id="B45">
  <title><p>Regularized sensitivity encoding ({SENSE}) reconstruction using
  {B}regman iterations</p></title>
  <aug>
    <au><snm>Liu</snm><fnm>B.</fnm></au>
    <au><snm>King</snm><fnm>K.</fnm></au>
    <au><snm>Steckner</snm><fnm>M.</fnm></au>
    <au><snm>Xie</snm><fnm>J.</fnm></au>
    <au><snm>Sheng</snm><fnm>J.</fnm></au>
    <au><snm>Ying</snm><fnm>L.</fnm></au>
  </aug>
  <source>Magnetic Resonance in Medicine</source>
  <pubdate>2008</pubdate>
  <volume>61</volume>
  <issue>1</issue>
  <fpage>145</fpage>
  <lpage>152</lpage>
</bibl>

<bibl id="B46">
  <title><p>Improved Time Series Reconstruction for Dynamic Magnetic Resonance
  Imaging</p></title>
  <aug>
    <au><snm>S{\"u}mb{\"u}l</snm><fnm>U.</fnm></au>
    <au><snm>Santos</snm><fnm>J. M.</fnm></au>
    <au><snm>Pauly</snm><fnm>J. M.</fnm></au>
  </aug>
  <source>IEEE Transactions on Medical Imaging</source>
  <pubdate>2009</pubdate>
  <volume>28</volume>
  <issue>7</issue>
  <fpage>1093</fpage>
  <lpage>-1104</lpage>
</bibl>

<bibl id="B47">
  <title><p>Fast reproducible identification and large-scale databasing of
  individual functional cognitive networks</p></title>
  <aug>
    <au><snm>Pinel</snm><fnm>P.</fnm></au>
    <au><snm>Thirion</snm><fnm>B.</fnm></au>
    <au><snm>M{\'e}riaux</snm><fnm>S.</fnm></au>
    <au><snm>Jobert</snm><fnm>A.</fnm></au>
    <au><snm>Serres</snm><fnm>J.</fnm></au>
    <au><snm>{Le Bihan}</snm><fnm>D.</fnm></au>
    <au><snm>Poline</snm><fnm>J. B.</fnm></au>
    <au><snm>Dehaene</snm><fnm>S.</fnm></au>
  </aug>
  <source>BMC Neuroscience</source>
  <pubdate>2007</pubdate>
  <volume>8</volume>
  <issue>1</issue>
  <fpage>91</fpage>
</bibl>

<bibl id="B48">
  <title><p>{T}en {L}ectures on {W}avelets</p></title>
  <aug>
    <au><snm>Daubechies</snm><fnm>I.</fnm></au>
  </aug>
  <publisher>Philadelphia: {S}ociety for {I}ndustrial and {A}pplied
  {M}athematics</publisher>
  <pubdate>1992</pubdate>
</bibl>

<bibl id="B49">
  <title><p>Cerebral bases of number processing and calculation</p></title>
  <aug>
    <au><snm>Dehaene</snm><fnm>S.</fnm></au>
  </aug>
  <source>The New Cognitive Neurosciences</source>
  <publisher>Cambridge,: MIT Press</publisher>
  <editor>M. Gazzaniga</editor>
  <section><title><p>68</p></title></section>
  <pubdate>1999</pubdate>
  <fpage>987</fpage>
  <lpage>-998</lpage>
</bibl>

<bibl id="B50">
  <title><p>{Controlling the Familywise Error Rate in Functional Neuroimaging:
  A Comparative Review}</p></title>
  <aug>
    <au><snm>Nichols</snm><fnm>T. E.</fnm></au>
    <au><snm>Hayasaka</snm><fnm>S.</fnm></au>
  </aug>
  <source>Statistical Methods in Medical Research</source>
  <pubdate>2003</pubdate>
  <volume>12</volume>
  <issue>5</issue>
  <fpage>419</fpage>
  <lpage>-446</lpage>
</bibl>

<bibl id="B51">
  <title><p>Introduction to Random Field Theory</p></title>
  <aug>
    <au><snm>Brett</snm><fnm>M.</fnm></au>
    <au><snm>Penny</snm><fnm>W.</fnm></au>
    <au><snm>Kiebel</snm><fnm>S.</fnm></au>
  </aug>
  <source>Human Brain Function</source>
  <publisher>Academic Press</publisher>
  <editor>R. S. J. Frackowiak and K. J. Friston and C. D. Fritch and R. J.
  Dolan and C. J. Price and W. D. Penny</editor>
  <edition>2</edition>
  <pubdate>2004</pubdate>
  <fpage>867</fpage>
  <lpage>-880</lpage>
</bibl>

<bibl id="B52">
  <title><p>A Hierarchical {B}ayesian Model For Frame
  Representation</p></title>
  <aug>
    <au><snm>Chaari</snm><fnm>L.</fnm></au>
    <au><snm>Pesquet</snm><fnm>J. C.</fnm></au>
    <au><snm>Tourneret</snm><fnm>J. Y.</fnm></au>
    <au><snm>Ciuciu</snm><fnm>P.</fnm></au>
    <au><snm>Benazza Benyahia</snm><fnm>A.</fnm></au>
  </aug>
  <source>IEEE Transactions on Signal Processing</source>
  <pubdate>2010</pubdate>
  <volume>58</volume>
  <issue>11</issue>
  <fpage>5560</fpage>
  <lpage>5571</lpage>
</bibl>

<bibl id="B53">
  <title><p>A Four-Dimensional Registration Algorithm With Application to Joint
  Correction of Motion and Slice Timing in f{MRI}</p></title>
  <aug>
    <au><snm>Roche</snm><fnm>A.</fnm></au>
  </aug>
  <source>IEEE Transactions on Medical Imaging</source>
  <pubdate>2011</pubdate>
  <volume>30</volume>
  <issue>8</issue>
  <fpage>1546</fpage>
  <lpage>1554</lpage>
</bibl>

<bibl id="B54">
  <title><p>A fully {B}ayesian approach to the parcel-based
  detection-estimation of brain activity in {fMRI}</p></title>
  <aug>
    <au><snm>Makni</snm><fnm>S.</fnm></au>
    <au><snm>Idier</snm><fnm>J.</fnm></au>
    <au><snm>Vincent</snm><fnm>T.</fnm></au>
    <au><snm>Thirion</snm><fnm>B.</fnm></au>
    <au><snm>Dehaene Lambertz</snm><fnm>G.</fnm></au>
    <au><snm>Ciuciu</snm><fnm>P.</fnm></au>
  </aug>
  <source>{{N}euroimage}</source>
  <pubdate>2008</pubdate>
  <volume>41</volume>
  <issue>3</issue>
  <fpage>941</fpage>
  <lpage>-969</lpage>
</bibl>

<bibl id="B55">
  <title><p>Spatially adaptive mixture modeling for analysis of {within-subject
  fMRI} time series</p></title>
  <aug>
    <au><snm>Vincent</snm><fnm>T.</fnm></au>
    <au><snm>Risser</snm><fnm>L.</fnm></au>
    <au><snm>Ciuciu</snm><fnm>P.</fnm></au>
  </aug>
  <source>IEEE Transactions on Medical Imaging</source>
  <pubdate>2010</pubdate>
  <volume>29</volume>
  <issue>4</issue>
  <fpage>1059</fpage>
  <lpage>-1074</lpage>
</bibl>

<bibl id="B56">
  <title><p>Group-level impacts of within- and between-subject hemodynamic
  variability in {fMRI}</p></title>
  <aug>
    <au><snm>Badillo</snm><fnm>S.</fnm></au>
    <au><snm>Vincent</snm><fnm>T.</fnm></au>
    <au><snm>Ciuciu</snm><fnm>P.</fnm></au>
  </aug>
  <source>NeuroImage</source>
  <pubdate>2013</pubdate>
  <volume>82</volume>
  <fpage>433</fpage>
  <lpage>448</lpage>
</bibl>

<bibl id="B57">
  <title><p>{WSPM}: Wavelet-based statistical parametric mapping</p></title>
  <aug>
    <au><snm>Van De Ville</snm><fnm>D.</fnm></au>
    <au><snm>Seghier</snm><fnm>M.</fnm></au>
    <au><snm>Lazeyras</snm><fnm>F.</fnm></au>
    <au><snm>Blu</snm><fnm>T.</fnm></au>
    <au><snm>Unser</snm><fnm>M.</fnm></au>
  </aug>
  <source>Neuroimage</source>
  <pubdate>2007</pubdate>
  <volume>37</volume>
  <issue>4</issue>
  <fpage>1205</fpage>
  <lpage>-1217</lpage>
</bibl>

<bibl id="B58">
  <title><p>Proximit{\'e} et dualit{\'e} dans un espace hilbertien</p></title>
  <aug>
    <au><snm>Moreau</snm><fnm>J. J.</fnm></au>
  </aug>
  <source>{B}ulletin de la {S}oci{\'e}t{\'e} {M}ath{\'e}matique de
  {F}rance</source>
  <pubdate>1965</pubdate>
  <volume>93</volume>
  <fpage>273</fpage>
  <lpage>-299</lpage>
</bibl>

<bibl id="B59">
  <title><p>A variational formulation for frame-based inverse
  problems</p></title>
  <aug>
    <au><snm>Chaux</snm><fnm>C.</fnm></au>
    <au><snm>Combettes</snm><fnm>P.</fnm></au>
    <au><snm>Pesquet</snm><fnm>J. C.</fnm></au>
    <au><snm>Wajs</snm><fnm>V. R.</fnm></au>
  </aug>
  <source>{I}nverse {P}roblems</source>
  <pubdate>2007</pubdate>
  <volume>23</volume>
  <issue>4</issue>
  <fpage>1495</fpage>
  <lpage>-1518</lpage>
</bibl>

<bibl id="B60">
  <title><p>Signal Recovery by proximal forward-backward splitting</p></title>
  <aug>
    <au><snm>Combettes</snm><fnm>P. L.</fnm></au>
    <au><snm>Wajs</snm><fnm>V. R.</fnm></au>
  </aug>
  <source>{M}ultiscale {M}odeling and {S}imulation</source>
  <pubdate>2005</pubdate>
  <volume>4</volume>
  <fpage>1168</fpage>
  <lpage>-1200</lpage>
</bibl>

<bibl id="B61">
  <title><p>Proximal splitting methods in signal processing</p></title>
  <aug>
    <au><snm>Combettes</snm><fnm>P. L.</fnm></au>
    <au><snm>Pesquet</snm><fnm>J. C.</fnm></au>
  </aug>
  <source>Fixed-Point Algorithms for Inverse Problems in Science and
  Engineering</source>
  <publisher>New York: {Springer Verlag}</publisher>
  <editor>H. H. Bauschke and R. Burachik and P. L. Combettes and V. Elser and
  D. R. Luke and H. Wolkowicz</editor>
  <section><title><p>1</p></title></section>
  <pubdate>2010</pubdate>
  <fpage>185</fpage>
  <lpage>212</lpage>
</bibl>

<bibl id="B62">
  <title><p>A {D}ouglas-{R}achford Splitting Approach to Nonsmooth Convex
  Variational Signal Recovery</p></title>
  <aug>
    <au><snm>Combettes</snm><fnm>P. L.</fnm></au>
    <au><snm>Pesquet</snm><fnm>J. C.</fnm></au>
  </aug>
  <source>IEEE Journal of Selected Topics in Signal Processing</source>
  <pubdate>2007</pubdate>
  <volume>1</volume>
  <issue>4</issue>
  <fpage>564</fpage>
  <lpage>-574</lpage>
</bibl>

<bibl id="B63">
  <title><p>Maximum likelihood from incomplete data via the {EM} algorithm
  (with discussion)</p></title>
  <aug>
    <au><snm>Dempster</snm><fnm>A. P.</fnm></au>
    <au><snm>Laird</snm><fnm>A. P.</fnm></au>
    <au><snm>Rubin</snm><fnm>D. B.</fnm></au>
  </aug>
  <source>Journal of the Royal Statistical Society, Series B</source>
  <pubdate>1977</pubdate>
  <volume>39</volume>
  <fpage>1</fpage>
  <lpage>38</lpage>
</bibl>

<bibl id="B64">
  <title><p>A simple technique to estimate partition functions and equilibrium
  constants from {M}onte {C}arlo simulations</p></title>
  <aug>
    <au><snm>Vieth</snm><fnm>M.</fnm></au>
    <au><snm>Kolinski</snm><fnm>A.</fnm></au>
    <au><snm>Skolnick</snm><fnm>J.</fnm></au>
  </aug>
  <source>Journal of Chemical Physics</source>
  <pubdate>1995</pubdate>
  <volume>102</volume>
  <fpage>6189</fpage>
  <lpage>6193</lpage>
</bibl>

<bibl id="B65">
  <title><p>Robust extrapolation scheme for fast estimation of {3D I}sing field
  partition functions. Application to within-subject f{MRI} data
  analysis.</p></title>
  <aug>
    <au><snm>Risser</snm><fnm>L.</fnm></au>
    <au><snm>Vincent</snm><fnm>T.</fnm></au>
    <au><snm>Ciuciu</snm><fnm>P.</fnm></au>
    <au><snm>Idier</snm><fnm>J.</fnm></au>
  </aug>
  <source>12th{P}roc. Medical Image Computing and Computer Assisted
  Intervention</source>
  <publisher>London, UK: Springer Verlag Berlin Heidelberg</publisher>
  <pubdate>2009</pubdate>
  <fpage>975</fpage>
  <lpage>-983</lpage>
</bibl>

<bibl id="B66">
  <title><p>{Min-max} extrapolation scheme for fast estimation of {3D Potts}
  field partition functions. Application to the joint detection-estimation of
  brain activity in {fMRI}.</p></title>
  <aug>
    <au><snm>Risser</snm><fnm>L.</fnm></au>
    <au><snm>Vincent</snm><fnm>T.</fnm></au>
    <au><snm>Forbes</snm><fnm>F.</fnm></au>
    <au><snm>Idier</snm><fnm>J.</fnm></au>
    <au><snm>Ciuciu</snm><fnm>P.</fnm></au>
  </aug>
  <source>{{J}ournal of {S}ignal {P}rocessing {S}ystems}</source>
  <pubdate>2011</pubdate>
  <volume>65</volume>
  <issue>3</issue>
  <fpage>325</fpage>
  <lpage>-338</lpage>
</bibl>

<bibl id="B67">
  <title><p>Hyperparameter estimation for satellite image restoration using a
  {MCMC} maximum likelihood method</p></title>
  <aug>
    <au><snm>Jalobeanu</snm><fnm>A.</fnm></au>
    <au><snm>Blanc F\'eraud</snm><fnm>L.</fnm></au>
    <au><snm>Zerubia</snm><fnm>J.</fnm></au>
  </aug>
  <source>Pattern Recognition</source>
  <pubdate>2002</pubdate>
  <volume>35</volume>
  <issue>2</issue>
</bibl>

<bibl id="B68">
  <title><p>A nonlinear {S}tein based estimator for multichannel image
  denoising</p></title>
  <aug>
    <au><snm>Chaux</snm><fnm>C.</fnm></au>
    <au><snm>Duval</snm><fnm>L.</fnm></au>
    <au><snm>Benazza Benyahia</snm><fnm>A.</fnm></au>
    <au><snm>Pesquet</snm><fnm>J. C.</fnm></au>
  </aug>
  <source>{IEEE T}ransactions on {S}ignal {P}rocessing</source>
  <pubdate>2008</pubdate>
  <volume>56</volume>
  <issue>8</issue>
  <fpage>3855</fpage>
  <lpage>3870</lpage>
</bibl>

<bibl id="B69">
  <title><p>Nonlinear programming, Second Edition</p></title>
  <aug>
    <au><snm>Bertsekas</snm><fnm>D. P.</fnm></au>
  </aug>
  <publisher>Belmont, {USA}: Athena Scientific</publisher>
  <pubdate>1995</pubdate>
</bibl>

</refgrp>
} 
